\documentstyle[amssym]{mn}

\font\japit = cmti10 at 10truept
\input{epsf.sty}

\title
     [Uncorrelated Modes of the Power Spectrum]
{\vglue-3.0truecm
\centerline{\japit Published in Monthly Notices}
\vglue 2.5truecm
\noindent
Uncorrelated Modes of the Nonlinear Power Spectrum
\author[A. J. S. Hamilton]
     {A. J. S. Hamilton \\
	JILA and Dept.\ Astrophysical \& Planetary Sciences,
	Box 440, U. Colorado, Boulder CO 80309, USA; \\
	\ Andrew.Hamilton@Colorado.EDU; http:$/\!/$casa.colorado.edu/$\sim$ajsh}
}

\newcommand{\rmn}{\rm}
\newcommand{\bmi}{\bmath}

\newcommand{\mx}{\sf}		

\newcommand{\be}{\begin{equation}}
\newcommand{\ee}{\end{equation}}
\newcommand{\ba}{\begin{eqnarray}}
\newcommand{\ea}{\end{eqnarray}}
\newcommand{\nn}{\nonumber \\}
\newcommand{\dd}{{\rmn d}}	
\newcommand{\e}{{\rmn e}}	
\newcommand{\im}{{\rmn i}}	
\newcommand{\PI}{{\rmn \pi}}	
\newcommand{\Partial}{{\rmn \partial}}	
\newcommand{\ddd}{\dd^3\!}

\newcommand{\Sym}[1]{\!\begin{array}[t]{c} {\rmn Sym} \\
                  [-1ex] \scriptstyle #1 \end{array}\!}

\newcommand{\transpose}{\top}
\newcommand{\skipp}{\vskip2pt \noindent}

\newcommand{\deltaD}{\delta_{3D}}
\newcommand{\nbar}{\bar n}

\newcommand{\k}{{\bmi k}}

\newcommand{\r}{{\bmi r}}

\newcommand{\cL}{{\cal L}}

\newcommand{\Arg}{{\rmn Arg}}
\newcommand{\FKP}{{\rmn FKP}}
\newcommand{\Mpc}{{\rmn Mpc}}

\newcommand{\true}{{\rmn true}}

\newcommand{\fB}{{\frak B}}
\newcommand{\fC}{{\frak C}}

\newcommand{\1}{{\mx 1}}
\newcommand{\bA}{{\mx A}}

\newcommand{\bE}{{\mx E}}

\newcommand{\bR}{{\mx R}}

\newcommand{\bZ}{{\mx Z}}
\newcommand{\vZ}{{\sf Z}}

\newcommand{\aap}[2]{A\&A, #1, #2}
\newcommand{\aaps}[2]{A\&AS, #1, #2}
\newcommand{\aj}[2]{AJ, #1, #2}
\newcommand{\apj}[2]{ApJ, #1, #2}
\newcommand{\apjs}[2]{ApJS, #1, #2}
\newcommand{\araa}[2]{ARAA, #1, #2}
\newcommand{\mn}[2]{MNRAS, #1, #2}
\newcommand{\pasp}[2]{PASP, #1, #2}
\newcommand{\prl}[2]{PRL, #1, #2}
\newcommand{\pr}[2]{Phys.\ Rev., #1, #2}

\newcommand{\ptsfig}{
    \begin{figure}
    \begin{center}
    \leavevmode
    \epsfxsize=2in
    \epsfbox{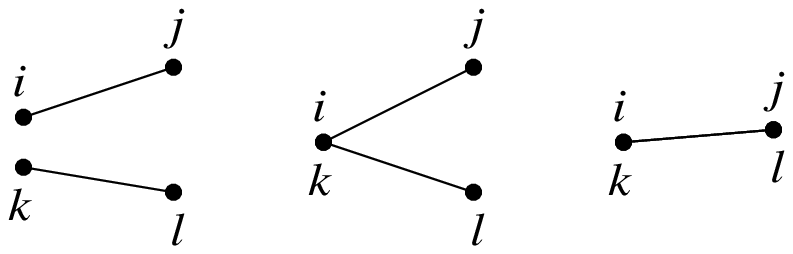}
    \end{center}
    \caption[1]{
Schematic illustration of
the 4-point, 3-point, and 2-point contributions
to the covariance $\fC_{ijkl}$ of pairs $ij$ with other pairs $kl$.
The 3-point and 2-point contributions are shot-noise contributions
in which one or both galaxies of the pair $ij$ are the same
as one or both of the pair $kl$.
    \label{pts}
    }
    \end{figure}
}

\newcommand{\Mfig}{
    \begin{figure}
    \begin{center}
    \leavevmode
    \epsfxsize=3in
    \epsfbox{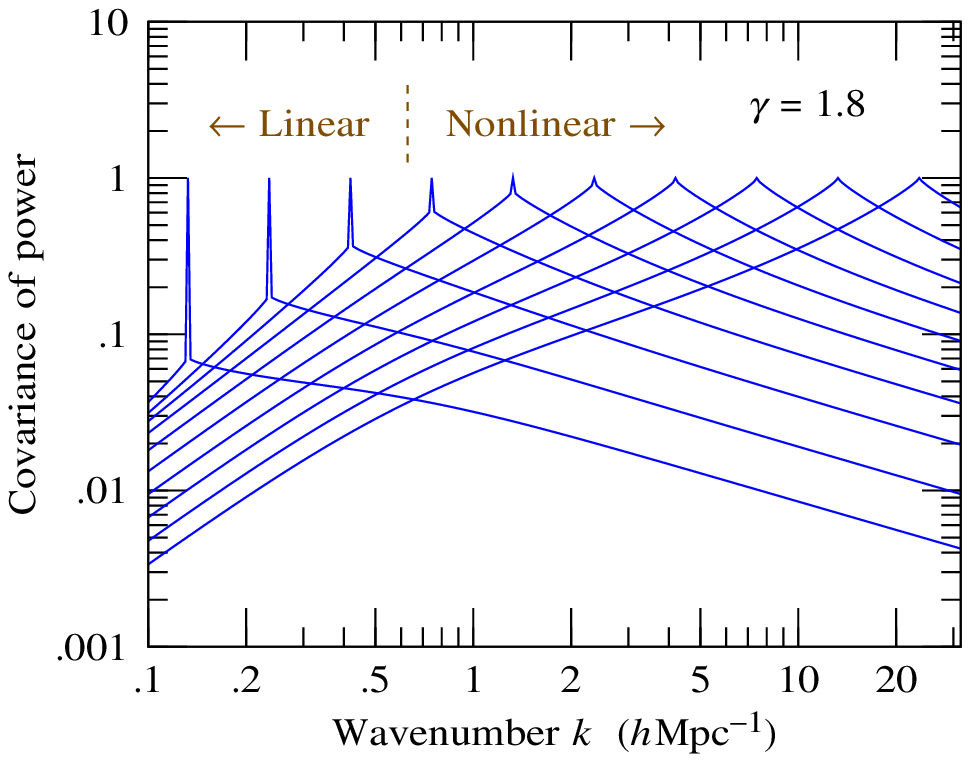}
    \end{center}
    \caption[1]{
Correlation coefficient
$K(k_\alpha,$\protect\discretionary{}{}{}$k_\beta)/$\protect\discretionary{}{}{}$[K(k_\alpha,$\protect\discretionary{}{}{}$k_\alpha) $\protect\discretionary{}{}{}$K(k_\beta,$\protect\discretionary{}{}{}$k_\beta)]^{1/2}$
of the 4-point contribution $K(k_\alpha,k_\beta)$ to the covariance
of the power
(i.e.\ the covariance without shot-noise)
in the case of a power law power spectrum with correlation function
$\xi(r) = (r/5 \, h^{-1} \Mpc)^{- 1.8}$.
Each line is the correlation coefficient for a fixed $k_\beta$,
and each line peaks at $k_\alpha = k_\beta$, whereat the value is unity.
The hierarchical amplitudes are $R_a = - R_b = 4.195$.
The resolution is $128$ points per decade, $\Delta\log k = 1/128$.
    \label{M}
    }
    \end{figure}
}

\newcommand{\Mwfig}{
    \begin{figure}
    \begin{center}
    \leavevmode
    \epsfxsize=3in
    \epsfbox{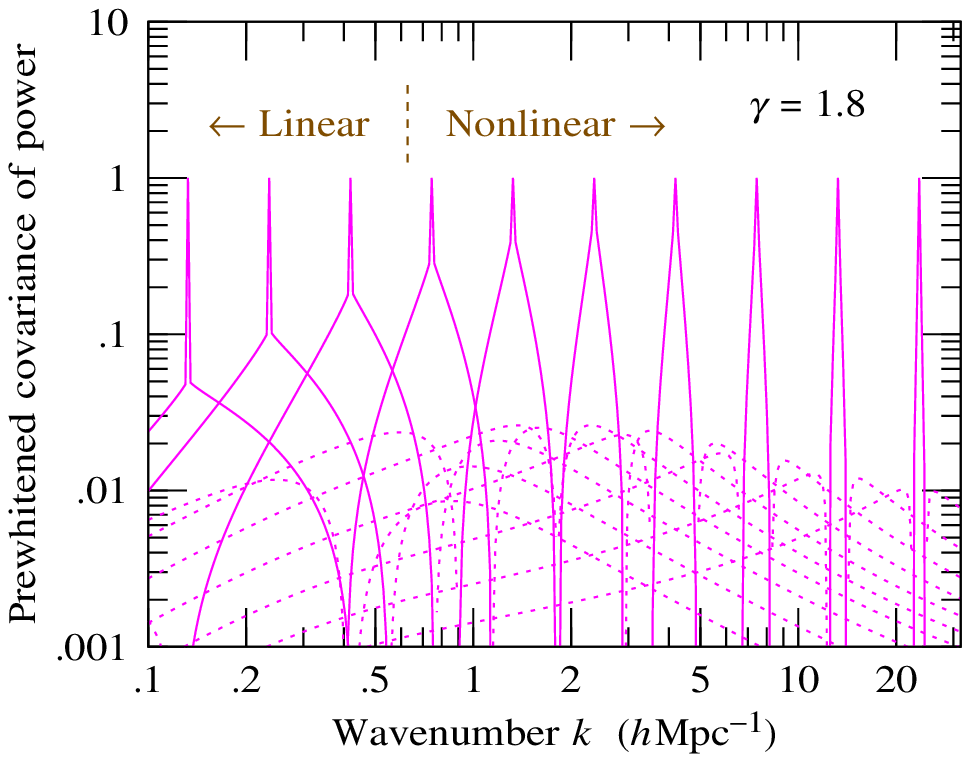}
    \end{center}
    \caption[1]{\small
Correlation coefficient
$M(k_\alpha,k_\beta)/$\protect\discretionary{}{}{}$[M(k_\alpha,k_\alpha)$\protect\discretionary{}{}{}$M(k_\beta,$\protect\discretionary{}{}{}$k_\beta)]^{1/2}$
of the 4-point contribution $M(k_\alpha,k_\beta)$ to the prewhitened covariance
of a power law power spectrum with correlation function
$\xi(r) = (r/5 \, h^{-1} \Mpc)^{- 1.8}$.
Lines are dotted where the correlation coefficient is negative.
This is the same as Figure~\protect\ref{M},
except that the covariance is prewhitened.
    }
    \label{Mw}
    \end{figure}
}

\newcommand{\Msfig}{
    \begin{figure*}
    \begin{minipage}{175mm}
    \begin{center}
    \leavevmode
    \epsfxsize=6in
    \epsfbox{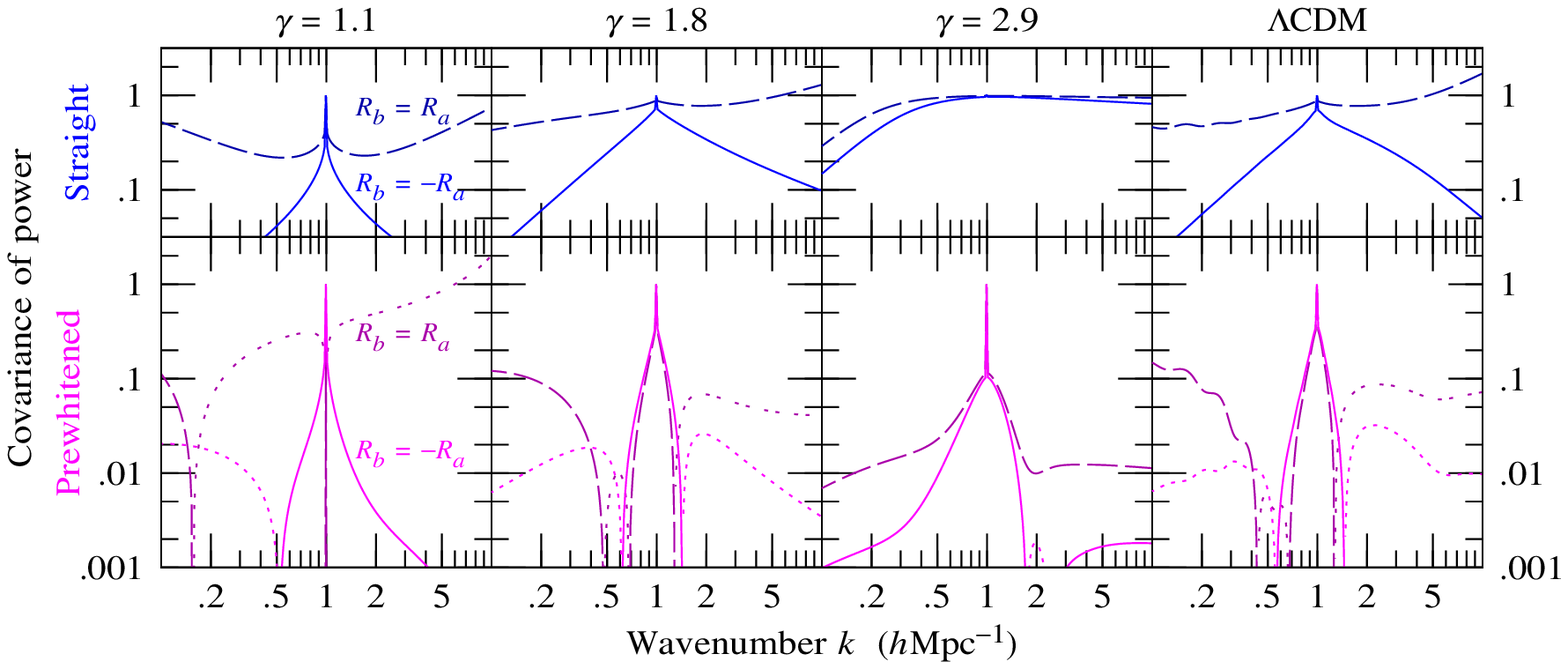}
    \end{center}
    \caption[1]{\small
Correlation coefficients
(top)
$K(k_\alpha,k_\beta)/$\protect\discretionary{}{}{}$[K(k_\alpha,k_\alpha) $\protect\discretionary{}{}{}$K(k_\beta,k_\beta)]^{1/2}$
of the covariance,
and (bottom)
$M(k_\alpha,k_\beta)/$\protect\discretionary{}{}{}$[M(k_\alpha,k_\alpha) $\protect\discretionary{}{}{}$M(k_\beta,k_\beta)]^{1/2}$
of the prewhitened covariance,
of the power spectrum in four different models of the power spectrum.
Each curve is the correlation coefficient at fixed
$k_\beta = 1 \, h \, \Mpc^{-1}$, plotted as a function of $k_\alpha$.
The three sets of panels starting from the left are for power law power spectra
with correlation functions $\xi(r) = (r/5 \, h^{-1} \Mpc)^{-\gamma}$
with indices $\gamma = 1.1$, $1.8$, and $2.9$,
while the rightmost panel is for the $\Lambda$CDM power spectrum of
Eisenstein \& Hu (1998) with
$\Omega_\Lambda = 0.7$,
$\Omega_m = 0.3$,
$\Omega_b h^2 = 0.02$,
and $h = 0.65$,
nonlinearly evolved by the procedure of Peacock \& Dodds (1996).
The two lines on each graph are for 4-point hierarchical amplitudes
(solid) $R_b = - R_a$,
and (long-dash) $R_b = R_a$.
Lines are dotted where the correlation coefficient is negative.
The Schwarz inequality, which requires that the correlation coefficient
be $\leq 1$, is violated by the hierarchical model with $R_b = R_a$
at values $k \ll k'$ and $k \gg k'$.
The resolution is $128$ points per decade,
the same as in Figures~\protect\ref{M} and \protect\ref{Mw}.
    \label{Ms}
    }
    \end{minipage}
    \end{figure*}
}

\newcommand{\Mwthreefig}{
    \begin{figure}
    \begin{center}
    \leavevmode
    \epsfxsize=3in
    \epsfbox{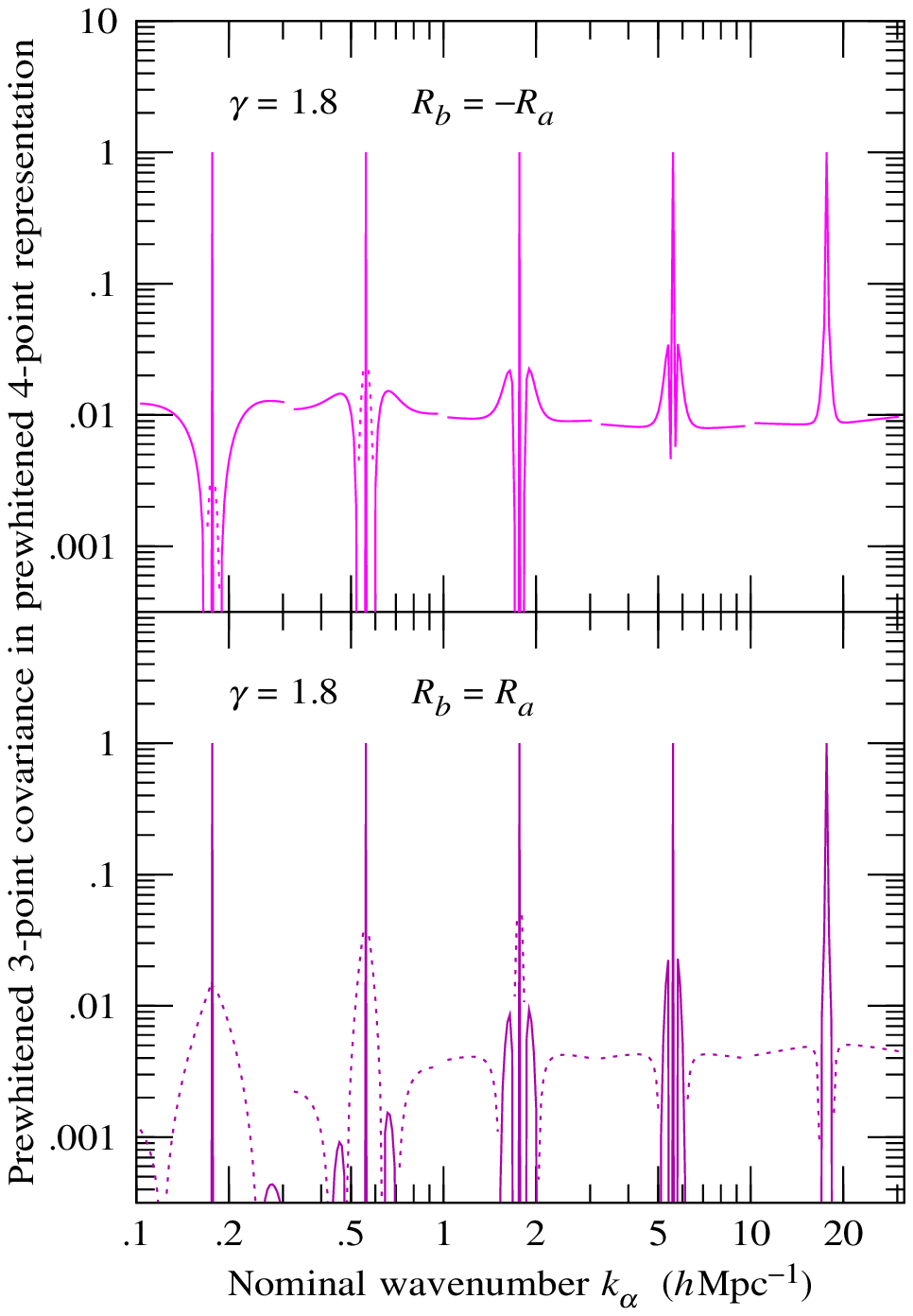}
    \end{center}
    \caption[1]{\small
Correlation coefficient
$L_{\alpha\beta}/$\protect\discretionary{}{}{}$(L_{\alpha\alpha} $\protect\discretionary{}{}{}$L_{\beta\beta})^{1/2}$
of the prewhitened 3-point covariance $L_{\alpha\beta}$
in the representation of eigenfunctions $\phi_\alpha$
of the prewhitened 4-point covariance $M_{\alpha\beta}$,
for a power law power spectrum with correlation function
$\xi(r) = (r/5 \, h^{-1} \Mpc)^{- 1.8}$.
Each line is the correlation coefficient at a fixed nominal wavenumber
$k_\beta$, plotted against the nominal wavenumber $k_\alpha$,
which labels the 4-point eigenfunctions $\phi_\alpha$ ordered by eigenvalue.
Each line peaks at $k_\alpha = k_\beta$,
whereat the correlation coefficient is unity.
The upper panel is
for 4-point hierarchical amplitudes $R_b = - R_a$;
the lower panel is for $R_b = R_a$.
Lines are dotted where the correlation coefficient is negative.
The resolution is $\Delta\log k = 1/128$.
    \label{Mw3}
    }
    \end{figure}
}

\newcommand{\Mwthreesfig}{
    \begin{figure*}
    \begin{minipage}{175mm}
    \begin{center}
    \leavevmode
    \epsfxsize=6in
    \epsfbox{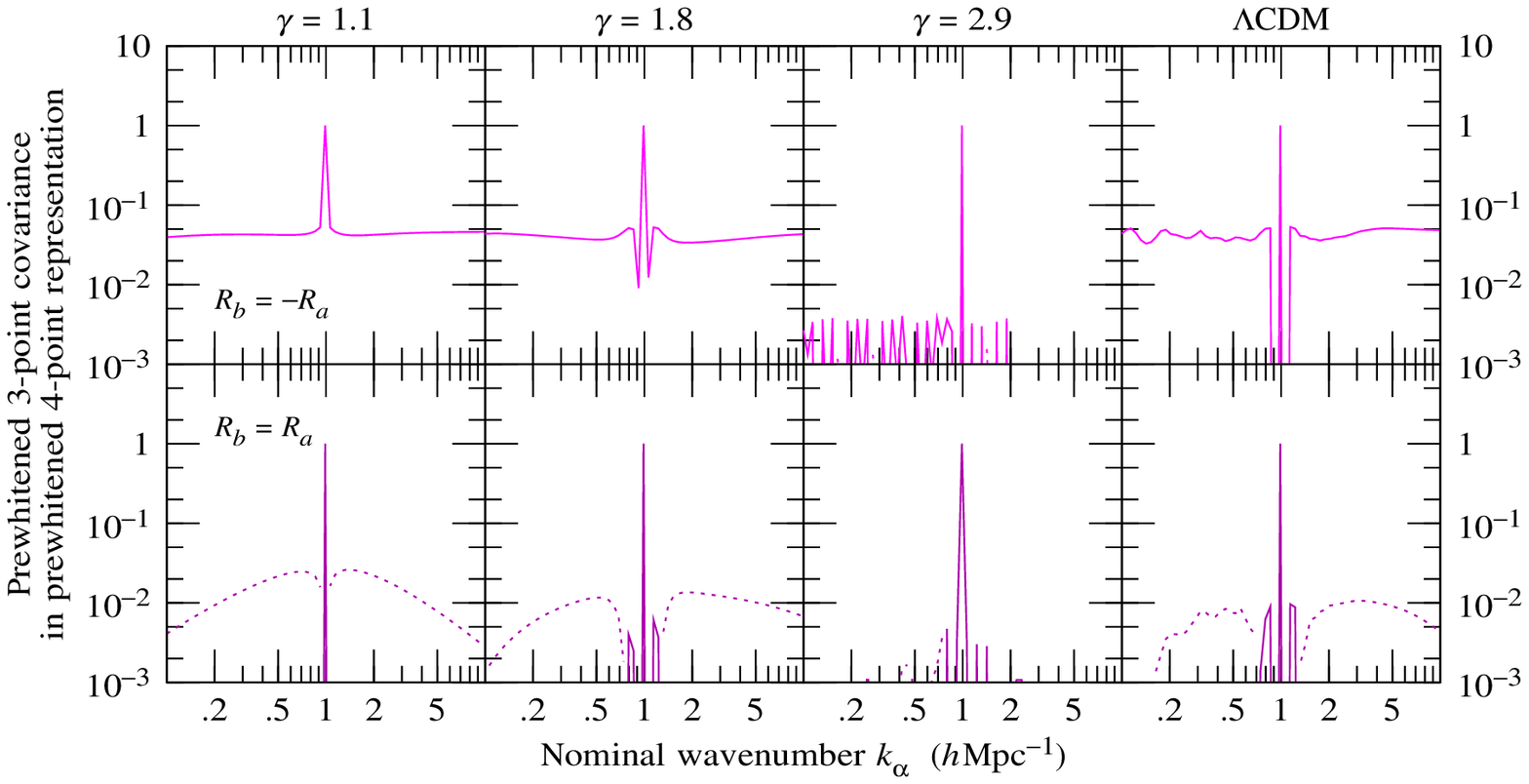}
    \end{center}
    \caption[1]{\small
Correlation coefficient
$L_{\alpha\beta}/$\protect\discretionary{}{}{}$(L_{\alpha\alpha} $\protect\discretionary{}{}{}$L_{\beta\beta})^{1/2}$
of the prewhitened 3-point covariance $L_{\alpha\beta}$
in the representation of eigenfunctions $\phi_\alpha$
of the prewhitened 4-point covariance $M_{\alpha\beta}$,
at a representative nominal wavenumber $k_\beta = 1 \, h \, \Mpc^{-1}$.
The horizontal axis is the nominal wavenumber $k_\alpha$,
which labels the 4-point eigenfunctions $\phi_\alpha$ ordered by eigenvalue.
The three sets of panels starting from the left are for power law power spectra
with correlation functions $\xi(r) = (r/5 \, h^{-1} \Mpc)^{-\gamma}$
with indices $\gamma = 1.1$, $1.8$, and $2.9$,
while the rightmost panel is for the $\Lambda$CDM power spectrum of
Eisenstein \& Hu (1998)
with 
$\Omega_\Lambda = 0.7$,
$\Omega_m = 0.3$,
$\Omega_b h^2 = 0.02$,
and $h = 0.65$.
Upper panels are
for 4-point hierarchical amplitudes $R_b = - R_a$,
lower panels for $R_b = R_a$.
Lines are dotted where the correlation coefficient is negative.
The resolution is $\Delta\log k = 1/32$,
four times coarser than that of Figure~\protect\ref{Mw3}.
What appears to be noise in the curve for $\gamma = 2.9$
results from a degeneracy of eigenvalues that mixes
the correspondence between eigenfunctions $\phi_\alpha$
and nominal wavenumbers $k_\alpha$.
    \label{Mw3s}
    }
    \end{minipage}
    \end{figure*}
}

\newcommand{\Xfig}{
    \begin{figure}
    \begin{center}
    \leavevmode
    \epsfxsize=3in
    \epsfbox{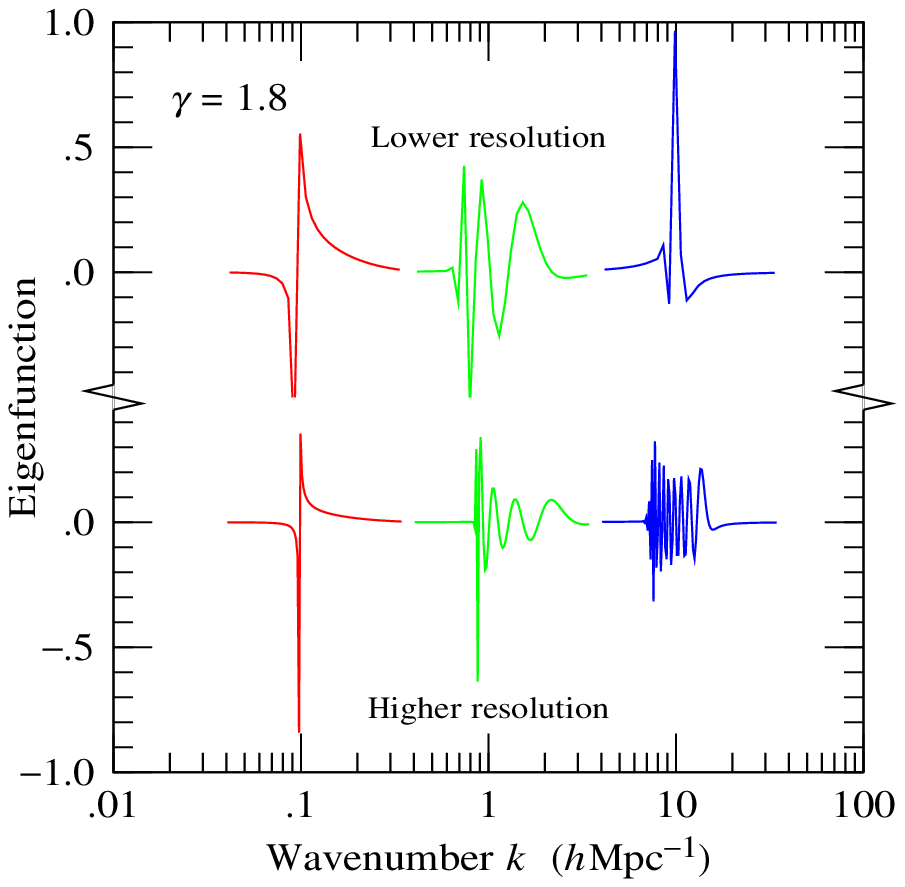}
    \end{center}
    \caption[1]{\small
Sample of (discretized) eigenfunctions
$\phi_\alpha(k)$\protect\discretionary{}{}{}$[4\PI k^3$\protect\discretionary{}{}{}$\Delta\ln k$\protect\discretionary{}{}{}$/(2\PI)^3]^{1/2}$
of the prewhitened 4-point covariance $M_{\alpha\beta}$
for a power law power spectrum with correlation function
$\xi(r) = (r/5 \, h^{-1} \Mpc)^{-1.8}$,
and $R_b = - R_a$,
at resolutions of
(top) $\Delta\log k = 1/32$
and (bottom) $\Delta\log k = 1/128$.
The eigenfunctions have nominal wavenumbers $k_\alpha$ of
$0.1$, $1$, and $10 \, h \, \Mpc^{-1}$.
In the hierarchical regime,
the eigenfunctions grow wigglier as the resolution increases.
    \label{X}
    }
    \end{figure}
}

\newcommand{\diagfig}{
    \begin{figure}
    \begin{center}
    \leavevmode
    \epsfxsize=3in
    \epsfbox{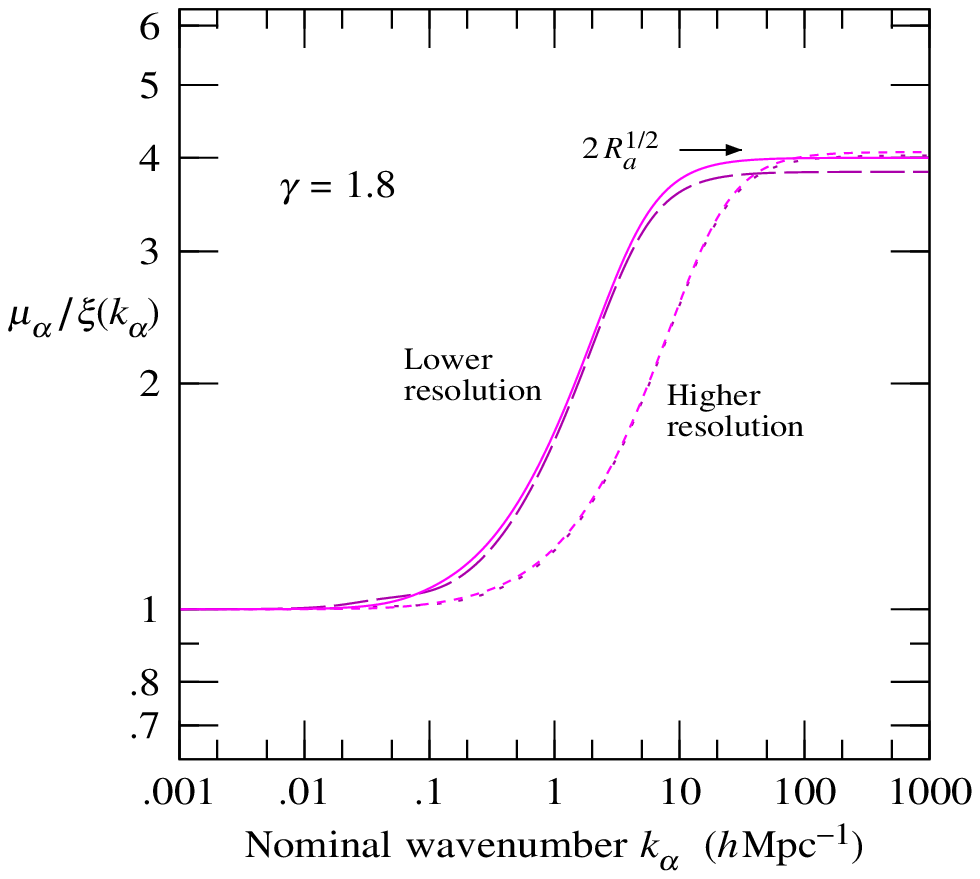}
    \end{center}
    \caption[1]{\small
Ratio $\mu_\alpha/\xi(k_\alpha)$ of the eigenvalue $\mu_\alpha$
of the 4-point prewhitened covariance matrix $M$
to the nonlinear power spectrum $\xi(k_\alpha)$,
as a function of the nominal wavenumber $k_\alpha$,
which labels the 4-point eigenfunctions $\phi_\alpha$ ordered by eigenvalue,
for a power law power spectrum with correlation function
$\xi(r) = (r/5 \, h^{-1} \Mpc)^{- 1.8}$.
The relation between eigenvalue and nominal wavenumber varies with resolution.
The low resolution case has $\Delta\log k = 1/32$;
the solid line is for $R_b = - R_a$,
the long-dashed line for $R_b = R_a$.
The high resolution case has $\Delta\log k = 1/128$;
here the (dashed) $R_b = - R_a$ and (dotted) $R_b = R_a$ curves
lie practically on top of each other.
    \label{diag}
    }
    \end{figure}
}

\newcommand{\diagthreefourfig}{
    \begin{figure}
    \begin{center}
    \leavevmode
    \epsfxsize=3in
    \epsfbox{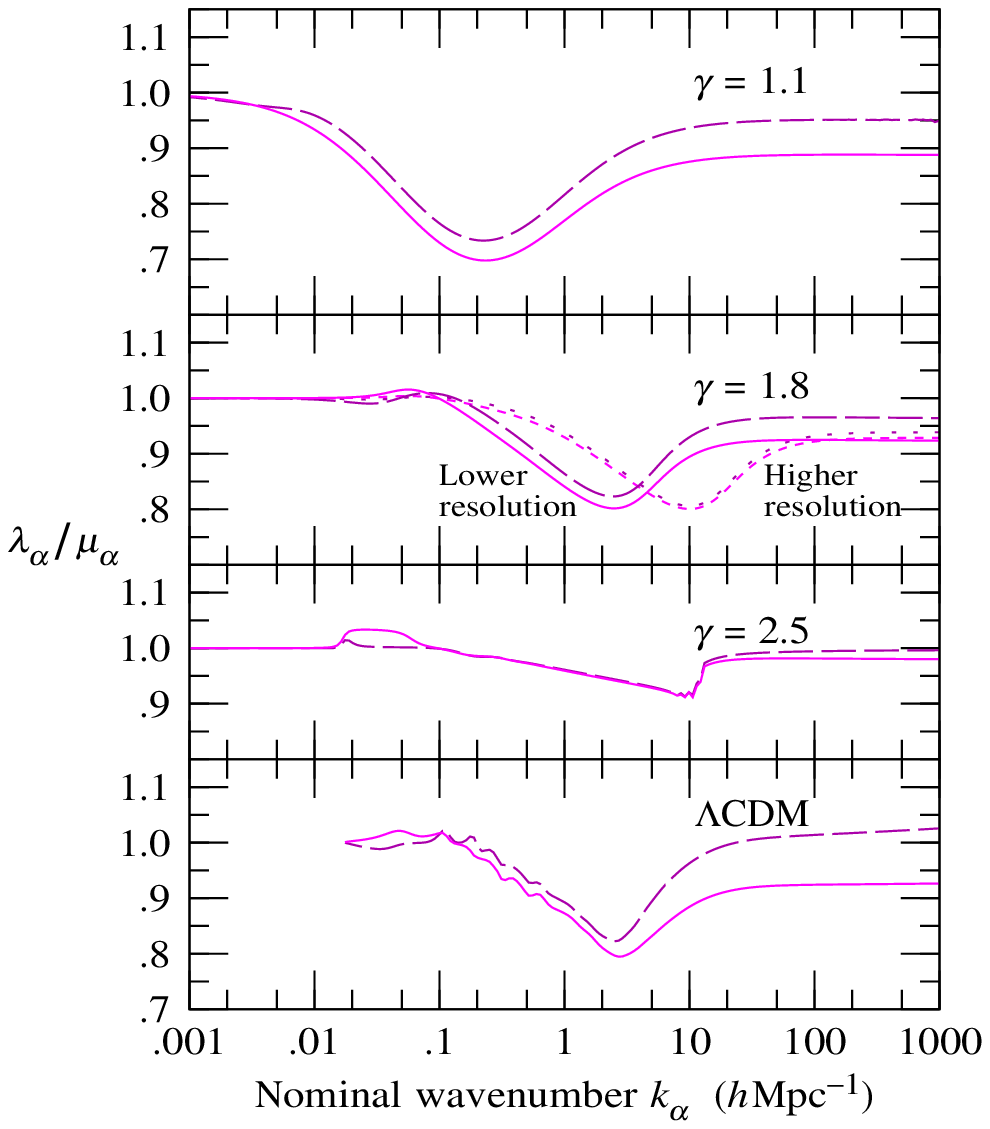}
    \end{center}
    \caption[1]{\small
Ratio $\lambda_\alpha/\mu_\alpha$ of the eigenvalues
of the 3-point and 4-point prewhitened covariance matrices,
for various power spectra.
The two lines in each case are for 4-point hierarchical amplitudes
(solid) $R_b = - R_a$,
and (long-dash) $R_b = R_a$.
The horizontal axis is the nominal wavenumber $k_\alpha$,
which labels the 3-point and 4-point eigenfunctions
$\varphi_\alpha$ and $\phi_\alpha$ ordered by eigenvalue.
The relation between eigenvalue and nominal wavenumber varies with
resolution.
The resolution is $\Delta\log k = 1/32$,
except for a high resolution case shown for $\gamma = 1.8$,
where $\Delta\log k = 1/128$.
    \label{diag34}
    }
    \end{figure}
}

\newcommand{\xikfig}{
    \begin{figure*}
    \begin{minipage}{175mm}
    \begin{center}
    \leavevmode
    \epsfxsize=6in
    \epsfbox{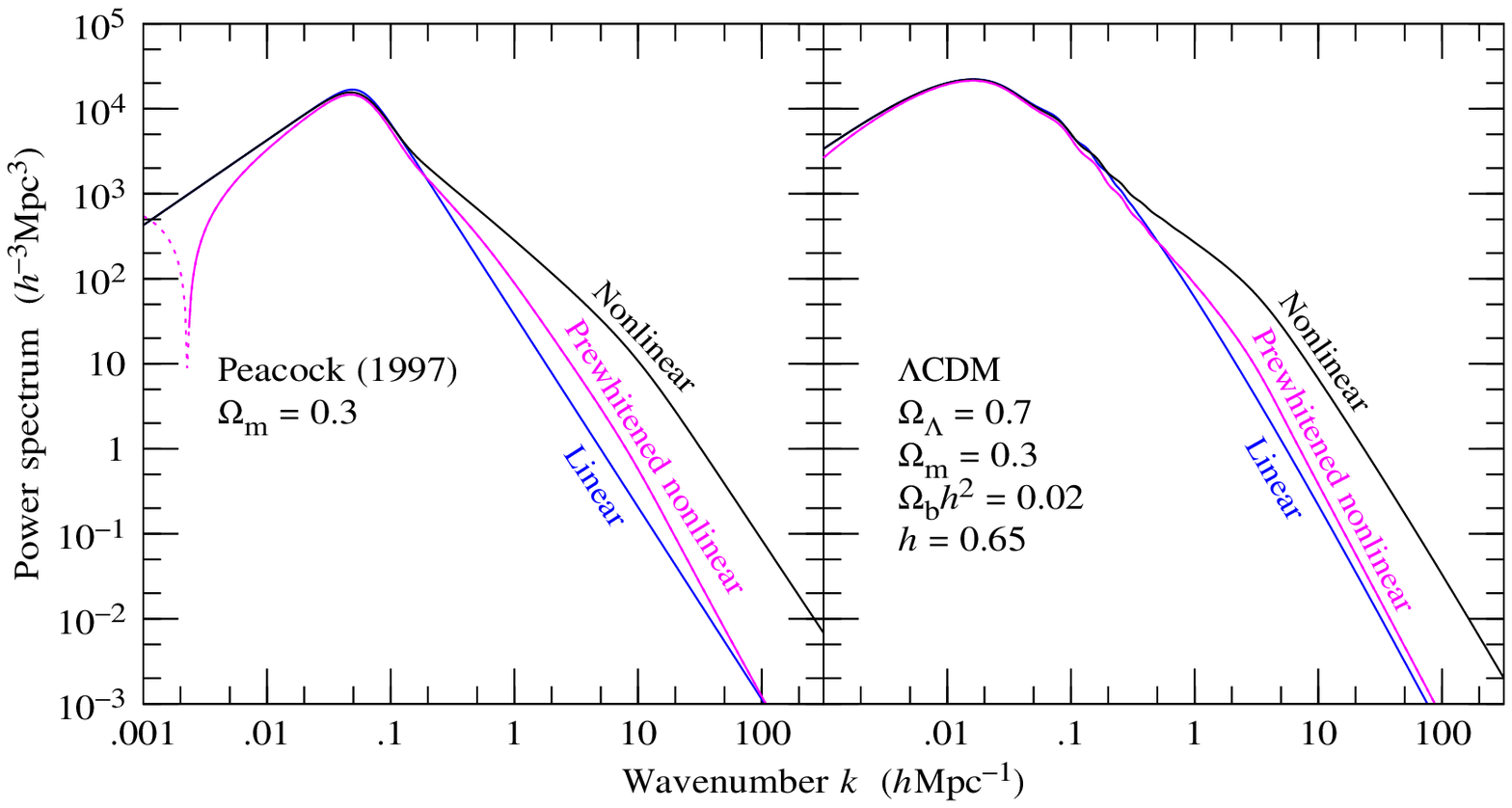}
    \end{center}
    \caption[1]{\small
Linear power spectrum $\xi_L(k)$,
nonlinear power spectrum $\xi(k)$,
and prewhitened nonlinear power spectrum $X(k)$ for
(left) the $\Omega_m = 0.3$ power spectrum derived from observations
by Peacock (1997),
and (right) the COBE-normalized $\Lambda$CDM power spectrum
from the fitting formulae of Eisenstein \& Hu (1998),
with parameters as listed on the graph.
The nonlinear power spectra were computed from the linear power spectra
according to the formula of Peacock \& Dodds (1996).
The $\Lambda$CDM power spectrum is the one used in
Figures~\protect\ref{Ms}, \protect\ref{Mw3s}, \protect\ref{diag34},
\protect\ref{mueff}, and \protect\ref{xir}.
    \label{xik}
    }
    \end{minipage}
    \end{figure*}
}

\newcommand{\muefffig}{
    \begin{figure}
    \begin{center}
    \leavevmode
    \epsfxsize=3in
    \epsfbox{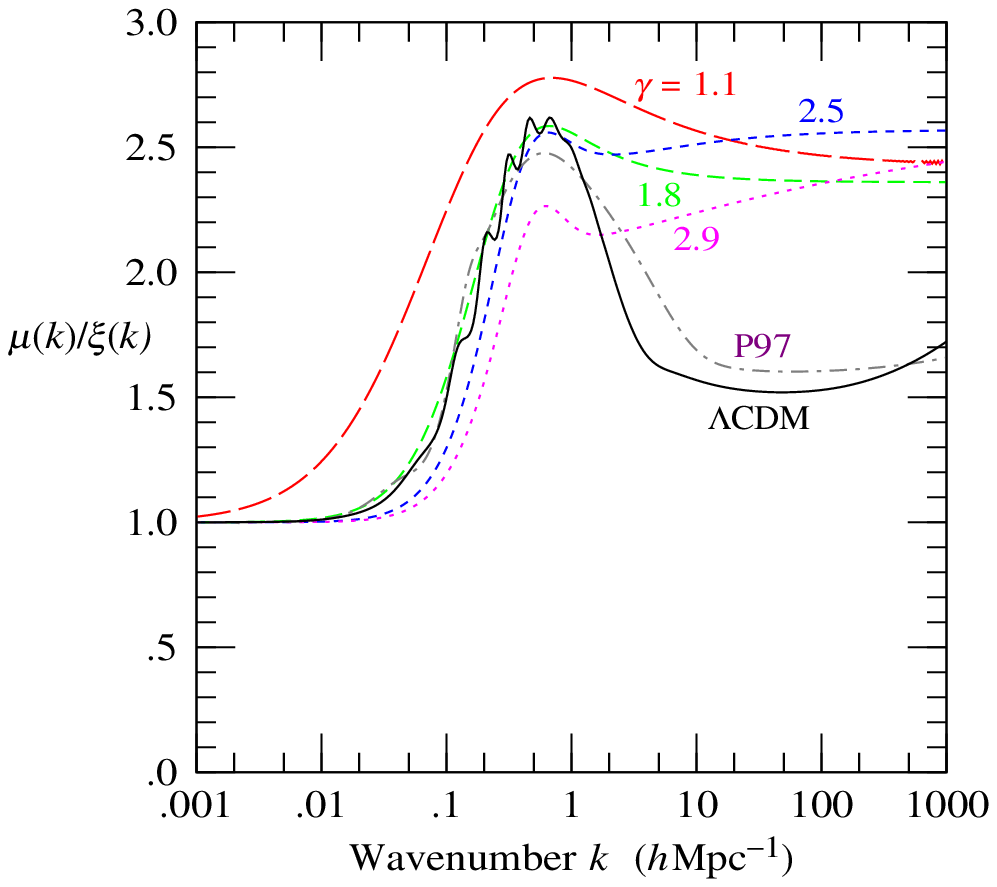}
    \end{center}
    \caption[1]{\small
Ratio $\mu(k)/\xi(k)$ of the effective FKP constant $\mu(k)$
to the nonlinear power spectrum $\xi(k)$,
as a function of wavenumber $k$,
for several different power spectra.
The ratios should be regarded as indicative rather than definitive,
because they depend on the validity of the hierarchical model
(see text).
The numbered curves are for power law power spectra
with correlation functions
$\xi(r) = (r/5 \, h^{-1} \Mpc)^{-\gamma}$,
the number label being the index $\gamma$.
The curve labelled P97 is for the $\Omega_m = 0.3$ power spectrum
derived from observations by Peacock (1997),
while that labelled $\Lambda$CDM is for the $\Lambda$CDM power spectrum
of Eisenstein \& Hu (1998).
    \label{mueff}
    }
    \end{figure}
}

\newcommand{\xirfig}{
    \begin{figure}
    \begin{center}
    \leavevmode
    \epsfxsize=3in
    \epsfbox{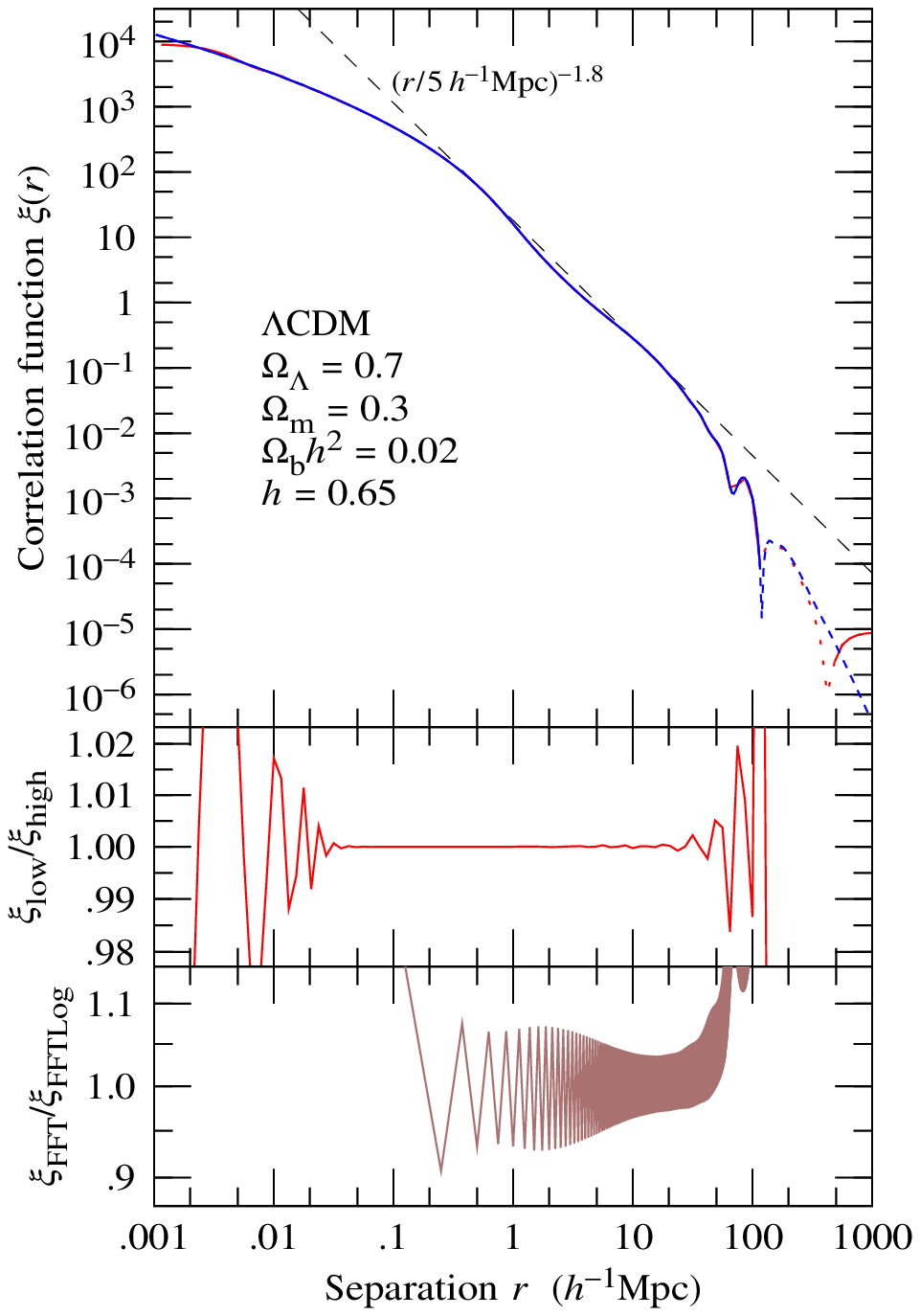}
    \end{center}
    \caption[1]{\small
Correlation function $\xi(r)$ corresponding to
the nonlinear COBE-normalized $\Lambda$CDM power spectrum
of Eisenstein \& Hu (1998), Figure~\protect\ref{xik}.
The top panel shows the correlation function computed with FFTLog
at two different resolutions, plotted on top of each other:
(a)
with 96 points
over the range $r = 10^{-3}$ to $10^3 \, h^{-1} \Mpc$
(low resolution),
and
(b)
with 768 points
over the range $r = 10^{-6}$ to $10^6 \, h^{-1} \Mpc$
(high resolution).
The lines are dashed where the correlation function is negative,
at separations $r > 119 \, h^{-1} \Mpc$.
The low and high resolution curves are almost indistinguishable except at
$r \ga 200  \, h^{-1} \Mpc$:
the high resolution curve is the one that declines as
a power law $\sim r^{-4}$ at large $r$.
The straight dashed line shows the canonical power law
$(r/5 \, h^{-1} \Mpc)^{-1.8}$ for reference.
The middle panel shows the ratio $\xi_{\rmn low}/\xi_{\rmn high}$
of the low to high resolution correlation functions.
The bottom panel shows the ratio $\xi_{\rmn FFT}/\xi_{\rmn FFTLog}$
of the correlation function $\xi_{\rmn FFT}$
computed with a normal FFT (sine transform) with 1023 points over the range
$r = 0.125$ to $128  \, h^{-1} \Mpc$,
to the high resolution correlation function $\xi_{\rmn FFTLog}$
computed with FFTLog.
The FFT'd correlation function $\xi_{\rmn FFT}$ rings
at the $\pm 5$~percent level.
    \label{xir}
    }
    \end{figure}
}

\newcommand{\fftlogexamplefig}{
    \begin{figure}
    \begin{center}
    \leavevmode
    \epsfxsize=3in
    \epsfbox{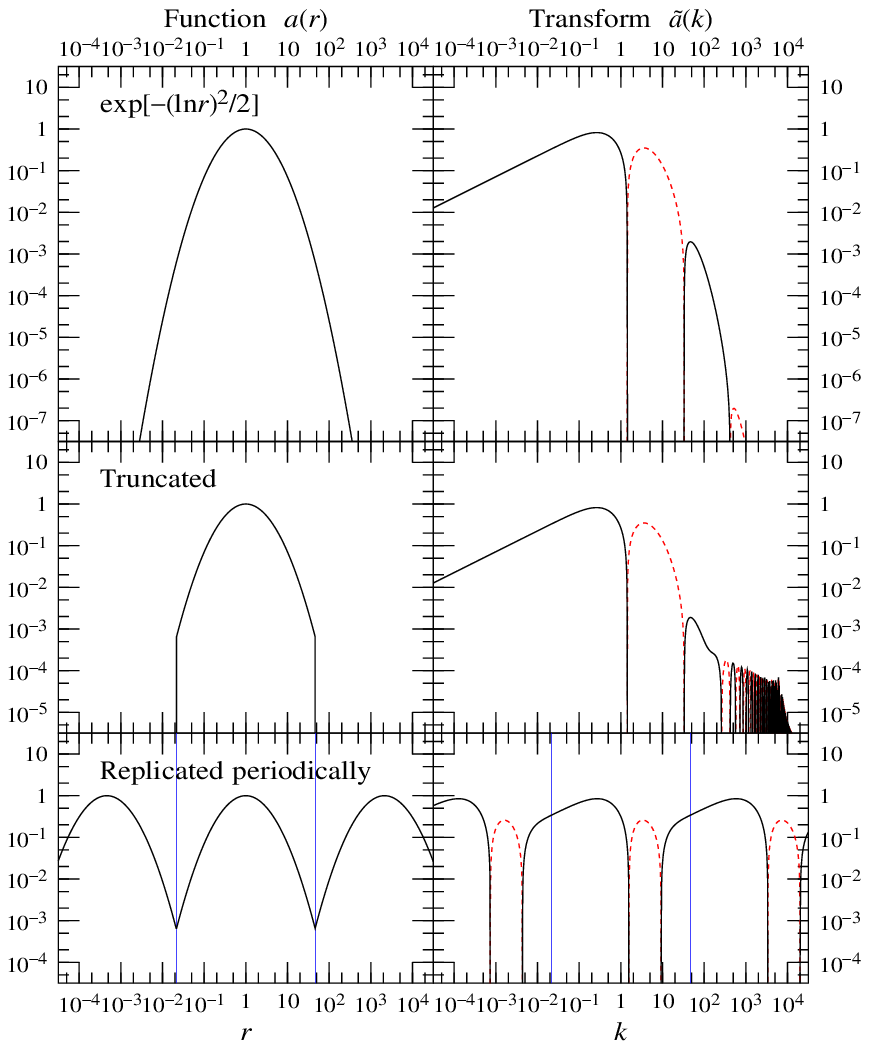}
    \end{center}
    \caption[1]{\small
Illustrating the ringing and aliasing that occurs
when the continuous Hankel transform of a function
is approximated by the discrete Hankel transform of a finite segment
of the function.
Lines are dashed where values are negative.
The function $a(r)$ is shown to the left,
and its corresponding Hankel transform $\tilde a(k)$ to the right.
The panels from top to bottom are:
(top)
the original function $a(r)$ and its Hankel transform $\tilde a(k)$;
(middle)
the truncated function $a(r)$ and its Hankel transform $\tilde a(k)$,
which rings at high frequencies $k$;
and
(bottom)
the truncated, periodically replicated function $a(r)$
and its corresponding periodically replicated Hankel transform $\tilde a(k)$,
which is aliased.
Vertical lines in the bottom panels demarcate periodic intervals.
    \label{fftlogexample}
    }
    \end{figure}
}

\begin{document}

\maketitle

\begin{abstract}

Nonlinear evolution causes the galaxy power spectrum
to become broadly correlated over different wavenumbers.
It is shown that prewhitening the power spectrum
-- transforming the power spectrum in such a way that
the noise covariance becomes proportional to the unit matrix --
greatly narrows the covariance of power.
The eigenfunctions of the covariance of the prewhitened nonlinear power spectrum
provide a set of almost uncorrelated nonlinear modes
somewhat analogous to the Fourier modes of the power spectrum
itself in the linear, Gaussian regime.
These almost uncorrelated modes make it possible
to construct a near minimum variance estimator and Fisher matrix
of the prewhitened nonlinear power spectrum
analogous to the Feldman-Kaiser-Peacock estimator of the linear power spectrum.
The paper concludes with summary recipes,
in gourmet, fine, and fastfood versions, of how to measure the prewhitened
nonlinear power spectrum from a galaxy survey in the FKP approximation.
An Appendix presents FFTLog,
a code for taking the fast Fourier or Hankel transform
of a periodic sequence of logarithmically spaced points,
which proves useful in some of the manipulations.
\end{abstract}



\section{Introduction}
\label{intro}

Most of the information about cosmological parameters
bottled inside current\footnote{
For a review of redshift surveys of galaxies see Strauss (1999)
and references therein.
Recent surveys include:
Updated Zwicky Catalog (UZC) (Falco et al.\ 1999); 
IRAS Point Source Catalogue Redshift Survey (PSCz) (Sutherland et al.\ 1999);
Redshift Survey of Zwicky Catalog Galaxies in a $2^h \times 15 \deg$ Region around 3C~273 (Grogin, Geller \& Huchra 1998); 
Durham/UKST (Ratcliffe et al.\ 1998); 
Southern Sky Redshift Survey (SSRS) (da Costa et al.\ 1998); 
ESO Slice Project (ESP) (Vettolani et al.\ 1998); 
Muenster Redshift Project (MRSP) (Schuecker et al.\ 1998); 
CNOC2 Field Galaxy Redshift Survey (Carlberg et al.\ 1998); 
Century Survey (Geller et al.\ 1997); 
Norris Survey of the Corona Borealis Supercluster (Small, Sargent \& Hamilton 1997); 
Stromlo-APM (Loveday et al.\ 1996); 
Las Campanas Redshift Survey (LCRS) (Shectman et al.\ 1996); 
Hawaii Deep Fields (Cowie et al.\ 1996); 
Canada-France Redshift Survey (CFRS) (Lilly et al.\ 1995).
}
and coming galaxy surveys,
notably the Two-Degree Field Survey (2dF) (Colless 1998; Folkes et al.\ 1999) 
and the Sloan Digital Sky Survey (SDSS) (Gunn \& Weinberg 1995; Margon 1998), 
lies in the nonlinear regime.
Even in the linear regime, nonlinearities perturb.

At large, linear scales, the power spectrum
-- the covariance of the density field,
expressed in the Fourier representation --
is the preeminent measure of large scale structure.
It is a generic, though by no means universal, prediction of inflation
(Turner 1997)
that linear density fluctuations should be Gaussian.
More generally,
primordial fluctuations should be Gaussian
whenever they result from superpositions of many independent processes,
thanks to the central limit theorem.
Observations of large scale structure are consistent with
linear density fluctuations being Gaussian
(Bouchet et al.\ 1993;
Juszkiewicz, Bouchet, \& Colombi 1993;
Gazta\~{n}aga 1994;
Gazta\~{n}aga \& Frieman 1994;
Nusser, Dekel \& Yahil 1995;
Stirling \& Peacock 1996;
Colley 1997;
Chiu, Ostriker \& Strauss 1998;
Frieman \& Gazta\~{n}aga 1999)
although the evidence is not definitive
(White 1999).
If linear density fluctuations are Gaussian,
then the 3-point and higher irreducible moments are zero,
so that the covariance of the density field contains complete information
about the statistical properties of the field,
hence all information about cosmological parameters.
Compared to other measures of covariance such as the correlation function,
the power spectrum has the additional advantage that
estimates of power at different wavenumbers are uncorrelated,
for Gaussian fluctuations.
This asset of the power spectrum is intimately related
to the assumption that the field is statistically translation invariant,
and to the fact that Fourier modes are eigenfunctions
of the translation operator.

At smaller, nonlinear scales, the power spectrum loses some of its glow.
Nonlinear evolution drives the density field away from Gaussianity,
coupling Fourier modes, feeding higher order moments,
and causing power at different wavenumbers to become correlated.
The broad extent of the correlation of the nonlinear power spectrum
has been emphasized by
Meiksin \& White (1999)
and Scoccimarro, Zaldarriaga \& Hui (1999),
and is illustrated in Figure~\ref{M} of the present paper.

The purpose of the present paper is to show how to unfold the nonlinear
power spectrum into a set of nearly uncorrelated modes,
somewhat analogous to the Fourier modes of the power spectrum itself
in the linear, Gaussian regime.
The present paper is a natural successor to
Hamilton (1997a,b, hereafter Papers~1 and 2),
which showed how to derive the minimum variance estimator
and Fisher matrix of the power spectrum
of a galaxy survey in the Feldman, Kaiser \& Peacock (1994, hereafter FKP)
approximation, for Gaussian fluctuations.
Section~5.2 of Paper~1 posed, but was unable to solve,
the non-Gaussian problem solved in the present paper.
A following paper (Hamilton \& Tegmark 2000, hereafter Paper~4),
describes how to complete the processing of the power spectrum
into fully decorrelated band-powers.

It turns out that a key to solving the non-Gaussian problem is to
`prewhiten' the power spectrum -- to transform the nonlinear power spectrum
in such a way that the (2-point) shot-noise contribution
to the covariance matrix is proportional to the unit matrix.
The properties of the prewhitened nonlinear power spectrum appear empirically
to be sweeter than might reasonably have been expected.

This paper is devoted entirely to the problem of nonlinearity.
It ignores the equally important problem of redshift distortions
(Hamilton 1998),
and the problematic question of light-to-mass bias
(Coles 1993;			
Fry \& Gazta\~naga 1993;	
Mo, Jing \& White 1997;		
Mann, Peacock \& Heavens 1998;	
Tegmark \& Peebles 1998;	
Moscardini et al.\ 1998;	
Scherrer \& Weinberg 1998;	
Dekel \& Lahav 1999;		
Col\'{\i}n et al.\ 1999;	
Cen \& Ostriker 1999;		
Narayanan, Berlind \& Weinberg 1999;	
Blanton  et al.\ 1999;		
Benson et al.\ 1999;		
Bernardeau \& Schaeffer 1999;	
Coles, Melott \& Munshi 1999).	
It further assumes that uncertainties arising
either from the selection function
(Binggeli, Sand\-age \& Tammann 1988;
Willmer 1997; Tresse 1999)
or from evolution in the cosmological volume element or the galaxy population,
are negligible.

Several authors have recently published
estimates of how well measurements of the power spectrum
from future galaxy surveys will constrain cosmological parameters
(Tegmark 1997b;
Goldberg \& Strauss 1998;
Hu, Eisenstein \& Tegmark 1998;
Eisenstein, Hu \& Tegmark 1998, 1999).
The procedures described in the present paper should assist this enterprise.

The aims of the present paper are complementary to those of
Bond, Jaffe \& Knox (1998b).
The question Bond et al.\ considered was:
If the power spectrum
(of the Cosmic Microwave Background, specifically)
is quadratically compressed
(Tegmark 1997a; Tegmark et al.\ 1997, 1998)
into a set of band-powers,
then what is the best way to use those band-powers in Maximum Likelihood
estimation of parameters?
For example,
one general procedure is to use not the band-powers themselves,
but rather functions of the band-powers arranged such that their
variances remain constant as the prior power is varied.
Bond et al.\ argued that the likelihood function is then more nearly Gaussian.
The purpose of this paper and Paper~4 is rather to arrive at the point
where one has decorrelated band-powers to work with
in the first place.

The plan of this paper is as follows.
Section~\ref{preliminaries}
sets up the notation
and defines reference material needed in subsequent sections.
Section~\ref{problems}
goes through the difficulties one meets in attempting
to measure the nonlinear power spectrum in minimum variance fashion,
and describes how to overcome them.
Section~\ref{4+3point}
reveals the unexpectedly nice properties of the prewhitened covariance
of the power spectrum,
key to the whole enterprise of this paper.
Section~\ref{prewhiten}
defines the prewhitened power spectrum.
Sections~\ref{fisher} and \ref{power}
show how the approximations motivated in previous sections
lead to a practical way
to evaluate the Fisher matrix of the prewhitened nonlinear power,
and to measure the prewhitened nonlinear power spectrum from a galaxy survey.
Section~\ref{full}
discusses how to evaluate the Fisher matrix and nonlinear power spectrum
using the FKP approximation alone, without any additional approximation.
Section~\ref{recipe}
summarizes the results of previous sections into recipes,
in gourmet, fine, and fastfood versions,
for measuring nonlinear power, the end product being
a set of uncorrelated prewhitened nonlinear band-powers, with error bars,
over some prescribed grid of wavenumbers.
Section~\ref{conclusions}
summarizes the conclusions.
Appendix~B gives details of FFTLog,
a code for taking the fast Fourier or Hankel transform
of a periodic sequence of logarithmically spaced points.

\section{Preliminaries}
\label{preliminaries}

This section contains reference material needed in subsequent sections.
The reader interested in new results may like to skip to the next section,
\S\ref{problems},
referring back to the present section as needed.

\subsection{Data, parameters}

`He will, of course, use maximum likelihood because his textbooks
have told him that'
-- E. T. Jaynes (1996, p.~624).

According to Bayes' theorem,
the probability distribution of parameters $\theta_\alpha$
given data $\delta_i$ is,
up to a normalization factor,
the product of the prior probability
with the likelihood function $\cL(\delta_i | \theta_\alpha)$.
The data $\delta_i$ in a galaxy survey
can be taken to be overdensities $\delta(\r)$
at positions $\r$ in the survey
\be
\label{delta}
  \delta(\r) \equiv {n(\r) - \nbar(\r) \over \nbar(\r)}
\ee
where $n(\r)$ is the observed number density of galaxies,
and $\nbar(\r)$ is the selection function.
The parameters $\theta_\alpha$ are, for the present purpose,
some parametrization of the galaxy power spectrum;
the focus of this paper is on the case where
the parameters are the power spectrum $\xi_\alpha$ itself.

This paper conforms to the common convention used by cosmologists
to relate the power spectrum $\xi(k)$ in Fourier space
to the correlation function $\xi(r)$ in real space,
notwithstanding the extraneous factors of $2\PI$ that result:
\be
 \xi(k) = \int \e^{\im \k.\r} \xi(r) \, \ddd r
    = \int_0^\infty \! j_0(k r) \xi(r) \, 4\PI r^2 \dd r
\ee
\be
 \xi(r) = \int \e^{- \im \k.\r} \xi(k) \, {\ddd k \over (2\PI)^3}
    = \int_0^\infty \! j_0(k r) \xi(k)
      \, {4\PI k^2 \dd k \over (2\PI)^3}
\ee
where $j_0(x) = \sin x / x$ is a spherical Bessel function.

%

\subsection{Hilbert space}

As in Paper~1,
it is convenient to adopt a notation in which Latin indices $i$, $j$, $...$,
refer to 3-dimensional positions,
while Greek indices $\alpha$, $\beta$, $...$,
run over the space of parameters,
and more specifically over the 1-dimensional space of
wavenumbers or pair separations.

For generality, brevity, and ease of manipulation,
it is convenient to treat quantities such as the data vector $\delta_i$,
or the power spectrum $\xi_\alpha$,
as vectors in a Hilbert space
(for a didactic exposition, see Hamilton 1998 \S3.3).
Such vectors have a meaning independent of the particular basis,
i.e.\ complete set of linearly independent functions,
with respect to which they might be expressed.
For example,
the data vector has components
$\delta_\r$ [$= \delta(\r)$]
when expressed in real space,
or components
$\delta_\k$
[$= \delta(\k) = \int \e^{\im \k.\r} \delta(\r) \, \ddd r$]
when expressed in Fourier space,
but from a Hilbert space point of view these are the same vector,
and in this paper they are both denoted by the same symbol $\delta_i$.

Similarly the power spectrum $\xi_\alpha$ has components
$\xi_k$ [$= \xi(k)$] when expressed in Fourier space,
or $\xi_r$ [$= \xi(r)$] when expressed in real space,
but again from a Hilbert space point of view these are the same vector,
and in this paper they are both denoted by the same symbol $\xi_\alpha$.

Latin indices $i$, $j$, $...$, on vectors and matrices
run over the 3-dimensional space of positions $\r$,
or more generally over any 3-dimensional basis of the Hilbert space.
Unless stated otherwise,
repeated pairs of indices signify the inner product in Hilbert space,
as in
\be
\label{aibi}
  a^i b_i = \int a^\ast\!(\r) b(\r) \, \ddd r
    = \int a^\ast\!(\k) b(\k) \, \ddd k/(2\PI)^3
  \ .
\ee
By definition, the inner product is a scalar,
the same quantity independent of the choice of basis.
The raised index $a^i$ denotes the Hermitian conjugate
(if the basis is orthonormal) of the vector $a_i$.
One of the indices in an inner product is always raised, the other lowered.
In this paper, all vectors in the Hilbert space are real-valued
when expressed in real space,
so that $a^\ast\!(\r) = a(\r)$
and $a^\ast\!(\k) = a(-\k)$.

Adhering to the  raised/\discretionary{}{}{}lowered index convention
serves as a useful reminder that one of the pair of vectors
in an inner product is a Hermitian conjugate
(if the basis is orthonormal).
In Fourier space, for example,
this means using $-\k$ for one index (raised) and $+\k$ for the other
index (lowered) of an inner product.

Greek indices $\alpha$, $\beta$, $...$,
run over the space of 1-dimensional pair separations $r$,
or wavenumbers $k$,
or more generally over any 1-dimensional basis in the associated Hilbert space.
Again, unless stated otherwise, repeated indices signify the inner product
\be
\label{aaba}
  a^\alpha b_\alpha = \int a^\ast\!(r) b(r) \, 4\PI r^2 \dd r
    = \int a^\ast\!(k) b(k) \, 4\PI k^2 \dd k/(2\PI)^3
\ee
which is again a scalar,
the same quantity independent of the choice of basis.
Again, in this paper all vectors in the Hilbert space are real-valued
in real space, so
$a^\ast\!(r) = a(r)$ and $a^\ast\!(k) = a(k)$.
Although there is no distinction in this case between
vectors with raised and lowered indices in either real or Fourier space,
adhering to the raised/\discretionary{}{}{}lowered index convention
again serves as a useful reminder.

The unit matrix $\1_\alpha^\beta$ in any representation
is defined such that its inner product with any vector $a_\beta$
leaves the vector unchanged,
\be
  \1_\alpha^\beta \, a_\beta = a_\beta \, \1^\beta_\alpha = a_\alpha
  \ .
\ee
In the continuous real representation, the unit matrix is
\be
  \1_{r_\alpha}^{r_\beta} =
  \deltaD(r_\alpha-r_\beta)
\ee
where $\deltaD(r_\alpha-r_\beta)$ denotes the 3-dimensional
Dirac delta-function, defined such that
\be
  \int \deltaD(r_\alpha-r_\beta) \, 4\PI r_\alpha^2 \dd r_\alpha = 1
  \ .
\ee
In the continuous Fourier representation, the unit matrix is
\be
  \1_{k_\alpha}^{k_\beta} =
  (2\PI)^3 \deltaD(k_\alpha-k_\beta)
\ee
again a 3-dimensional Dirac delta-function.

\subsection{Discretization of matrices}
\label{discrete}

Many of the operations in this paper involve manipulations
of matrices in the 1-dimensional space of separations.
Continuous matrices must be discretized to manipulate them numerically.
Discretization should be done in such a way as to preserve
the inner product~(\ref{aaba}),
so that integration over the volume element,
$4\PI r^2 \dd r$ in real space,
or $4\PI k^2 \dd k/(2\PI)^3$ in Fourier space,
translates into summation in the corresponding discrete space.
This ensures that matrix operations such multiplication,
diagonalization, and inversion can be done in the usual fashion.

Most of the manipulations in this paper are done in Fourier space
on a logarithmically spaced grid of wavenumbers $k_\alpha$.
In this case, a continuous vector $a(k_\alpha)$ is discretized
by multiplying it by
$[ 4\PI k_\alpha^3 \Delta\ln k / (2\PI)^3 ]^{1/2}$
\be
  a(k_\alpha) \rightarrow
  {\sf a}_{k_\alpha} =
  a(k_\alpha)
  \left[ 4\PI k_\alpha^3 \Delta\ln k / (2\PI)^3 \right]^{1/2}
\ee
and a continuous matrix $A(k_\alpha,k_\beta)$ is discretized
by multiplying it by
$4\PI (k_\alpha k_\beta)^{3/2} \Delta\ln k / (2\PI)^3$
\be
  A(k_\alpha,k_\beta) \rightarrow
  \bA_{k_\alpha k_\beta} =
  A(k_\alpha,k_\beta) \,
  4\PI (k_\alpha k_\beta)^{3/2} \Delta\ln k / (2\PI)^3
  \ .
\ee
The unit matrix
$(2\PI)^3 \deltaD(k_\alpha-k_\beta)$
in the continuous Fourier representation
translates to the unit matrix $\1_{\alpha\beta}$ in the discrete case
\be
  (2\PI)^3 \deltaD(k_\alpha-k_\beta) \rightarrow
  \1_{\alpha\beta}
  \ .
\ee

Similarly, a continuous vector $a(r_\alpha)$ in real space
is discretized on to a logarithmically spaced grid of separations $r_\alpha$
by multiplying the vector by
$\left[ 4\PI r_\alpha^3 \Delta\ln r \right]^{1/2}$
\be
  a(r_\alpha) \rightarrow
  {\sf a}_{r_\alpha} =
  a(r_\alpha)
  \left[ 4\PI r_\alpha^3 \Delta\ln r \right]^{1/2} 
\ee
and a continuous matrix $A(r_\alpha,r_\beta)$ is discretized
by multiplying it by
$4\PI (r_\alpha r_\beta)^{3/2} \Delta\ln r$
\be
  A(r_\alpha,r_\beta) \rightarrow
  \bA_{r_\alpha r_\beta} =
  A(r_\alpha,r_\beta) \,
  4\PI (r_\alpha r_\beta)^{3/2} \Delta\ln r
  \ .
\ee
The unit matrix
$\deltaD(r_\alpha-r_\beta)$
in the continuous real representation
translates to the unit matrix $\1_{\alpha\beta}$ in the discrete case
\be
\label{deltaD}
  \deltaD(r_\alpha-r_\beta) \rightarrow
  \1_{\alpha\beta}
  \ .
\ee

The transformation between Fourier and real space
for logarithmically spaced wavenumbers $k_\alpha$ and separations $r_\alpha$
may be accomplished with FFTLog (Appendix~B).

%

\subsection{Gaussian density field}
\label{gaussian}

If the density distribution $\delta(\r)$ were Gaussian
-- which is {\em not\/} true in the present case --
then one would have the luxury of being able to write down
an explicit Gaussian likelihood function
\be
\label{Llin}
  \cL \propto {1 \over | C |^{1/2}}
  \exp \left( - \frac{1}{2} \delta_i C^{-1ij} \delta_j \right)
\ee
where $|C|$ and $C^{-1}$ are the determinant and inverse
of the covariance matrix $C$ of overdensities
\be
  C_{ij} \equiv \langle \delta_i \delta_j \rangle
  \ .
\ee
Angle-brackets here and throughout this paper signify 
averages over possible data sets $\delta_i$ predicted by the likelihood function
\be
  \langle t \rangle \equiv
    \int t \, \cL(\delta_i | \theta_\alpha) \, \dd[\delta_i]
  \ .
\ee
Maximum Likelihood (ML) estimates $\hat\theta_\alpha$
of the parameters $\theta_\alpha$
(the hat distinguishing the estimate $\hat\theta_\alpha$
from the true value $\theta_\alpha$)
are given by the vanishing of the vector of partial derivatives
of the log-likelihood function
\be
\label{dLdth}
  {\Partial \ln \cL \over \Partial \theta_\alpha}
  = \frac{1}{2}
    {\Partial C_{ij} \over \Partial \theta_\alpha}
    C^{-1ik} C^{-1jl}
    ( \delta_k \delta_l - C_{kl} )
\ee
\be
\label{thML}
  \left. {\Partial \ln \cL \over \Partial \theta_\alpha}
     \right|_{\theta_\alpha = \hat\theta_\alpha}
  = 0
  \ .
\ee
The covariance
$\langle \Delta \hat\theta_\alpha \Delta \hat\theta_\beta \rangle$
of the estimated parameters is given approximately by
the inverse of the Fisher information matrix $F^{\alpha\beta}$,
defined to be minus the expectation value of the matrix of second partial
derivatives of the log-likelihood function
\be
  F^{\alpha\beta} \equiv
  - \left\langle
    {\Partial^2 \ln \cL \over \Partial \theta_\alpha \Partial \theta_\beta}
    \right\rangle
  = \frac{1}{2}
    {\Partial C_{ij} \over \Partial \theta_\alpha}
    C^{-1ik} C^{-1jl}
    {\Partial C_{kl} \over \Partial \theta_\beta}
\ee
\be
\label{covth}
  \langle \Delta \hat\theta_\alpha \Delta \hat\theta_\beta \rangle
  \approx F^{-1}_{\alpha\beta}
  \ .
\ee
The approximation~(\ref{covth}) is exact if the estimated parameters
$\hat\theta_\alpha$ are Gaussianly distributed about their expectation values.
The central limit theorem asserts that the parameters become Gaussianly
distributed in the asymptotic limit of a large amount of data.

It is commonly assumed,
and the same assumption is adopted here,
that the dominant source of variance in a galaxy survey is a combination of
cosmic (sample) variance
and shot-noise arising from the discrete sampling of galaxies.
If the sampling of galaxies is random -- a Poisson process --
then the covariance $C_{ij}$ is a sum of
the cosmic covariance $\xi_{ij}$
with Poisson sampling noise $N_{ij}$
\be
\label{C}
  C_{ij} = \xi_{ij} + N_{ij}
  \ .
\ee
In the real representation,
the cosmic covariance $\xi_{ij}$ is the correlation function
\be
  \xi_{ij} = \xi(|\r_i-\r_j|)
\ee
and the noise matrix $N_{ij}$ is the diagonal matrix
\be
\label{N}
  N_{ij} = {\deltaD(\r_i-\r_j) \over \nbar(\r_i)}
\ee
with $\deltaD(\r_i-\r_j)$ a 3-dimensional Dirac delta-function.
In the Fourier representation
the cosmic covariance $\xi_{ij}$ is the diagonal matrix
\be
  \xi_{ij} = (2\PI)^3 \deltaD(\k_i+\k_j) \, \xi(k_i)
\ee
whose eigenvalues $\xi(k_i)$ constitute the power spectrum.

The focus of this paper is on the case
where the parameters $\theta_\alpha$
are the power spectrum $\xi_\alpha$ itself
(in this paper the cosmic covariance function $\xi_\alpha$,
expressed in an arbitrary representation,
will often be referred to as the `power spectrum',
even though this name is commonly reserved for the covariance $\xi(k)$
expressed in Fourier space; no confusion should result).
In this case the covariance $C_{ij}$ is a linear function of
the parameters $\xi_\alpha$
\be
\label{Clin}
  C_{ij} = D^\alpha_{ij} \xi_\alpha + N_{ij}
\ee
where
in real space $\xi_\alpha=\xi(r_\alpha)$ is the correlation function, and
\be
\label{Dr}
  D^\alpha_{ij} = \deltaD(|\r_i-\r_j| - r_\alpha)
\ee
is a 3-dimensional Dirac delta-function, equation~(\ref{deltaD}),
while in Fourier space $\xi_\alpha=\xi(k_\alpha)$
is the thing commonly called the power spectrum, and
\be
\label{Dk}
  D^\alpha_{ij}
  = (2\PI)^6 \deltaD(\k_i + \k_j) \deltaD(k_i - k_\alpha)
  \ .
\ee

It follows from equations~(\ref{dLdth}) and (\ref{thML}) that
the ML estimator $\hat\xi_\alpha$ of the power spectrum,
for Gaussian fluctuations,
is that solution of
\be
\label{xiML}
  \hat\xi_\alpha
  = \frac{1}{2} F^{-1}_{\alpha\beta} D^\beta_{ij}
    C^{-1ik} C^{-1jl}
    (\delta_k \delta_l - N_{kl})
\ee
for which the estimate is equal to the prior, $\hat\xi_\alpha = \xi_\alpha$.
The variance of the ML estimator is
\be
  \langle \Delta \hat\xi_\alpha \Delta \hat\xi_\beta \rangle
  \approx F^{-1}_{\alpha\beta}
\ee
and the Fisher matrix is
\be
  F^{\alpha\beta}
  = \frac{1}{2} D^\alpha_{ij} C^{-1ik} C^{-1jl} D^\beta_{kl}
  \ .
\ee

If the prior power $\xi_\alpha$ is regarded as fixed,
then equation~(\ref{xiML}) yields an estimated power $\hat\xi_\alpha$
that is quadratic in overdensities $\delta_i$.
If this estimated power is folded back into the prior,
then equation~(\ref{xiML}) with the revised prior
yields another estimate of power.
Iterated to convergence, the result is the ML estimator of the power.
It is to be noted that even without iteration,
equation~(\ref{xiML}) yields a measurement of power that
(as long as the prior is at least roughly correct)
should already be a good approximation, since
`if the prior matters, then you are not learning much from the data',
to quote one of the refrains from the 1997 Aspen workshop
on Precision Measurement of Large Scale Structure.

The question of how to apply quadratic estimators
(such as given by equation~[\ref{xiML}])
to measure the power spectrum is addressed by
Tegmark et al.\ (1998) for galaxies,
and by Tegmark (1997a), Tegmark et al.\ (1997),
and Bond, Jaffe \& Knox (1998a,b) for the CMB.

\subsection{Non-Gaussian density field}
\label{nongauss}

Ultimately, one might look forward to a wondrous $N$-body machine
able to compute the probability distribution of linear initial conditions
given noisy and incomplete data from a survey
(Narayanan \& Weinberg 1998;
Monaco \& Efstathiou 1999;
and references therein).

In the meantime it is far from clear what to write down as a
likelihood function for the nonlinear density field
(Dodelson, Hui \& Jaffe 1999).
Certainly it would be a bad idea to use a Gaussian likelihood function
for a non-Gaussian density field,
since that would lead to a serious underestimate of the true uncertainty
in the measured nonlinear power spectrum.

An alternative procedure is to seek a minimum variance unbiased estimator
of power.
Now the power spectrum is by definition a covariance of overdensities,
and by the presumption of Poisson sampling, any a priori weighted
sum of quantities quadratic in observed overdensities
(with self-terms excluded, to eliminate shot-noise)
provides an unbiased estimate of the power spectrum
linearly windowed in some fashion.
It was shown in \S2.3 of Paper~1 that,
amongst estimators quadratic in observed overdensities $\delta_i$,
the unbiased estimator $\hat\xi_\alpha$ of the power spectrum
having minimum variance is
\be
\label{ximv}
  \hat\xi_\alpha = F^{-1}_{\alpha\beta} D^\beta_{ij} \fC^{-1ijkl}
    (\delta_k \delta_l - \hat N_{kl})
\ee
with variance
\be
\label{varmv}
  \langle \Delta \hat\xi_\alpha \Delta \hat\xi_\beta \rangle
  = F^{-1}_{\alpha\beta}
\ee
where $F^{\alpha\beta}$ is the Fisher matrix
\be
\label{Fmv}
  F^{\alpha\beta} = D^\alpha_{ij} \fC^{-1ijkl} D^\beta_{kl}
  \ ,
\ee
$\fC_{ijkl}$ is the covariance of shot-noise-subtracted
products of overdensities
\be
\label{CC}
  \fC_{ijkl} \equiv
    \left\langle (\delta_i \delta_j - \hat N_{ij} - \xi_{ij})
    \, (\delta_k \delta_l - \hat N_{kl} - \xi_{kl}) \right\rangle
\ee
and $\fC^{-1ijkl}$ is its inverse,
meaning
$\fC_{ijmn} \fC^{-1mnkl}
= {\rmn Sym}_{(kl)} \1_i^k \1_j^l$.
The symbol ${\rmn Sym}_{(ij)}$ signifies symmetrization over its underscripts,
as in
\be
  \Sym{(ij)} A_{ij} \equiv \frac{1}{2} ( A_{ij} + A_{ji} )
  \ .
\ee

The quantity $\hat N_{kl}$ in equations~(\ref{ximv}) and (\ref{CC})
is the `actual' shot-noise,
the contribution to $\delta_k \delta_l$ from self-pairs of galaxies,
pairs consisting of a galaxy and itself.
The actual shot-noise $\hat N_{kl}$ in a survey is to be distinguished
from its expectation value $N_{kl} \equiv \langle \hat N_{kl} \rangle$.
If the expected shot-noise $N_{kl}$ is used in equation~(\ref{ximv})
in place of the actual shot-noise,
then additional terms
(given in eq.~[8] of Paper~1)
appear in the covariance matrix
$\fC_{ijkl}$, increasing the variance of the estimator.
Why does the ML estimator $\hat\xi_\alpha$ in the Gaussian case,
equation~(\ref{xiML}), involve the expected shot-noise $N_{kl}$
rather than the actual shot-noise $\hat N_{kl}$?
Because a discretely sampled Gaussian field is not really Gaussian,
except in the limit where a cubic wavelength contains many galaxies,
so the assumption of a Gaussian likelihood function is not strictly correct.
In fact it is plain that the Gaussian ML estimator $\hat\xi_\alpha$
would also be improved if the actual shot-noise $\hat N_{kl}$ were used
in place of the expected shot-noise $N_{kl}$ in equation~(\ref{xiML}),
since using the actual shot-noise exploits additional information about
the character of the Poisson sampling that is discarded by the Gaussian
likelihood.
However, as discussed by Tegmark et al.\ (1998 Appendix~A),
the gain from subtracting the actual versus the expected shot-noise
is in practice small at linear scales,
where a cubic wavelength is likely to contain many galaxies.

\ptsfig

In the same Poisson sampling approximation as equation~(\ref{C}),
the covariance $\fC_{ijkl}$ of shot-noise-subtracted products of overdensities,
equation~(\ref{CC}), is,
in the real representation with no implicit summation,
\ba
\label{Cquad}
  \fC_{ijkl}
  \!\!\!&=&\!\!\!
    \xi_{ik} \xi_{jl} + \xi_{il} \xi_{jk} + \eta_{ijkl}
  \nn
  && \!\!\!\!\!\!
    \mbox{} + \bigl[ N_{ik} (\xi_{jl} + \zeta_{ijl})
      + (i \leftrightarrow j, k \leftrightarrow l) \bigr] (\mbox{4 terms})
  \nn
  && \!\!\!\!\!\!
    \mbox{} + ( N_{ik} N_{jl} + N_{il} N_{jk} ) (1 + \xi_{ij})
\ea
in which the top line is the 4-point,
the middle the 3-point,
and the bottom line the 2-point contribution to the covariance,
as illustrated in Figure~\ref{pts}.
For Gaussian density fluctuations
equation~(\ref{Cquad}) reduces to
\be
\label{CCgau}
  \fC_{ijkl} =
    2 \Sym{(kl)} C_{ik} C_{jl}
\ee
with inverse
\be
\label{CCinvgau}
  \fC^{-1ijkl}
  = \frac{1}{2} \Sym{(kl)} C^{-1ik} C^{-1jl}
  \ .
\ee
It follows from equation~(\ref{CCinvgau})
that for Gaussian fluctuations the minimum variance
estimator of the power spectrum, equation~(\ref{ximv}),
is the same as the ML estimator, equation~(\ref{xiML}),
if the estimate is folded back into the prior and iterated to convergence
(modulo the comments about shot-noise in the previous paragraph).

\section{Problems}
\label{problems}


\subsection{FKP approximation}
\label{FKP}

Calculating the minimum variance estimate $\hat\xi_\alpha$
of the power spectrum,
equation~(\ref{ximv}),
involves the formidable problem
of inverting the pair covariance $\fC_{ijkl}$,
a rank 4 matrix of 3-dimensional quantities.
Whereas for Gaussian fluctuations the rank 4 matrix
$\fC_{ijkl}$
factorizes into a product of rank 2 matrices, equation~(\ref{CCgau}),
for non-Gaussian fluctuations it does not factorize.
Again, whereas for Gaussian fluctuations it may be possible,
at least at the largest scales,
to pixelize a survey into large enough pixels
that brute force numerical inversion is feasible,
for non-Gaussian fluctuations brute force inversion is quite impossible.

A natural way to simplify the problem is to adopt the
Feldman, Kaiser \& Peacock (1994, FKP) approximation,
where the selection function $\nbar(\r)$ is taken to be locally constant.
The FKP approximation
is expected to be valid at wavelengths much smaller than the
characteristic size of the survey.
Section~5 of Paper~1 terms this the `classical' approximation,
since it is valid to the extent that the position and wavelength
of a density mode can be measured simultaneously.
While the FKP approximation is liable to break down
at larger scales,
particularly for pencil beam or slice
surveys,
it should be a good approximation at smaller, nonlinear scales,
especially in surveys with broad contiguous sky coverage.

Even if the selection function $\nbar$ is taken to be constant,
the general problem of inverting
the rank 4 matrix $\fC_{ijkl}$ remains intractible.
Notice however that $\fC^{-1ijkl}$ appears multiplied
in both equations~(\ref{ximv}) and (\ref{Fmv}) by the matrix
$D^\alpha_{ij}$.
Now $D^\alpha_{ij}$ has translation and rotation symmetry,
and in the FKP approximation the matrix $\fC_{ijkl}$ also has
translation and rotation symmetry,
the selection function $\nbar$ being constant.
Indeed,
inspection of equation~(\ref{Cquad}) reveals that
the matrix $\fC_{ijkl}$ remains translation and rotation invariant
even if the selection functions $\nbar_i$ and $\nbar_j$ at positions $i$ and $j$
are two different constants.
It follows that the combination $\fC_{ijkl} D_\alpha^{kl}$
is likewise translation and rotation symmetric,
which implies that it can be expressed in the form
\be
\label{Cab}
  \fC_{ijkl} D_\alpha^{kl}
  = \fC_{\alpha\beta}(\nbar_i,\nbar_j) \, D^\beta_{ij}
\ee
for some matrix $\fC_{\alpha\beta}$,
which can be termed the `reduced' covariance matrix.
Equation~(\ref{Cab}) is the FKP approximation,
expressed in concise mathematical form;
additional details of the justification of this equation are provided
in Appendix~A.
The reduced matrix is written in equation~(\ref{Cab}) as
$\fC_{\alpha\beta} (\nbar_i,\nbar_j)$ to emphasize the fact that
it is a function of the selection functions $\nbar_i$ and $\nbar_j$ at
positions $i$ and $j$;
note that no implicit summation over $i$ or $j$ is intended
on the right hand side of equation~(\ref{Cab}).
Inspection of equation~(\ref{Cquad}) for $\fC_{ijkl}$
shows that the reduced covariance $\fC_{\alpha\beta}(\nbar_i,\nbar_j)$
takes the form
\be
\label{CKJI}
  \fC_{\alpha\beta}(\nbar_i,\nbar_j)
  = 2 \left[
    K_{\alpha\beta}
    + (\nbar_i^{-1}\!\!+\!\nbar_j^{-1}) J_{\alpha\beta}
    + \nbar_i^{-1} \nbar_j^{-1} H_{\alpha\beta}
    \right]
\ee
a linear combination of 4-point, 3-point, and 2-point contributions
$K_{\alpha\beta}$,
$J_{\alpha\beta}$,
and $H_{\alpha\beta}$.
Multiplying equation~(\ref{Cab}) by $\fC^{-1\gamma\alpha} \fC^{-1mnij}$
shows that the inverse of $\fC_{ijkl}$ is similarly related
to the inverse of the reduced matrix $\fC_{\alpha\beta}$
\be
\label{Cabinv}
  \fC^{-1ijkl} D^\alpha_{kl}
  = \fC^{-1\alpha\beta}(\nbar_i,\nbar_j) \, D_\beta^{ij}
  \ .
\ee

Physically,
to the extent that the selection functions $\nbar_i$ and $\nbar_j$
at positions $i$ and $j$ are constants,
the minimum variance pair-weighting attached to a pair $ij$
should be a function only of the separation $\alpha$ of the pair,
not of their position or orientiation.
Just as $\fC_{ijkl}$ is the covariance between
a pair $ij$ and another pair $kl$,
so the reduced covariance matrix $\fC_{\alpha\beta}$
is the covariance between
a pair separated by $\alpha$ and another pair separated by $\beta$.

In the FKP approximation given by equation~(\ref{Cabinv}),
the minimum-variance estimate~(\ref{ximv}) of the power spectrum is
\be
\label{ximvFKP}
  \hat\xi_\alpha
  = F^{-1}_{\alpha\beta} \fC^{-1\beta\gamma}
    D_\gamma^{ij} (\delta_i \delta_j - \hat N_{ij})
\ee
and the associated Fisher matrix~(\ref{Fmv}) is
\be
\label{FmvFKPasym}
  F^{\alpha\beta} =
    \fC^{-1\alpha\gamma} D_\gamma^{ij} D^\beta_{ij}
  \ .
\ee
Notice that the approximate Fisher matrix given by this
equation~(\ref{FmvFKPasym}) is not symmetric,
whereas the original Fisher matrix, equation~(\ref{Fmv}), was symmetric.
The asymmetry results from the asymmetry of the FKP approximation,
equation~(\ref{Cab}).
The approximate expression~(\ref{FmvFKPasym})
would be symmetric if the FKP approximation were exact,
and in practice it should be nearly symmetric;
if not, it is a signal that the FKP approximation is breaking down.

To ensure symmetry of the Fisher matrix,
one might be inclined at this point to symmetrize equation~(\ref{FmvFKPasym}),
since after all an equally good approximation to the Fisher matrix
would be the same expression~(\ref{FmvFKPasym})
with the indices swapped on the right hand side,
$\alpha \leftrightarrow \beta$.
However,
it is desirable that the FKP estimator $\hat\xi_\alpha$,
equation~(\ref{ximvFKP}), should be unbiased, meaning that
\be
  \langle \hat\xi_\alpha \rangle = \xi_\alpha
  \ .
\ee
Averaging equation~(\ref{ximvFKP}) gives, since
$\langle \delta_i \delta_j - \hat N_{ij} \rangle = D^\alpha_{ij} \xi_\alpha$
according to equation~(\ref{Clin}),
\be
  \langle \hat\xi_\alpha \rangle
  = F^{-1}_{\alpha\beta} \fC^{-1\beta\gamma}
    D_\gamma^{ij} D^\epsilon_{ij} \, \xi_\epsilon
\ee
which shows that the FKP estimator $\hat\xi_\alpha$ is unbiased only if
the Fisher matrix in equation~(\ref{ximvFKP})
is interpreted as satisfying the asymmetric expression~(\ref{FmvFKPasym}).
A detailed discussion of this issue is deferred to \S\ref{power}.
Here it suffices to remark that,
to the extent that the FKP approximation is valid,
the variance of the FKP estimator $\hat\xi_\alpha$ is equal to the inverse
of the symmetrized Fisher matrix given by equation~(\ref{FmvFKPasym})
\be
\label{xiaxib}
  \langle \Delta\hat\xi_\alpha \Delta\hat\xi_\beta \rangle
  =
  F^{-1}_{(\alpha\beta)}
\ee
where $F^{(\alpha\beta)} \equiv {\rmn Sym}_{(\alpha\beta)} F^{\alpha\beta}$
denotes the symmetrized Fisher matrix,
and $F^{-1}_{(\alpha\beta)}$ its inverse.

\subsection{Hierarchical model}
\label{hierarchical}

The pair covariance matrix $\fC_{ijkl}$, equation~(\ref{Cquad}),
hence also the reduced covariance matrix $\fC_{\alpha\beta}$,
equation~(\ref{Cab}),
involves the 3-point and 4-point correlation functions
$\zeta_{ijk}$ and $\eta_{ijkl}$.
The problem here is that these correlation functions are not known precisely.

Available observational and $N$-body evidence
(see for example the summaries by
Scoccimarro \& Frieman 1999
and
Hui \& Gazta\~{n}aga 1999)
is consistent
with a hierarchical model in which the 3-point and 4-point functions are,
in the real representation with no implicit summation,
\be
  \zeta_{ijk}
    = Q \left( \xi_{ij} \xi_{jk} + \xi_{jk} \xi_{ki} + \xi_{ki} \xi_{ij} \right)
\ee
\ba
  \eta_{ijkl}
    \!\!\!&=&\!\!\!
      R_a \bigl[
      \xi_{ij} \xi_{jk} \xi_{kl}  + \mbox{cyclic (12 snake terms)} \bigr]
  \nn
  && \!\!\!\!\!\!
    \mbox{} + 
      R_b \bigl[
      \xi_{ij} \xi_{ik} \xi_{il}  + \mbox{cyclic (4 star terms)} \bigr]
\ea
with approximately constant hierarchical amplitudes $Q$, $R_a$, and $R_b$.
On the other hand
it is clear that the hierarchical amplitudes do vary at some level,
both as a function of scale and configuration shape.

In the translinear regime,
perturbation theory predicts that the hierarchical amplitudes should
vary (somewhat) with both scale and configuration,
for density fluctuations growing by gravity from Gaussian initial conditions
(Fry 1984;
Scoccimarro et al.\ 1998).

In the deeply nonlinear regime,
predictions for the behaviour of the hierarchical amplitudes
are more empirical.
Scoccimarro \& Frieman (1999) have recently suggested an ansatz,
which they dub hyperextended perturbation theory (HEPT),
that the hierarchical amplitudes in the highly nonlinear regime
go over to the values predicted by perturbation theory
for configurations collinear in Fourier space.
For power law power spectra $\xi(k) \propto k^n$,
HEPT predicts a 3-point amplitude
\be
\label{Q3}
  Q = {4 - 2^n \over 1 + 2\ 2^n}
\ee
and 4-point amplitudes $R_a = R_b = Q_4$ with
\be
\label{Q4}
  Q_4 = {54 - 27\ 2^n + 2\ 3^n + 6^n \over 2 \, (1 + 6\ 2^n + 3\ 3^n + 6\ 6^n)}
 \ .
\ee

For simplicity, the present paper adopts the hierarchical model,
with constant hierarchical amplitudes set equal to the HEPT values~(\ref{Q3})
and (\ref{Q4}).
For reasons to be discussed shortly
(namely that the Schwarz inequality is violated otherwise),
most of the calculations shown take
\be
  R_a = - R_b = Q_4
\ee
although where possible results are also shown for
\be
  R_a = R_b = Q_4
  \ .
\ee
In addition to power law power spectra,
the present paper shows results for
the power spectrum derived from observations by Peacock (1997),
and for an observationally concordant $\Lambda$CDM model
from the fitting formulae of Eisenstein \& Hu (1998),
nonlinearly evolved according to the procedure of Peacock \& Dodds (1996).
In these cases
the adopted amplitudes are those corresponding to $n = -1.2$,
i.e.\ a correlation function with slope $\gamma = n + 3 = 1.8$,
for which $Q = 1.906$ and $Q_4 = 4.195$.

In the hierarchical model with constant hierarchical amplitudes,
the 4-point, 3-point, and 2-point contributions to
the reduced covariance matrix $\fC_{\alpha\beta}$,
equation~(\ref{CKJI}), are,
in the Fourier representation with no implicit summation,
\ba
  K(k_\alpha,k_\beta)
    \!\!\!&=&\!\!\!
      (2\PI)^3 \deltaD(k_\alpha\!-\!k_\beta) \, \xi(k_\alpha)^2
  \nn
  &&
    \mbox{} +
      R_a [ \xi(k_\alpha) + \xi(k_\beta) ]^2 A(k_\alpha,k_\beta)
  \nn
  &&
    \mbox{} +
      R_b \, \xi(k_\alpha) \xi(k_\beta) [ \xi(k_\alpha) + \xi(k_\beta) ]
\label{Khier}
\ea
\ba
  J(k_\alpha,k_\beta)
    \!\!\!&=&\!\!\!
      (2\PI)^3 \deltaD(k_\alpha\!-\!k_\beta) \, \xi(k_\alpha)
  \nn
  &&
    \mbox{} +
      Q [ \xi(k_\alpha) + \xi(k_\beta) ] A(k_\alpha,k_\beta)
  \nn
  &&
    \mbox{} +
      Q \, \xi(k_\alpha) \xi(k_\beta)
\label{Jhier}
\ea
\be
  I(k_\alpha,k_\beta)
    =
      (2\PI)^3 \deltaD(k_\alpha\!-\!k_\beta) + A(k_\alpha,k_\beta)
\label{Ihier}
\ee
where
in the real space representation
the matrix $A_{\alpha\beta}$ is the diagonal matrix
\be
\label{Ar}
  A(r_\alpha,r_\beta)
    = \deltaD(r_\alpha\!-\!r_\beta) \, \xi(r_\alpha)
\ee
while in the Fourier representation $A_{\alpha\beta}$ is
\be
\label{Ak}
  A(k_\alpha,k_\beta)
    = {1 \over 2 \, k_\alpha k_\beta}
      \int_{|k_\alpha-k_\beta|}^{k_\alpha+k_\beta} \!\!\!
      \xi(k) \, k \dd k
  \ .
\ee

Convergence of $A(k_\alpha,k_\beta)$ at $k_\alpha = k_\beta$
requires that $\xi(k) \sim k^n$ with $n > -2$ at small wavenumber $k$.
Convergence of $\int \xi(r) \, 4\PI r^2 \dd r$
at small $r$ requires that $\xi(r) \sim r^{-\gamma}$
with $\gamma < 3$ at small separation $r$.
Thus for power law power spectra $\xi(k) \propto k^n$
(this is the evolved, nonlinear power spectrum,
not the original, linear power spectrum),
equivalent to power law correlation functions $\xi(r) \propto r^{-\gamma}$
with $\gamma = n + 3$,
the hierarchical model is consistent only for
\be
  -2 < n < 0
  \quad \mbox{or equivalently} \quad
  1 < \gamma < 3
  \ .
\ee

It is straightforward to determine that,
for power law power spectra $\xi(k) \propto k^n$ in the hierarchical limit
(where the Gaussian contribution becomes negligible),
the correlation coefficient of the 4-point contribution $K_{\alpha\beta}$
to the reduced covariance $\fC_{\alpha\beta}$ is,
for $k_\alpha \gg k_\beta$,
\be
  {K_{\alpha\beta} \over ( K_{\alpha\alpha} K_{\beta\beta} )^{1/2}}
  \rightarrow
    {(R_a + R_b) \over
    2 \left( \displaystyle{2^{n+2} \over n+2} R_a + R_b \right)}
    \left( {k_\beta \over k_\alpha} \right)^{n/2}
\ee
which diverges as $k_\alpha/k_\beta \rightarrow \infty$
(for $-2 < n < 0$) unless $R_b = - R_a$.
Thus the Schwarz inequality,
which requires that the absolute value of the correlation coefficient
be less than or equal to unity,
is violated unless $R_b = - R_a$.
This problem has been remarked and discussed by
Scoccimarro et al.\ (1999 \S3.3).
Scoccimarro et al.\ show from $N$-body simulations that
the traditional relation $R_a \approx R_b$ holds approximately
for $k_\alpha \approx k_\beta$,
but that indeed $R_a + R_b$ decreases systematically
as $k_\alpha$ and $k_\beta$ become more and more separated.
Scoccimarro et al.\ conclude that the simple hierarchical model with
constant amplitudes is not a good description of the 4-point
function in the highly nonlinear regime.

For simplicity,
the present paper adopts the hierarchical model with constant amplitudes,
and either $R_b = - R_a$ or $R_b = R_a$.
Ultimately, the latter choice leads to unphysically huge variances,
plainly a consequence of the violation of the Schwarz inequality.
Thus the canonical models in this paper have $R_b = - R_a$.
However, where possible, intermediate results are also shown for $R_b = R_a$.

\subsection{Prewhitening}
\label{prewhitening}

The minimum variance estimator $\hat\xi_\alpha$
and associated Fisher matrix $F^{\alpha\beta}$,
equations~(\ref{ximvFKP}) and (\ref{FmvFKPasym}),
involve 6-dimensional integrals of
$\fC^{-1\alpha\beta}(\nbar_i,\nbar_j)$
over all pairs $ij$ of volume elements in a survey.
This is actually quite a feasible numerical problem.
The reduced covariance matrix $\fC_{\alpha\beta}(\nbar_i,\nbar_j)$
is a rank 2 matrix of 1-dimensional quantities,
so is straightforward to invert numerically for any particular
values of the selection functions $\nbar_i$ and $\nbar_j$.
If, as is typical, the selection function separates into the product
of an angular mask and a radial selection function,
then the angular integrals can be done analytically
(Hamilton 1993),
leaving a double integral of $\fC^{-1\alpha\beta}(\nbar_i,\nbar_j)$
over the radial directions, which is doable.
This direct procedure is discussed further in \S\ref{full},
and forms the basis of the gourmet recipe summarized in \S\ref{gourmet}.
Still, the integration is burdensome,
and it is enlightening to explore whether further simplification is possible.

Ideally what one would like is that
there would exist a representation in which
$\fC_{\alpha\beta}(\nbar_i,\nbar_j)$
were simultaneously diagonal for arbitrary values of
the selection function $\nbar$.
Precisely this situation obtains in the case of Gaussian fluctuations,
for which the reduced covariance matrix $\fC_{\alpha\beta}$
is diagonal in Fourier space
\be
  \fC_{\alpha\beta}(\nbar_i,\nbar_j) =
    2 \, (2\PI)^3 \deltaD(k_\alpha\!-\!k_\beta)
    \bigl[ \xi(k_\alpha) + \nbar_i^{-1} \bigr]
    \bigl[ \xi(k_\alpha) + \nbar_j^{-1} \bigr]
\ee
regardless of the values $\nbar_i$ and $\nbar_j$ of the selection function.

For non-Gaussian fluctuations,
the reduced covariance $\fC_{\alpha\beta}(\nbar_i,\nbar_j)$
is a linear combination of 4-point, 3-point, and 2-point matrices
$K_{\alpha\beta}$,
$J_{\alpha\beta}$,
and $H_{\alpha\beta}$,
according to equation~(\ref{CKJI}).
Finding a representation in which
$\fC_{\alpha\beta}(\nbar_i,\nbar_j)$
is diagonal for any $\nbar_i$ and $\nbar_j$,
thus means diagonalizing the three matrices
$K$, $J$, and $H$ simultaneously.
This is of course generically impossible.

However, it is possible to diagonalize two
($K$ and $H$)
of the three matrices simultaneously
by the trick of prewhitening,
and to cross one's fingers on the third matrix ($J$).
The term prewhitening refers to the operation of multiplying
a signal by a function in such a way that the noise becomes white,
or constant
(Blackman \& Tukey 1959 \S11).
Prewhitening is commonly used in the construction of
Karhunen-Lo\`{e}ve modes (signal-to-noise eigenmodes),
in order to allow a signal and its noise to be diagonalized simultaneously
(Vogeley \& Szalay 1996;
Tegmark, Taylor \& Heavens 1997;
Tegmark et al.\ 1998).

Define the prewhitened reduced covariance $\fB_{\alpha\beta}$ to be
\be
\label{B}
  \fB \equiv H^{-1/2} \fC \, H^{-1/2}
\ee
and similarly define the prewhitened 4-point and 3-point matrices
$M_{\alpha\beta}$ and $L_{\alpha\beta}$ to be
\be
  M \equiv H^{-1/2} K \, H^{-1/2}
\ee
\be
  L \equiv H^{-1/2} J \, H^{-1/2}
  \ .
\ee
By construction, the prewhitened 2-point matrix is the unit matrix,
$H^{-1/2} H \, H^{-1/2} = \1$.
In terms of the prewhitened 4-point and 3-point matrices
$M_{\alpha\beta}$ and $L_{\alpha\beta}$,
the prewhitened reduced covariance $\fB_{\alpha\beta}(\nbar_i,\nbar_j)$ is
(compare eq.~[\ref{CKJI}])
\be
\label{BML1}
  \fB_{\alpha\beta}(\nbar_i,\nbar_j)
  = 2 \left[
    M_{\alpha\beta}
    + (\nbar_i^{-1}\!\!+\!\nbar_j^{-1}) L_{\alpha\beta}
    + \nbar_i^{-1} \nbar_j^{-1} \1_{\alpha\beta}
    \right]
  \ .
\ee

The properties of the prewhitened 4-point and 3-point matrices
$M$ and $L$ are examined in \S\ref{4+3point}.

\subsection{FFTLog}

Several of the manipulations described in this paper
involve transforming between real and Fourier space.
Ideally, one would like to be able to cover
several orders of magnitude in separation or wavenumber.
The SDSS, for example, should be able to probe scales
from $10^{-2} \, h^{-1} \Mpc$ to $10^3 \, h^{-1} \Mpc$,
a range of $10^5$.
If the Fourier transforms were done using standard Fast Fourier Transform
(FFT) techniques, which require lineary spaced points,
covering such a range would require $10^5$ points.
The trouble with this is that one would then have to manipulate
$10^5 \times 10^5$ matrices.
Clearly this is a problem of the shoe not fitting the foot;
that is, a linear spacing of points
is not well suited to the case at hand:
while the difference between separations of
$0.01 \, h^{-1} \Mpc$ and $0.02 \, h^{-1} \Mpc$
may be significant, the difference between
$1000.01 \, h^{-1} \Mpc$ and $1000.02 \, h^{-1} \Mpc$
is practically irrelevant.

The problem may be solved by using an FFT method
originally proposed by Talman (1978),
that works for logarithmically spaced points,
and which I have implemented in a code FFTLog.
FFTLog is analogous to the normal FFT
in that it gives the exact Fourier transform of a discrete sequence
that is uniformly spaced and periodic in logarithmic space.
More generally,
FFTLog yields Fast Hankel (= Fourier-Bessel) Transforms
of arbitrary order, including both integral and $1/2$-integral orders.
FFTLog, like the normal FFT,
suffers from the usual problems of
ringing (response to sudden steps)
and aliasing (periodic folding of frequencies),
but under appropriate circumstances and with suitable precautions,
discussed in Appendix~B,
it yields reliable Fourier transforms
covering ranges of many orders of magnitude with modest numbers of points.

Appendix~B gives further details of FFTLog.
The code may be downloaded from
http:\discretionary{}{}{}$/\!/\discretionary{}{}{}$casa\discretionary{}{}{}.colorado\discretionary{}{}{}.edu\discretionary{}{}{}/$\sim$ajsh\discretionary{}{}{}/FFTLog/\ .

\section{Prewhitened 4-point and 3-point covariance matrices}
\label{4+3point}

\subsection{Computation}
\label{comp4+3point}

Before showing pictures,
it is helpful to comment on the numerical computation of the
4-point and 3-point covariance matrices $K_{\alpha\beta}$ and $J_{\alpha\beta}$
and their prewhitened counterparts $M_{\alpha\beta}$ and $L_{\alpha\beta}$.

Equations~(\ref{Khier}) and (\ref{Jhier})
give expressions for the 4-point and 3-point matrices
$K(k_\alpha,k_\beta)$ and $J(k_\alpha,k_\beta)$
in Fourier space,
for the hierarchical model with constant hierarchical amplitudes.
These are discretized as described in \S\ref{discrete}.
An issue here is the calculation of the subsidiary matrix
$A(k_\alpha,k_\beta)$.
This matrix $A_{\alpha\beta}$ is diagonal in real space
with diagonal entries $\xi(r_\alpha)$, equation~(\ref{Ar}),
so one way to calculate $A(k_\alpha,k_\beta)$ is to start with the diagonal
matrix $A(r_\alpha,r_\beta)$ in real space,
and then Fourier transform it into Fourier space.
Unfortunately the resulting Fourier transformed matrix $A(k_\alpha,k_\beta)$
shows evident signs of ringing and aliasing,
which is true whether the wavenumbers $k_\alpha$ are linearly spaced (FFT)
or logarithmically spaced (FFTLog).
Part of the difficulty is that the diagonal matrix $A(r_\alpha,r_\beta)$
is liable to vary by several orders of magnitude along the diagonal;
since the FFT (or FFTLog) assumes that the matrix is periodic,
the matrix appears to have a sharp step at its boundary.
These problems can be reduced by padding the matrix,
and in the case of FFTLog by biasing the matrix with a suitable power law
(see Appendix~B).
Still, artefacts from the FFT remain a concern.

A more robust procedure, the one used in this paper,
is to avoid FFTs altogether,
and to calculate the matrix $A(k_\alpha,k_\beta)$ directly from
its Fourier expression~(\ref{Ak}).

A similar issue arises when prewhitening the 4-point and 3-point matrices
$K$ and $J$.
The prewhitening matrix $H^{-1/2} = (\1 + A)^{-1/2}$ is again diagonal
in real space, with diagonal entries $[1+\xi(r)]^{-1/2}$.
Thus one way to prewhiten $K$ (say) is to start with $K(k_\alpha,k_\beta)$
in Fourier space,
Fourier transform it into real space,
prewhiten
$M(r_\alpha,r_\beta)
= [1+\xi(r_\alpha)]^{-1/2}$\discretionary{}{}{}$K(r_\alpha,r_\beta)$\discretionary{}{}{}$[1+\xi(r_\beta)]^{-1/2}$,
and then Fourier transform back into Fourier space.
Once again the resulting matrix
$M(k_\alpha,k_\beta)$
shows signs of ringing and aliasing.

Again, a more robust procedure, the one used in this paper,
is to avoid FFTs,
and to calculate the prewhitening matrix $H^{-1/2} = (\1 + A)^{-1/2}$
directly in Fourier space.
Specifically,
take the Fourier expression~(\ref{Ak}) for $A(k_\alpha,k_\beta)$,
add the unit matrix $\1$ to form $H$,
and evaluate the inverse positive square root $H^{-1/2}$
via an intermediate diagonalization.
This yields the prewhitening matrix $H^{-1/2}$ in Fourier space,
which can be used directly to prewhiten the
4-point and 3-point covariances matrices $K$ or $J$
in Fourier space.
This manner of constructing $H^{-1/2}$ guarantees that
the prewhitened 2-point covariance matrix
$H^{-1/2} H \, H^{-1/2}$ is numerically equal to the unit matrix $\1$,
as it should be.
Although this procedure is slower than using FFTs,
it yields results that are robust with respect to range, resolution,
and linear or logarithmic binning,
and consistent with the results from FFTs
if due care is taken with the latter.

\Mfig

\Mwfig

\Msfig

\subsection{Prewhitened 4-point covariance matrix}
\label{4point}

Figure~\ref{M} shows the correlation coefficient
$K_{\alpha\beta}/(K_{\alpha\alpha} K_{\beta\beta})^{1/2}$
(no implicit summation)
of the 4-point contribution $K_{\alpha\beta}$
to the (unprewhitened) reduced covariance matrix $\fC_{\alpha\beta}$,
equation~(\ref{CKJI}),
for the case of a power law power spectrum
having correlation function $\xi(r) = (r/5 \, h^{-1} \Mpc)^{-1.8}$.
Physically,
the quantity plotted is the (correlation coefficient of) the covariance
of estimates of power in the case of a perfect survey with no shot-noise,
$\nbar \rightarrow \infty$.

The correlation coefficient offers a good way to visualize the covariance,
since a value of $(-)1$ means two quantities are perfectly
(anti-)correlated,
and the Schwarz inequality requires that the absolute value of the
correlation coefficient always be less than or equal to unity.

The Gaussian spikes evident in the curves on the leftward, linear, side
of Figure~\ref{M}
reflect the fact that the covariance of power becomes diagonal in the linear,
Gaussian regime.
In the nonlinear regime,
the hierarchical contribution to the covariance dominates,
and the covariance of power becomes quite broad,
a point previously made by
Meiksin \& White (1999)
and Scoccimarro et al.\ (1999).

It should be borne in mind that the shape of the correlation coefficient
shown in Figure~\ref{M} depends on the resolution in wavenumber $k$,
a point emphasized by Scoccimarro et al.\ (1999).
In Figure~\ref{M} the points are logarithmically spaced
with 128 points per decade, so $\Delta\log k = 1/128$.
However, the correlation coefficient varies in an unsurprising way:
as the resolution increases, the Gaussian spikes gets spikier,
tending in principle to a Dirac delta-function
in the limit of infinite resolution.

Figure~\ref{Mw} shows the correlation coefficient
$M_{\alpha\beta}$\discretionary{}{}{}$/$\discretionary{}{}{}$(M_{\alpha\alpha} M_{\beta\beta})^{1/2}$
of the 4-point contribution $M_{\alpha\beta}$
to the prewhitened reduced covariance $\fB_{\alpha\beta}$,
equations~(\ref{B}) and (\ref{BML1}),
again for the case of a power law power spectrum
having correlation function $\xi(r) = (r/5 \, h^{-1} \Mpc)^{-1.8}$.
The only difference between this Figure and Figure~\ref{M}
is that the covariance is now prewhitened.

The prewhitened covariance $M$ plotted in Figure~\ref{Mw}
appears to be remarkably narrow,
certainly substantially narrower than the covariance
shown in Figure~\ref{M}.
The Gaussian spikes again show up in the linear regime,
and again the hierarchical contribution to the prewhitened covariance
dominates in the nonlinear regime.
The hierarchical contribution appears empirically to have
a constant width of $\Delta k \approx \PI/r_0 \approx 1 \, h \, \Mpc^{-1}$,
where $r_0 = 5 \, h^{-1} \Mpc$ is the correlation length.
Thus the prewhitened covariance appears to become relatively narrower
at large wavenumber $k$.

Figure~\ref{Ms} shows the correlation coefficients
of the covariance of the power,
both straight $K$ and prewhitened $M$,
for several other power spectra.
In each case the covariance of power with the power at
$k = 1 \, h \, \Mpc^{-1}$
is plotted,
which is essentially the `worst case',
where the prewhitened covariance $M$ is relatively broadest.

The solid lines in Figure~\ref{Ms}
are for 4-point hierarchical amplitudes $R_b = - R_a$,
while the dashed lines are for $R_b = R_a$.
As discussed in \S\ref{hierarchical},
the hierarchical model violates
the Schwarz inequality at $k_\alpha \gg k_\beta$ (or $k_\alpha \ll k_\beta$)
unless $R_b = - R_a$.

Figure~\ref{Ms} illustrates that the pattern
encountered in Figures~\ref{M} and \ref{Mw}
is remarkably robust over different power spectra.
That is, while the covariance of the power is itself broad,
in all cases the covariance of the prewhitened power is substantially narrower,
at least for $R_b = - R_a$ (solid lines).
Note that the power law power spectra illustrated in Figure~\ref{Ms}
cover essentially the full range of indices, $1 < \gamma < 3$,
allowed by the hierarchical model.

\Mwthreefig

The situation for $R_b = R_a$ is muddier.
Although the core of the prewhitened covariance is for the most part
reasonably narrow also in this case,
the off-diagonal covariances at
$k_\alpha \gg k_\beta$ (or $k_\alpha \ll k_\beta$)
are starting to become worrying large in several cases.
Some of this behaviour is undoubtedly inherited from the unphysical
(Schwarz-inequality-violating) behaviour of the ordinary covariance,
and is surely not realistic.
Here I leave the problem with the comment that
further investigation is clearly required,
along the lines being pioneered by
Scoccimarro et al.\ (1999).

\subsection{Prewhitened 3-point covariance matrix}
\label{3point}

As discussed in \S\ref{prewhitening},
it would be ideal if the prewhitened 3-point contribution
$L_{\alpha\beta}$ to the covariance of power
were diagonal in the same representation as the 4-point contribution
$M_{\alpha\beta}$.

\Mwthreesfig

Figure~\ref{Mw3}
shows the correlation coefficient of the prewhitened 3-point covariance
$L_{\alpha\beta}$ in the representation of eigenfunctions of the
prewhitened 4-point covariance $M_{\alpha\beta}$,
for the case of $\xi(r) = (r/5 \, h^{-1} \Mpc)^{-1.8}$.
The horizontal axis here is a nominal wavenumber $k_\alpha$
labelling each eigenfunction $\phi_\alpha(k)$
of the prewhitened 4-point covariance $M_{\alpha\beta}$.
In practice, the eigenfunctions are simply ordered by eigenvalue,
which in most cases (see below) yields a satisfactory ordering by
wavenumber, in the sense that the corresponding eigenfunctions
$\phi_\alpha(k)$ have their largest components around $k \approx k_\alpha$.

At first sight, the correlation coefficient plotted in Figure~\ref{Mw3}
looks astonishingly diagonal at all wavenumbers,
for both $R_b = - R_a$ and $R_b = R_a$.
However, as Scoccimarro et al.\ (1999) emphasize,
off-diagonal elements, though they may be small, are many.
The resolution in Figure~\ref{Mw3} is 128 points per decade,
and the off-diagonal elements in the case $R_b = - R_a$
are down at the level of $\sim 1/100$,
which means that cumulative off-diagonal covariance over a decade
of wavenumber would be comparable to the diagonal variance.
Curiously,
the off-diagonal elements are somewhat smaller for $R_b = R_a$
than for $R_b = - R_a$.

In \S\ref{fisher} and thereafter
the approximation will be made, equation~(\ref{Bnice}),
that the 3-point matrix $L$ is indeed
diagonal in the representation of 4-point eigenfunctions.
If $L$ is not precisely diagonal, then
the `minimum variance' pair-weighting that emerges from assuming
diagonality will not be precisely minimum variance.
But a linear error in the pair-weighting
will raise the variance quadratically from its minimum,
so the pair-weighting should be close to minimum variance
as long as $L$ is not too far from being diagonal.
In any case, as discussed in \S\ref{estimate},
the estimate of power remains unbiased
whatever approximations are made.

Figure~\ref{Mw3s}
shows the correlation coefficient of the prewhitened 3-point covariance
$L_{\alpha\beta}$ in the representation of eigenfunctions of the
prewhitened 4-point covariance $M_{\alpha\beta}$
for a number of different power spectra,
at a representative nominal wavenumber $k_\alpha = 1 \, h \, \Mpc^{-1}$.
The Figure illustrates that this correlation coefficient
remains remarkably diagonal for all power spectra.
Again, the range of power law power spectra shown
covers essentially the full range $1 < \gamma < 3$
allowed by the hierarchical model.

In the case $\gamma = 2.9$,
the off-diagonal elements of the correlation coefficient
shown in Figure~\ref{Mw3s}
appear to bounce around,
even though taken as a whole the correlation correlation coefficient
appears more diagonal in this case than any other.
The apparent noise is caused by a near degeneracy of eigenvalues.
Such degeneracy is not too surprising,
since in the limit $\gamma \rightarrow 3$,
the prewhitened 3-point and 4-point matrices
$L_{\alpha\beta}$ and $M_{\alpha\beta}$
are both expected to become proportional to the unit matrix.
Numerically,
for both 3-point and 4-point matrices,
there is a degeneracy of eigenvalues between eigenfunctions
at small and large wavenumbers
(in the sense that eigenfunctions with nearly the same eigenvalue
may have their largest components at either small or large wavenumbers):
the eigenvalues are larger at small and large wavenumber,
and go through a minimum at intermediate wavenumber.
The degeneracy causes mixing of the eigenfunctions at small and
large wavenumber,
making the correspondence between eigenvalue and nominal wavenumber
ambiguous,
and resulting in the oscillations in the off-diagonal components
apparent in Figure~\ref{Mw3s}.

\subsection{4-point and 3-point eigenvalues}
\label{eigenvalues}

\diagfig

Denote the eigenvalues of the 4-point and 3-point prewhitened
covariance matrices $M$ and $L$ by
\be
\label{mu}
  M \, \phi_\alpha = \mu_\alpha^2 \phi_\alpha
\ee
\be
\label{lambda}
  L \, \varphi_\alpha = \lambda_\alpha \varphi_\alpha
\ee
(no implicit summation on the right hand side)
so that for Gaussian fluctuations the eigenvalues
$\mu_\alpha$ and $\lambda_\alpha$ would be
$\mu_\alpha = \lambda_\alpha = \xi(k_\alpha)$.

Figure~\ref{diag}
shows the ratio $\mu_\alpha/\xi(k_\alpha)$ of the 4-point eigenvalues
$\mu_\alpha$ to the nonlinear power spectrum $\xi(k_\alpha)$,
plotted as a function of the nominal wavenumber $k_\alpha$,
which labels the eigenfunctions $\phi_\alpha$ ordered by eigenvalue,
for a power law power spectrum with correlation function
$\xi(r) = (r/5 \, h^{-1} \Mpc)^{- 1.8}$.
The eigenvalue is comparable to the power spectrum at all wavenumbers,
$\mu_\alpha \sim \xi(k_\alpha)$.
In the Gaussian, small $k_\alpha$ regime the eigenvalue is equal to the
power spectrum, $\mu_\alpha = \xi(k_\alpha)$, as expected,
while in the hierarchical, large $k_\alpha$ regime
the eigenvalue asymptotes to close to $2 \, R_a^{1/2}$ times the power spectrum,
$\mu_\alpha \approx 2 \, R_a^{1/2} \xi(k_\alpha)$.
Similar behaviour is found for other power spectra (not plotted),
and for the 3-point eigenvalue $\lambda_\alpha$,
which in the hierarchical regime asymptotes to
$\lambda_\alpha \approx 2 \, Q \, \xi(k_\alpha)$.

\diagthreefourfig

Figure~\ref{diag34}
shows the ratio $\lambda_\alpha/\mu_\alpha$ of
3-point to 4-point eigenvalues,
as a function of the nominal wavenumber $k_\alpha$,
for several power spectra.
Remarkably,
the ratio $\lambda_\alpha/\mu_\alpha$ of eigenvalues is quite close to unity
at all wavenumbers and for all power spectra.
The case $\gamma = 2.9$ is not plotted,
in part because of the same problem of mixing of eigenfunctions shown
in Figure~\ref{Mw3s}.
In any case,
for $\gamma = 2.9$ the ratio $\lambda_\alpha/\mu_\alpha$
differs from unity by less than 1~percent at all wavenumbers.
Analytically,
the ratio is expected to equal one in the limit $\gamma \rightarrow 3$.

In the $\Lambda$CDM model,
the eigenfunctions $\phi_\alpha$ (and $\varphi_\alpha$)
mix where the eigenvalues $\mu_\alpha$ (and $\lambda_\alpha$)
are degenerate, which happens because the $\Lambda$CDM power spectrum
goes through a maximum at $k \approx 0.017 \, h \, \Mpc^{-1}$.
For the purpose of plotting the ratio $\lambda_\alpha/\mu_\alpha$
for the $\Lambda$CDM model in the bottom panel of Figure~\ref{diag34},
this mixing was avoided by the device of truncating
the matrices $M_{\alpha\beta}$ and $L_{\alpha\beta}$
at a wavenumber close to the peak.
Mixing causes no problems for the evaluation of the
minimum variance estimator and Fisher matrix of the prewhitened power spectrum
in \S\ref{fisher} and \S\ref{power}
(so there is no need to truncate the matrices in general),
but mixing does muddy the physical interpretation of the eigenfunctions.

Curiously,
the ratios $\mu_\alpha/\xi(k_\alpha)$ and $\lambda_\alpha/\xi(k_\alpha)$,
regarded as functions of the nominal wavenumber $k_\alpha$,
vary with the resolution $\Delta\log k$ of the matrix,
as illustrated in Figures~\ref{diag} and \ref{diag34}
for the case $\gamma = 1.8$.
In the Gaussian limit of small $k_\alpha$,
the ratios do not change with resolution,
but in the hierarchical limit of large $k_\alpha$,
the ratios seems to shift
(to the right on the Figures, as the resolution increases)
in such a way that the ratios are functions of the product
$k_\alpha \Delta\log k$.
At intermediate $k_\alpha$, the shift is intermediate.
Now the wavenumber $k_\alpha$ is only a nominal wavenumber,
a labelling of the eigenfunctions ordered by eigenvalue,
and it is only in the Gaussian regime that
the eigenmodes are Fourier modes and
the correspondence between nominal and true wavenumber is precise.
Still, the shift seems surprising;
for example, in the limit of infinite resolution $\Delta\log k \rightarrow 0$,
the ratio $\mu_\alpha/\xi(k_\alpha)$ plotted in Figure~\ref{diag}
would shift to the right so far that $\mu_\alpha/\xi(k_\alpha)$
would equal 1 at all finite wavenumbers.
Similarly,
the ratio $\lambda_\alpha/\mu_\alpha$ plotted in Figure~\ref{diag34}
would shift to the right so far that $\lambda_\alpha/\mu_\alpha$
would equal 1 at all finite wavenumbers.
Numerically, to the limit that I have tested it ($\Delta\log k = 1/1024$),
this is indeed what seems to happen:
both $\mu_\alpha/\xi(k_\alpha)$ and $\lambda_\alpha/\xi(k_\alpha)$,
hence also their ratio $\lambda_\alpha/\mu_\alpha$,
shift to the right together as the resolution increases,
for all power spectra.

This does not appear to be a numerical error,
because `observable' quantities computed
via the eigenfunctions $\phi_\alpha$ and their eigenvalues $\mu_\alpha$,
such as the error bars attached to the prewhitened power spectrum $\hat X(k)$
in Fourier space (\S\ref{power}),
appear robust against changes in resolution.

\Xfig

Examination of the eigenfunctions of the 4-point and 3-point
matrices $M$ and $L$ reveals at least part of the reason why
their eigenvalues seem to shift as the resolution increases.
Figure~\ref{X}
shows a sampling of eigenfunctions $\phi_\alpha$ of the
4-point matrix $M$ for the case $\xi(r) = (r/5 \, h^{-1} \Mpc)^{-1.8}$,
at two different resolutions,
$\Delta\log k = 1/32$ and $1/128$.
Whereas in the Gaussian, small $k_\alpha$ regime
the eigenfunctions go over to delta-functions in Fourier space,
in the hierarchical, large $k_\alpha$ regime
the eigenfunctions grow ever wigglier as the resolution increases.
What seems to happen is that,
as the resolution increases,
eigenfunctions at neighbouring nominal wavenumbers $k_\alpha$ strive to remain
orthogonal to each other,
which they accomplish by becoming wigglier and wigglier.
To the limit that I have tested it numerically,
there seems to be no end to the wiggliness.
Given that there is no asymtotic limit to which the eigenfunctions
appear to tend, perhaps it is not surprising that their eigenvalues
should shift systematically too.
However, it would be nice to have a better understanding of what is going on.

\xikfig

\section{Prewhitened power spectrum}
\label{prewhiten}

\subsection{Definition}
\label{Xa}

Given the nice properties of the prewhitened covariance of power
established in the previous section, \S\ref{4+3point},
it makes sense to define a prewhitened power spectrum $X_\alpha$,
and a corresponding estimator $\hat X_\alpha$ thereof,
with the property that the covariance of the prewhitened power
equals the prewhitened covariance of power.

Define, therefore, the prewhitened power spectrum $X_\alpha$ by,
in the real space representation,
\be
\label{Xr}
  X(r) \equiv {2 \, \xi(r) \over 1 + [1+\xi(r)]^{1/2}}
  \ .
\ee
The expression~(\ref{Xr}) is equivalent to
$X(r) \equiv 2 \, [1+\xi(r)]^{1/2} - 2$,
but the former expression~(\ref{Xr}) is numerically stabler to evaluate
when $\xi(r)$ is small.
Similarly,
define an estimator $\hat X_\alpha$ of the prewhitened power
in terms of the minimum variance estimator $\hat\xi_\alpha$,
equation~(\ref{ximv}), of the power spectrum by,
again in the real space representation,
\be
\label{Xrhat}
  \hat X(r) \equiv {2 \, \hat\xi(r) \over 1 + [1+\hat\xi(r)]^{1/2}}
\ee
which by construction has the property that for small $\Delta \hat X(r)$,
as should be true in the limit of a large amount of data
(the following equation is essentially the derivative of eq.~[\ref{Xrhat}]),
\be
  \Delta \hat X(r) = {\Delta \hat\xi(r) \over [ 1 + \hat\xi(r) ]^{1/2}}
  \ .
\ee

The covariance of the estimate $\hat X_\alpha$
of the prewhitened power spectrum is given by
\be
  \langle \Delta \hat X_\alpha \Delta \hat X_\beta \rangle
  =
    (H^{-1/2})_\alpha^\gamma \,
    \langle \Delta \hat\xi_\gamma \Delta \hat\xi_\delta \rangle \,
    (H^{-1/2})_\beta^\delta
  = E_{\alpha\beta}^{-1}
\ee
where the Fisher matrix $E^{\alpha\beta}$ of the prewhitened power
equals the prewhitened Fisher matrix of the power,
equation~(\ref{Fmv}),
\be
\label{E}
  E^{\alpha\beta}
  = (H^{1/2})^\alpha_\gamma \, F^{\gamma\delta} \, (H^{1/2})_\delta^\beta
  \ .
\ee

In \S\ref{power} it will be found convenient to deal with another
prewhitened estimator $\hat Y_\alpha$ defined by
\be
\label{Y}
  \hat Y_\alpha \equiv
    (H^{-1/2})_\alpha^\beta \, \hat\xi_\beta
  \ .
\ee
The prewhitened estimator $\hat Y_\alpha$
has the same covariance as $\hat X_\alpha$
\be
  \langle \Delta \hat Y_\alpha \Delta \hat Y_\beta \rangle
  =
  \langle \Delta \hat X_\alpha \Delta \hat X_\beta \rangle
  = E_{\alpha\beta}^{-1}
  \ .
\ee
So why not define $\hat Y_\alpha$ to be the prewhitened power?
The problem with the estimator $\hat Y_\alpha$ is that it depends
explicitly on the prior power spectrum $\xi_\alpha$.
That is, $\hat Y_\alpha$ in real space is
\be
\label{Yr}
  \hat Y(r) = {\hat\xi(r) \over [ 1 + \xi(r) ]^{1/2}}
\ee
which involves an estimated quantity $\hat\xi(r)$ in the numerator
and the prior quantity $\xi(r)$ in the denominator.
Imagine plotting $\hat Y_\alpha$ on a graph.
What is this quantity supposed to be an estimate of?
Obviously $\hat Y_\alpha$ is an estimate of
$Y_\alpha \equiv \langle \hat Y_\alpha \rangle
= (H^{-1/2})_\alpha^\beta \, \xi_\beta$.
But if one wanted to attach error bars to the estimate,
then to be fair one should include the full covariance
of the quantity being estimated, including the covariance that
arises from the denominator $[ 1 + \xi(r) ]^{1/2}$ in equation~(\ref{Yr}),
not just the covariance
$\langle \Delta \hat Y_\alpha \Delta \hat Y_\beta \rangle$
with the denominator held fixed.
Indeed, if one goes through the usual ML cycle
of permitting the data to inform the prior,
so that the estimated $\hat\xi(r)$ is inserted into the denominator
of equation~(\ref{Yr}),
then it becomes abundantly evident that it would be correct to include
covariance arising from the denominator.

To avoid confusion,
it should be understood that the quantities $\hat Y_\alpha$
are of course perfectly fine for carrying out ML estimation of parameters.
In ML estimation,
`error bars are attached to the model, not to the data',
to quote another of the refrains from the 1997 Aspen workshop
on Precision Measurement of Large Scale Structure.
Whereas in ML parameter estimation with $\hat X_\alpha$
one might form a likelihood from the `data' quantities
$\Delta\hat X(r) = 2 \, \hat\xi(r)/\{ 1 + [ 1 + \hat\xi(r) ]^{1/2}\} -$\discretionary{}{}{}$\, 2 \, \xi(r)/\{ 1 + [ 1 + \xi(r) ]^{1/2} \}$,
in ML parameter estimation with $\hat Y_\alpha$
one would instead form a likelihood from the `data' quantities
$\Delta\hat Y(r) = [\hat\xi(r) - \xi(r)]/$\discretionary{}{}{}$[ 1 + \xi(r) ]^{1/2}$.

But, for the purpose of plotting quantities on a graph,
plainly it is the prewhitened power spectrum $\hat X_\alpha$
defined by equation~(\ref{Xrhat}) that should be plotted,
not $\hat Y_\alpha$.

\subsection{Picture}


Figure~\ref{xik}
shows prewhitened nonlinear power spectra $X(k)$,
along with linear and nonlinear power spectra $\xi_L(k)$ and $\xi(k)$,
for the observationally derived power spectrum
of Peacock (1997) with $\Omega_m = 0.3$,
and for a $\Lambda$CDM model of Eisenstein \& Hu (1998)
with observationally concordant parameters as indicated on the graph.

The nonlinear power spectra $\xi(k)$ were constructed from
the linear power spectra $\xi_L(k)$
according to the formula of Peacock \& Dodds (1996).
Amongst other things, the Peacock \& Dodds formula
depends on the logarithmic slope of the linear power spectrum.
Now the Eisenstein \& Hu power spectrum contains baryonic wiggles,
causing the slope to oscillate substantially,
whereas what Peacock \& Dodds had in mind was a rough average slope.
For the slope of the $\Lambda$CDM model in the Peacock \& Dodds formula,
I therefore used the slope of the `no-wiggle' power spectrum
provided by Eisenstein \& Hu as a smooth fit through the baryonic wiggles.
The alternative of using the wiggly slope has the additional demerit
that it amplifies baryonic wiggles in the nonlinear regime,
which is opposite to the suppression of baryonic wiggles in the nonlinear
regime observed in $N$-body simulations by Meiksin, White \& Peacock (1999).

The prewhitened power spectra $X(k)$ shown in Figure~\ref{xik}
were computed
by transforming the nonlinear power spectrum $\xi(k)$ into real space
using FFTLog (see Appendix~B, Fig.~\ref{xir}),
constructing the prewhitened power $X(r)$ from $\xi(r)$ according
to equation~(\ref{Xr}),
and Fourier transforming back.

The prewhitened power spectra shown in Figure~\ref{xik}
appear to be interestingly close to the linear power spectra,
$X(k) \approx \xi_L(k)$,
another one-eyebrow-raising property of the prewhitened power spectrum.
But surely this is just coincidence,
since for a primordial power spectrum $\xi(k) \propto k^n$
the prewhitened correlation in the highly nonlinear regime
should go as $X(r) \approx 2 \, \xi(r)^{1/2}
\propto r^{-3 (n + 3)/ 2 (n + 5)}$
assuming stable clustering
(Peebles 1980, eq.~[73.12]),
whereas the linear power spectrum would go as $r^{-(n + 3)}$,
whose power law exponents agree only in the limiting case $n \rightarrow -3$.
Still, the coincidence is curious.


Figure~\ref{xik}
points up one defect of the prewhitened power spectrum,
which is that, surprisingly enough,
it does not reproduce the linear power spectrum
at the very largest scales (small $k$).
Indeed the prewhitened power goes negative
in the Peacock (1997) case at
$k \approx 0.0023 \, h \, \Mpc^{-1}$,
and in the $\Lambda$CDM case at
$k \approx 0.00021 \, h \, \Mpc^{-1}$.
This turns out to be a generic feature of the prewhitened power spectrum
if the true power spectrum goes to zero at zero wavenumber,
as is true for Harrison-Zel'dovich models,
$\xi(k) \propto k$ as $k \rightarrow 0$.
For if it is true that the power spectrum $\xi(k)$
goes to zero at zero wavenumber $k$
\be
  \lim_{k \rightarrow 0} \xi(k) =
  \int_0^\infty \xi(r)
  \, 4 \PI r^2 \dd r
  = 0
\ee
then it follows that the prewhitened power must go to a negative constant
at zero wavenumber
\be
  \lim_{k \rightarrow 0} X(k) =
  \int_0^\infty {2 \, \xi(r) \over 1 + [1+\xi(r)]^{1/2}}
  \, 4 \PI r^2 \dd r
  < 0
\ee
since the factor $2/\{1 + [1+\xi(r)]^{1/2}\}$ in the integrand is
less than one for all positive $\xi(r)$,
and greater than one for all negative $\xi(r)$.
It is not clear what to do about this,
if indeed anything needs to be done.
Adding a constant to $X(k)$ and $\hat X(k)$
(which would leave $\Delta \hat X$, hence the covariance
$\langle \Delta \hat X \Delta \hat X \rangle$, unchanged)
would spoil the nice behaviour of the prewhitened power
in the nonlinear regime.

\section{Fisher matrix of prewhitened nonlinear power in a survey}
\label{fisher}

It was found in \S\ref{4+3point}
that the prewhitened reduced covariance $\fB$ of power appears to have some
unexpectedly pleasant properties:
first, the prewhitened covariance is surprisingly narrow in Fourier space;
second, the 4-point and 3-point contributions $M$ and $L$,
equation~(\ref{BML1}),
to the prewhitened reduced covariance $\fB$ are almost simultaneously diagonal
(the 2-point contribution is by construction the unit matrix,
so is automatically diagonal in any representation);
third, the 4-point and 3-point eigenvalues $\mu_\alpha$ and $\lambda_\alpha$,
as defined by equations~(\ref{mu}) and (\ref{lambda}),
are approximately equal;
and fourth,
all these results hold for all power spectra tested.

It should be emphasized that the pleasant properties of the prewhitened
power are not perfect,
and that they are premised on the validity of the hierarchical model
with constant hierarchical amplitudes,
which as discussed in \S\ref{4point}
is certainly wrong at some level.

These pretty properties
lead to an approximate expression, equation~(\ref{Emvnice}),
for the Fisher matrix of the prewhitened nonlinear power spectrum
of a galaxy survey,
which looks the same as the FKP approximation to the Fisher matrix
of the power in the linear, Gaussian case,
with the difference that the eigenmodes of
the prewhitened covariance $M$ of the nonlinear power
take the place of the Fourier modes in the linear case.

\subsection{Fisher matrix}

To the extent that
the prewhitened 4-point and 3-point matrices $M$ and $L$
are simultaneously diagonal,
the prewhitened reduced covariance matrix
$\fB_{\alpha\beta}(\nbar_i,\nbar_j)$ is
diagonal in the representation of eigenfunctions $\phi_\alpha$ of $M$ and $L$,
with
\be
  \fB_{\alpha\beta}(\nbar_i,\nbar_j)
  \approx
    2 \, \1_{\alpha\beta}
    \left[
    \mu_\alpha^2
    + (\nbar_i^{-1} \!\!+\! \nbar_j^{-1}) \lambda_\alpha
    + \nbar_i^{-1} \nbar_j^{-1}
    \right]
  \ .
\ee
To the further extent that $\lambda_\alpha \approx \mu_\alpha$,
the prewhitened covariance matrix
$\fB_{\alpha\beta}(\nbar_i,\nbar_j)$ is just
\be
\label{Bnice}
  \fB_{\alpha\beta}(\nbar_i,\nbar_j)
  \approx
    2 \, \1_{\alpha\beta} \,
    (\mu_\alpha + \nbar_i^{-1}) (\mu_\alpha + \nbar_j^{-1})
  \ .
\ee

The Fisher matrix $F^{\alpha\beta}$ of the power spectrum is given
in the FKP approximation by equation~(\ref{FmvFKPasym}).
In terms of the prewhitened reduced covariance $\fB_{\alpha\beta}$,
the Fisher matrix $F^{\alpha\beta}$ is
\be
  F^{\alpha\beta}
  =
    (H^{-1/2})^\alpha_\delta \,
    \fB^{-1\delta\epsilon}
    (H^{-1/2})^\gamma_\epsilon \,
    D_\gamma^{ij} D^\beta_{ij}
  \ .
\ee
Now $(H^{-1/2})^\gamma_\epsilon$ commutes with $D_\gamma^{ij} D^\beta_{ij}$,
since both are simultaneously diagonal in real space.
It follows that the Fisher matrix
$E \equiv H^{1/2} F \, H^{1/2}$
of the prewhitened power, equation~(\ref{E}),
is, in the FKP approximation,
\be
\label{Emvasym}
  E^{\alpha\beta}
  = \fB^{-1\alpha\gamma}
    D_\gamma^{ij} D^\beta_{ij}
  \ .
\ee
Like $F^{\alpha\beta}$,
the prewhitened Fisher matrix $E^{\alpha\beta}$ is asymmetric,
inheriting its asymmetry from the FKP approximation, equation~(\ref{Cab}).

To the extent that the approximation~(\ref{Bnice}) to $\fB$ is true,
it follows from equation~(\ref{Emvasym}) that
the Fisher matrix $E_{\alpha\beta}$ of prewhitened power
in the FKP approximation is,
in the representation of eigenfunctions $\phi_\alpha$ of the prewhitened
covariance,
\be
\label{Emvnice}
  E_{\alpha\beta}
  =
    \int_0^\infty \phi_\alpha(r) \phi_\beta(r) R(r;\mu_\alpha) \, 4\PI r^2 \dd r
\ee
where $R(r;\mu_\alpha)$ are FKP-weighted pair integrals
(commonly denoted $\langle R R \rangle$ in the literature, for Random-Random)
\be
\label{R}
  R(r;\mu_\alpha)
    = \int
      {\deltaD(r_{ij}-r)
      \over 2 \,
      [\mu_\alpha + \nbar(\r_i)^{-1}] [\mu_\alpha + \nbar(\r_j)^{-1}]}
      \, \ddd r_i \ddd r_j
\ee
the integration being taken over all pairs of volume elements $ij$ separated by
$r_{ij} \equiv |\r_i-\r_j| = r$ in the survey.

The FKP approximation to the Fisher matrix $E_{\alpha\beta}$
of prewhitened power, equation~(\ref{Emvnice}),
takes the same form as the FKP approximation to the Fisher matrix
of the power spectrum for Gaussian fluctuations
derived in \S5 of Paper~1 and computed in \S3 of Paper~2.
The difference is that the eigenfunctions $\phi_\alpha(r)$
and their eigenvalues $\mu_\alpha$ here
take the place of the Fourier eigenfunctions $j_0(k_\alpha r)$
and their eigenvalues $\xi(k_\alpha)$ in the Gaussian case.

\subsection{Numerics}
\label{compE}

Equation~(\ref{Emvnice}) for $E_{\alpha\beta}$ involves the eigenfunctions
$\phi_\alpha(r)$ of the prewhitened 4-point matrix $M$ in real space,
whereas in \S\ref{comp4+3point} it was suggested that the most robust
way to compute $M$ is in Fourier space.
The problem is that FFTing the matrix $M$ from Fourier into real space
is liable to introduce ringing and aliasing, which one would like to avoid.

A more robust procedure is not to FFT $M$ into real space,
but rather to FFT the pair integrals $R(r;\mu_\alpha)$ into Fourier space;
this is the same procedure adopted in \S3 of Paper~2
(except that $R$ here is $1/2$ that of Paper~2).
If $\mu_\alpha$ is treated, temporarily, as a constant,
then equation~(\ref{Emvnice}) can be transformed into real space to yield
the diagonal matrix
\be
\label{Errm}
  E(r,r';\mu_\alpha) = \deltaD(r-r') \, R(r;\mu_\alpha)
  \ .
\ee
Beware of equation~(\ref{Errm})!
It does not signify that the Fisher matrix is diagonal in real space,
because the constant $\mu_\alpha$ is different for each row
of the Fisher matrix $E_{\alpha\beta}$.
The Fourier transform of $E(r,r';\mu_\alpha)$ is
$E(k,k';\mu_\alpha)
= \int j_0(kr) j_0(k'r) \, R(r;\mu_\alpha) \, 4\PI r^2 \dd r$,
which simplifies to
\be
\label{Ekkm}
  E(k,k';\mu_\alpha)
  =
    {\PI \over k k'}
    \left[ \tilde R(k\!-\!k';\mu_\alpha) - \tilde R(k\!+\!k';\mu_\alpha) \right]
\ee
where $\tilde R(k;\mu_\alpha)$ is the 1-dimensional cosine transform
of $R(r;\mu_\alpha)$
\be
\label{Rk}
  \tilde R(k;\mu_\alpha) \equiv
    2 \int_0^\infty \! \cos(kr) \, R(r;\mu_\alpha) \, \dd r
  \ .
\ee
Transforming $E(k,k';\mu_\alpha)$ into $\phi_\alpha$-space gives
\be
\label{Emvnicek}
  E_{\alpha\beta}
  = \int
    \phi_\alpha(k) \phi_\beta(k') \,
    E(k,k';\mu_\alpha) \,
    {4\PI k^2 \dd k \, 4 \PI k'^2 \dd k' \over (2\PI)^6}
  \ .
\ee
The cosine transform $\tilde R(k;\mu_\alpha)$, equation~(\ref{Rk}),
can be done with either FFT or FFTLog; both work well.
To ensure that
$\tilde R(k;\mu_\alpha)$ remains accurate at large (and small) wavenumbers $k$,
it helps to extrapolate $R(r;\mu_\alpha)$ to small (and large) separations $r$
before transforming.
The transformation into $\phi_\alpha$ space, equation~(\ref{Emvnicek}),
is done by discrete summations.

Evaluating the Fisher matrix $E_{\alpha\beta}$ with
equations~(\ref{Ekkm})--(\ref{Emvnicek})
successfully eliminates ringing and aliasing,
but it introduces another problem.
The problem is that
equation~(\ref{Ekkm}) is liable to overestimate the value of
$E(k,k';\mu_\alpha)$
along the diagonal $k = k'$ if the gridding in $k$-space is too coarse
to resolve the diagonal properly,
as typically occurs at moderate and large $k$ with logarithmic gridding.
What is important is that the integral of $E(k,k';\mu_\alpha)$
over the diagonal be correct.
Integrating $E(k,k';\mu_\alpha)$ over $k'$ yields
\be
\label{Eintk}
  \int_0^\infty E(k,k';\mu_\alpha) \, k' \dd k'
  =
  {2\PI \over k} \int_0^k \tilde R(k';\mu_\alpha) \, \dd k'
  \ .
\ee
The integral on the right can be done conveniently and reliably
by sine transforming (with FFT or FFTLog) the pair integral
\be
\label{Rki}
  \int_0^k \tilde R(k';\mu_\alpha) \, \dd k'
  =
  2 \int_0^\infty \sin(kr) {R(r;\mu_\alpha) \over r} \, \dd r
  \ .
\ee
Discretized (\S\ref{discrete}) on a logarithmic grid of wavenumbers $k$,
the continuous matrix
$E(k,k';\mu_\alpha)$
becomes
$\bE_{k k'}(\mu_\alpha) =
E(k,k';\mu_\alpha)
\,$\discretionary{}{}{}$4\PI (k k')^{3/2} \Delta\ln k/(2\PI)^3$,
and equation~(\ref{Eintk}) becomes
\be
\label{Esumk}
  \sum_{k'} \, (k'/k)^{1/2} \, \bE_{k k'}(\mu_\alpha)
  = {1 \over \PI} \int_0^k \tilde R(k';\mu_\alpha) \, \dd k'
  \ .
\ee
Numerically,
if the left hand side of equation~(\ref{Esumk}),
with $\bE_{k k'}(\mu_\alpha)$ discretized from equation~(\ref{Ekkm}),
exceeds the right hand side of equation~(\ref{Esumk}),
evaluated by equation~(\ref{Rki}),
then the value of the diagonal element $\bE_{k k}(\mu_\alpha)$
should be reduced so that the sum is satisfied.
Ultimately, this procedure yields error bars on decorrelated band-powers
(Paper~4)
that are robust with respect to range, resolution,
and linear or logarithmic binning.

\subsection{Coarse gridding}
\label{coarse}

Typically the pair integral $R(r;\mu_\alpha)$ is broad in real space,
so its cosine transform $\tilde R(k;\mu_\alpha)$ is a narrow window
about $k \approx 0$
with a width comparable to the inverse scale length of the survey.
It follows that the matrix $E(k,k';\mu_\alpha)$ given by equation~(\ref{Ekkm})
is likewise narrow in $k$-space,
with a width comparable to the inverse scale length of the survey.
Moreover the sum in equation~(\ref{Esumk})
approximates $R(0;\mu_\alpha)$
at wavenumbers exceeding the inverse scale length of the survey,
which is to say at all except the largest accessible wavelengths:
\be
\label{Esumklim}
  \sum_{k'} \, (k'/k)^{1/2} \, \bE_{k k'}(\mu_\alpha)
  \approx R(0;\mu_\alpha)
  \quad \mbox{for} \quad
  k \gg \mbox{scale}^{-1}
\ee
where $R(0;\mu_\alpha)$ is the pair integral at zero separation
\be
  R(0;\mu_\alpha)
  = \int {\ddd r \over 2 \, [\mu_\alpha + \nbar(\r)^{-1}]^2}
  \ .
\ee

Thus if the matrix
$\bE_{k k'}(\mu_\alpha)$
is discretized on a grid that is coarse compared to the inverse scale length
of the survey, then it is approximately proportional to the unit matrix
\be
  \bE_{k k'}(\mu_\alpha) \approx
    \1_{k k'} \, R(0;\mu_\alpha)
  \ .
\ee
The resulting discrete Fisher matrix $\bE_{\alpha\beta}$,
equation~(\ref{Emvnicek}), is diagonal in the $\phi_\alpha$-representation
\be
\label{Eapprox}
  \bE_{\alpha\beta} \approx \1_{\alpha\beta} \, R(0;\mu_\alpha)
  \ .
\ee
The result~(\ref{Eapprox}) is analogous
to that obtained by FKP for Gaussian fluctuations.

Of course if this diagonal Fisher matrix, equation~(\ref{Eapprox}),
is transformed back into Fourier space, then it is no longer diagonal.
That is, equation~(\ref{Eapprox}) asserts that the Fisher matrix
of the prewhitened nonlinear power spectrum
is approximately diagonal in $\phi_\alpha$-space, not in Fourier space.

\section{Estimate of prewhitened nonlinear power in a survey}
\label{power}

\subsection{Unbiased estimate}
\label{estimate}

`In the case of a Gaussian distribution...
rather than removing the bias we should approximately double it,
in order to minimize the mean square sampling error'
-- E. T. Jaynes (1996, sentence containing eq.\ 17-13).

It is convenient to start out by considering the prewhitened estimator
$\hat Y_\alpha$ defined by equation~(\ref{Y}).
The minimum variance estimator $\hat\xi_\alpha$ of the power spectrum
in the FKP approximation is given by equation~(\ref{ximvFKP}).
Translating this equation into prewhitened quantities,
one concludes that the minimum variance prewhitened estimator $\hat Y_\alpha$
in the FKP approximation is,
in terms of the prewhitened reduced covariance $\fB$, equation~(\ref{B}), and
its associated Fisher matrix $E$, equation~(\ref{E}),
\be
\label{Ymv}
  \hat Y_\alpha =
    E^{-1}_{\alpha\beta} \,
    \fB^{-1\beta\gamma}
    (H^{-1/2})_\gamma^\epsilon \,
    D_\epsilon^{ij} (\delta_i \delta_j - \hat N_{ij})
  \ .
\ee
The estimator $\hat Y_\alpha$ is minimum variance if and only if
$\hat\xi_\alpha$ is minimum variance,
since $\hat Y_\alpha \equiv (H^{-1/2})_\alpha^\beta \, \hat\xi_\beta$
is a linear combination of $\hat\xi_\alpha$.

Now the estimator $\hat Y_\alpha$, equation~(\ref{Ymv}),
is intended to be an estimate of
$Y_\alpha \equiv (H^{-1/2})_\alpha^\beta \, \xi_\beta$.
But is that really true, given the various approximations?
It will be true provided that the estimator is unbiased,
meaning that the expectation value of the estimator is equal to the true value
\be
  \langle \hat Y_\alpha \rangle = Y_\alpha
  \ .
\ee
The expectation value of the estimator $\hat Y_\alpha$
given by equation~(\ref{Ymv}) is, since
$\langle \delta_i \delta_j - \hat N_{ij} \rangle = D^\alpha_{ij} \xi_\alpha$
according to equation~(\ref{Clin}),
\ba
  \langle \hat Y_\alpha \rangle
  &\!\!\!=\!\!\!&
    E^{-1}_{\alpha\beta} \,
    \fB^{-1\beta\gamma}
    (H^{-1/2})_\gamma^\epsilon \,
    D_\epsilon^{ij} D^\zeta_{ij} \,
    \xi_\zeta
  \nn
  &\!\!\!=\!\!\!&
    E^{-1}_{\alpha\beta} \,
    \fB^{-1\beta\gamma}
    D_\gamma^{ij} D^\epsilon_{ij} \,
    Y_\epsilon
\ea
where the second line follows because
$(H^{-1/2})_\gamma^\epsilon$ commutes with
$D_\epsilon^{ij} D^\zeta_{ij}$,
both being diagonal in real space.
It follows that the estimator $\hat Y_\alpha$
will be unbiased, $\langle \hat Y_\alpha \rangle = Y_\alpha$, provided that
the Fisher matrix $E^{\alpha\beta}$ is taken to satisfy
the asymmetric equation~(\ref{Emvasym}),
not, for example, a symmetrized version of that equation.

An important point to recognize here is that
an estimate $\hat Y_\alpha$ of the form~(\ref{Ymv})
will be unbiased for {\em any\/} a priori choice of the matrix $\fB$,
regardless of the choice of prior power $\xi(k)$,
regardless of the hierarchical model,
regardless of the FKP approximation,
and regardless of the approximation (such as eq.~[\ref{Bnice}]) to $\fB$,
just so long as the matrix $E$ in the estimator is interpreted as satisfying
the unsymmetrized equation~(\ref{Emvasym}).
Ultimately this property of being unbiased is inherited from the basic prior
assumption that galaxies constitute a random, Poisson sampling of an underlying
statistically homogeneous, isotropic density field,
so that the product of overdensities $\delta_i \delta_j$
at any pair of points $ij$ separated by $r_\alpha$
provides an unbiased estimate of the correlation function $\xi(r_\alpha)$.
Note that the presumption here is that the galaxies sampled are an unbiased
tracer of the galaxy density itself,
not necessarily of the mass density.

Interpreting the estimator $\hat Y_\alpha$, equation~(\ref{Ymv}),
as involving the asymmetric matrix $E^{\alpha\beta}$, equation~(\ref{Emvasym}),
should be regarded not as changing the estimator to make it unbiased,
but rather as interpreting the estimator correctly.
If instead the estimator $\hat Y_\alpha$ were interpreted
as involving the symmetrized Fisher matrix
$E^{(\alpha\beta)} \equiv {\rmn Sym}_{(\alpha\beta)} E^{\alpha\beta}$,
for example,
then the expectation value of the estimator would be
$\langle \hat Y_\alpha \rangle
= E^{-1}_{(\alpha\beta)} E^{\beta\gamma} Y_{\gamma}$,
which is not the same as $Y_\alpha$,
although of course it should be almost the same to the extent that
$E_{\alpha\beta}$ is almost symmetric.

It is convenient to introduce yet another estimator $\hat Z^\alpha$
related to the estimator $\hat Y_\alpha$ by
\be
\label{Z}
  \hat Y_\alpha = E_{\alpha\beta}^{-1} \hat Z^\beta
  \ .
\ee
In the FKP approximation, the estimator $\hat Z^\alpha$ is
\be
\label{Zmv}
  \hat Z^\alpha =
    \fB^{-1\alpha\beta}
    (H^{-1/2})_\beta^\gamma \,
    D_\gamma^{ij} (\delta_i \delta_j - \hat N_{ij})
  \ .
\ee

If the approximation~(\ref{Bnice}) to the prewhitened covariance $\fB$ is used
in the estimate~(\ref{Zmv}) of $\hat Z$, then,
in the representation of eigenfunctions $\phi_\alpha$,
\be
\label{Zmvnice}
  \hat Z_\alpha =
    \int_0^\infty
    {\phi_\alpha(r) \hat S(r;\mu_\alpha) \over [1+\xi(r)]^{1/2}}
    \, 4\PI r^2 \dd r
\ee
where
$\hat S(r;\mu_\alpha)$
is the FKP-weighted integral over pairs of overdensities
$\delta(\r_i) \delta(\r_j)$ at points $ij$ separated by
$r_{ij} \equiv |\r_i-\r_j| = r$
(commonly denoted
$\langle D D \rangle - 2 \langle D R \rangle + \langle R R \rangle$
in the literature, $D$ for data, $R$ for random)
\be
\label{S}
  \hat S(r;\mu_\alpha) =
    \int
    {\deltaD(r_{ij}-r) \delta(\r_i) \delta(\r_j)
    \over 2 \, [\mu_\alpha + \nbar(\r_i)^{-1}] [\mu_\alpha + \nbar(\r_j)^{-1}]}
    \, \ddd r_i \ddd r_j
  \ .
\ee
The shot-noise $\hat N_{ij}$ is excluded from equation~(\ref{S})
by excluding from the integration the contribution from self-pairs of galaxies,
which of course have zero separation.
The associated asymmetric Fisher matrix $E^{\alpha\beta}$ is given by
equation~(\ref{Emvnice}).

Equation~(\ref{S}) is expressed as an integral over pairs
of overdensities $\delta(\r_i) \delta(\r_j)$ in real space.
One could just as well express $\hat S$ as an integral over pairs
of overdensities $\delta(\k_i) \delta(\k_j)$ in Fourier space,
or pairs
of overdensities $\delta(k_i,\ell_i,m_i) \delta(k_j,\ell_j,m_j)$
in spherical harmonic space, if one found it more convenient.

\subsection{Numerics}
\label{compZ}

As in \S\ref{compE},
to avoid potential problems of ringing and aliasing,
it is probably better to evaluate the estimator $\hat Z_\alpha$,
equation~(\ref{Zmvnice}),
by means of an expression that involves the eigenfunctions $\phi_\alpha(k)$
in Fourier space rather than the eigenfunctions $\phi_\alpha(r)$ in real space.

If $\mu_\alpha$ is treated, temporarily, as a constant,
then transforming equation~(\ref{Zmvnice}) into real space yields
\be
\label{Zrm}
  \hat Z(r;\mu_\alpha) =
    {\hat S(r;\mu_\alpha) \over [1+\xi(r)]^{1/2}}
  \ .
\ee
The Fourier transform of this is
\be
\label{Zkm}
  \hat Z(k;\mu_\alpha) =
    \int_0^\infty j_0(kr) \, \hat Z(r;\mu_\alpha) \, 4\PI r^2 \dd r
\ee
in terms of which the estimator $\hat Z_\alpha$, equation~(\ref{Zmvnice}), is
\be
\label{Zmvnicek}
  \hat Z_\alpha =
    \int_0^\infty
    \phi_\alpha(k) \, \hat Z(k;\mu_\alpha)
    \, {4\PI k^2 \dd k \over (2\PI)^3}
  \ .
\ee
The transformation into $\phi_\alpha$ space, equation~(\ref{Zmvnicek}),
is done by discrete summation.

The advantage of equation~(\ref{Zmvnicek}) over
the nominally equivalent equation~(\ref{Zmvnice})
is that in equation~(\ref{Zmvnicek})
it is the data that are Fourier transformed,
$\hat Z(r;\mu_\alpha) \rightarrow \hat Z(k;\mu_\alpha)$,
equation~(\ref{Zkm}),
whereas in equation~(\ref{Zmvnice})
it is the eigenfunctions of the matrix $M$ that must be transformed,
$\phi_\alpha(k) \rightarrow \phi_\alpha(r)$.
While the two methods would yield
identical results for $\hat Z_\alpha$
if the same unitary Fourier transform were applied in both cases,
in reality it may be advantageous to have the freedom to
Fourier transform the data the best way one can,
without regard to the irrelevant question of how
the eigenfunctions $\phi_\alpha$ behave when Fourier transformed.

\subsection{The covariance of $\hat Z^\alpha$}
\label{varZ}

It will now be argued that the covariances
of the estimators $\hat Z^\alpha$ and $\hat Y^\alpha$
are approximately equal to, respectively,
the symmetrized Fisher matrix $E^{(\alpha\beta)}$, and its inverse,
equations~(\ref{ZaZbE}) and ~(\ref{YaYbE}).
It seems worthwhile to go through the arguments rather carefully.
As a general rule,
one should estimate error bars as accurately as possible;
but if some approximation is necessary,
then one would prefer to err on the conservative side of overestimating
the true errors.

Equation~(\ref{ZaZbE}) will now be derived,
commentary on the derivation being deferred to the end.
The covariance of the estimate $\hat Z^\alpha$ is,
from equation~(\ref{Zmv}),
\ba
\label{ZaZb}
  \langle \Delta \hat Z^\alpha \Delta \hat Z^\beta\rangle
  &\!\!\!=\!\!\!&
    \fB^{-1\alpha\gamma}\!(\nbar_i,\nbar_j) \,
    (H^{-1/2})_\gamma^\epsilon \,
    D_\epsilon^{ij} \,
    \fC_{ijkl}^\true \,
    D_\zeta^{kl}
  \nn
  &&
    (H^{-1/2})^\zeta_\eta \,
    \fB^{-1\eta\beta}\!(\nbar_k,\nbar_l)
\ea
in which $\fB$ is the approximate prewhitened reduced covariance
matrix~(\ref{Bnice})
used to construct the estimate $\hat Z^\alpha$, equation~(\ref{Zmvnice}),
while $\fC_{ijkl}^\true$
is the true covariance matrix, equation~(\ref{Cquad}).
To the extent that the FKP approximation, equation~(\ref{Cab}),
is valid for $\fC_{ijkl}^\true$,
equation~(\ref{ZaZb}) reduces to
\ba
\label{ZaZbFKP}
  \langle \Delta \hat Z^\alpha \Delta \hat Z^\beta\rangle
  &\!\!\!=\!\!\!&
    \fB^{-1\alpha\gamma}\!(\nbar_i,\nbar_j) \,
    (H^{-1/2})_\gamma^\epsilon \,
    D_\epsilon^{ij} \,
    D^\theta_{ij}
  \nn
  &&
    \fC_{\theta\zeta}^\FKP\!(\nbar_i,\nbar_j) \,
    (H^{-1/2})^\zeta_\eta \,
    \fB^{-1\eta\beta}\!(\nbar_i,\nbar_j)
  \nn
  &\!\!\!=\!\!\!&
    \fB^{-1\alpha\gamma}\!(\nbar_i,\nbar_j) \,
    D_\gamma^{ij} \,
    D^\epsilon_{ij}
  \nn
  &&
    \fB_{\epsilon\zeta}^\FKP\!(\nbar_i,\nbar_j) \,
    \fB^{-1\zeta\beta}\!(\nbar_i,\nbar_j)
  \nn
  &\!\!\!=\!\!\!&
    E^{\alpha\gamma} \,
    \fB_{\gamma\epsilon}^\FKP\!(\nbar_i,\nbar_j) \,
    \fB^{-1\epsilon\beta}\!(\nbar_i,\nbar_j)
\ea
where $\fC^\FKP$
is the FKP covariance, equation~(\ref{CKJI}),
and $\fB^\FKP \equiv H^{-1/2} \fC^\FKP H^{-1/2}$
is its prewhitened counterpart, equation~(\ref{BML1}).
Note that going from equation~(\ref{ZaZb}) to the second expression
in equation~(\ref{ZaZbFKP}) includes, as part of the FKP approximation,
the assumption that $\nbar_k$ and $\nbar_l$ in
$\fB^{-1\eta\beta}\!(\nbar_k,\nbar_l)$ are approximately constant.
The expressions on the right hand side of equation~(\ref{ZaZbFKP})
are not symmetric in $\alpha\beta$,
because of the asymmetry of the FKP approximation~(\ref{Cab}).
To the further extent that the prewhitened covariance $\fB^\FKP$
equals the approximation $\fB$, equation~(\ref{Bnice}),
the covariance $\langle \Delta \hat Z^\alpha \Delta \hat Z^\beta \rangle$
reduces to the asymmetric matrix $E$
given by equation~(\ref{Emvnice})
\be
\label{ZaZbEasym}
  \langle \Delta \hat Z^\alpha \Delta \hat Z^\beta \rangle
  = E^{\alpha\beta}
\ee
the asymmetry of the right hand side being inherited from the FKP approximation.
An equally good approximation to the covariance would be the same
expression~(\ref{ZaZbEasym}) with the indices swapped on the right hand side,
$\alpha \leftrightarrow \beta$.
Thus it seems reasonable to conclude that
the covariance $\langle \Delta \hat Z^\alpha \Delta \hat Z^\beta \rangle$
should be approximately equal to the the symmetrized Fisher matrix
$E^{(\alpha\beta)}$
\be
\label{ZaZbE}
  \langle \Delta \hat Z^\alpha \Delta \hat Z^\beta \rangle
  = E^{(\alpha\beta)} \equiv \Sym{(\alpha\beta)} E^{\alpha\beta}
  \ .
\ee

Several comments can be made about the accuracy of the approximations
made in the above derivation.

Firstly,
one partial test of the validity of the FKP approximation is
the degree of asymmetry of the asymmetric Fisher matrix $E^{\alpha\beta}$,
equation~(\ref{Emvnice}).
If the survey is broad in real space,
which is the condition for the FKP approximation to hold,
then the pair integral $R(r;\mu_\alpha)$ in the integrand
on the right hand side of equation~(\ref{Emvnice})
will be a slowly varying function of pair separation $r$,
so that the matrix $E^{\alpha\beta}$ will be nearly diagonal,
hence symmetric.
The test is not definitive because $E^{\alpha\beta}$ would be symmetric in any
case if $\mu_\alpha = \mu_\beta$.
But in practice $\mu_\alpha \approx \xi(k_\alpha)$
in both linear and nonlinear regimes, Figure~\ref{diag},
and realistically the power spectrum $\xi(k)$ varies substantially,
so the consistency test should be indicative.

Secondly,
one of the weaknesses of the FKP approximation
is that it fails to deal with sharp edges
-- as typically occur at the angular boundaries of a survey --
correctly.
The FKP approximation tends to overestimate the variance
contributed by regions near boundaries,
since it assumes that those regions are accompanied by more correlated
neighbours than is actually the case.
Thus, at least as regards edge effects,
the FKP approximate covariance, equation~(\ref{ZaZbFKP}),
should tend to overestimate the exact covariance, equation~(\ref{ZaZb}),
of the approximate estimate $\hat Z^\alpha$.

Thirdly,
it is possible to check the accuracy of the approximation
made in going from equation~(\ref{ZaZbFKP}) to equation~(\ref{ZaZbEasym}).
The approximation involves setting $\fB^\FKP \fB^{-1} = \1$,
whereas comparing equation~(\ref{BML1}) for $\fB^\FKP$ to
the approximation~(\ref{Bnice}) for $\fB$
shows that this quantity is in fact,
in the representation of eigenfunctions $\phi_\alpha$
of the 4-point matrix $M$,
\ba
\label{BB}
\lefteqn{
  \fB_{\alpha\gamma}^\FKP\!(\nbar_i,\nbar_j)
  \fB^{-1\gamma\beta}\!(\nbar_i,\nbar_j)
  =
} &&
  \nn
  &&
  \1_\alpha^\beta +
    {( \nbar_i^{-1} + \nbar_j^{-1} )
    \over (\mu_\alpha + \nbar_i^{-1}) (\mu_\alpha + \nbar_j^{-1})}
    \left( L_\alpha^\beta - \mu_\alpha \, \1_\alpha^\beta \right)
  \ .
\ea
The correction term on the right hand side of equation~(\ref{BB})
should be small to the extent that the 3-point matrix $L_{\alpha\beta}$
is near diagonal in this 4-point representation,
with eigenvalues $\lambda_\alpha \approx \mu_\alpha$,
as was found to be the case in \S\ref{4+3point}.

If desired, one could use the expression on the right hand side
of equation~(\ref{BB})
to compute a more accurate approximation to the covariance of $\hat Z^\alpha$,
based on equation~(\ref{ZaZbFKP}) rather than on equation~(\ref{ZaZbEasym}).
However,
if one were willing to go to the trouble of computing a correction
from equation~(\ref{BB}),
then one would probably be willing to revert to equation~(\ref{Zmv}),
and to integrate $\fB^{-1 \alpha\beta}(\nbar_i,\nbar_j)$
numerically over all pairs of volume elements $ij$ in the survey,
inverting $\fB$ numerically for each pair $\nbar_i$, $\nbar_j$
of values of the selection function.
This latter procedure is in fact the gourmet recipe of \S\ref{gourmet}.

Fourthly,
the approximation $\lambda_\alpha \approx \mu_\alpha$
adopted in the approximation~(\ref{Bnice}) to $\fB$
tends to overestimate the true eigenvalues $\lambda_\alpha$ of the
3-point matrix $L$, according to Figure~\ref{diag34}.
This should lead to a slight overestimate of the variance.
In the realistic $\Lambda$CDM case, Figure~\ref{diag34},
the approximation $\lambda_\alpha \approx \mu_\alpha$ overestimates
the true eigenvalues $\lambda_\alpha$
by at worst 20~percent, at moderately nonlinear wavenumbers $k$.
This 20~percent overestimate is diluted to at worst 10~percent
because the 3-point variance
contributes at most half of the combined 2-point, 3-point, and 4-point variance,
where the selection function satisfies $\nbar^{-1} = \mu_\alpha$.
The overestimate is further diluted because in practice the selection function
varies, and is unlikely to sit everywhere near the worst value.

The conclusion is that the covariance
of the approximate estimator $\hat Z_\alpha$,
equation~(\ref{Zmvnice}) or (\ref{Zmvnicek}),
should be given approximately, equation~(\ref{ZaZbE}),
by the symmetrized Fisher matrix $E^{(\alpha\beta)}$
of equation~(\ref{Emvnice}),
and that if anything this covariance is likely to be on the conservative
side of the true covariance.

\subsection{The covariance of $\hat Y_\alpha$}
\label{varY}

From the expression~(\ref{ZaZbE}) for the covariance of $\hat Z^\alpha$,
one might conclude (falsely) that the covariance of
$\hat Y_\alpha$, equation~(\ref{Z}), is
\be
\label{YaYbEEE}
  \langle \Delta \hat Y_\alpha \Delta \hat Y_\beta \rangle
  = E_{\alpha\gamma}^{-1} \,
    E^{(\gamma\delta)} \,
    E_{\beta\delta}^{-1}
  \ .
\ee
A more direct derivation of the covariance of $\hat Y_\alpha$,
along the lines of equations~(\ref{ZaZb})--(\ref{ZaZbE}),
leads to the same (false) conclusion.
The analogue of equation~(\ref{ZaZbEasym}) is
\be
\label{YaYbEasym}
  \langle \Delta \hat Y_\alpha \Delta \hat Y_\beta \rangle
  = E_{\beta\alpha}^{-1}
\ee
with the asymmetric matrix $E$ on the right hand side.
At this point one might be inclined to symmetrize this
equation~(\ref{YaYbEasym}),
as was done for
$\langle \Delta \hat Z^\alpha \Delta \hat Z^\beta \rangle$
in equation~(\ref{ZaZbE}),
writing
\be
\label{YaYbEi}
  \langle \Delta \hat Y_\alpha \Delta \hat Y_\beta \rangle
  = \Sym{(\alpha\beta)} E_{\alpha\beta}^{-1}
  \ .
\ee
The symmetrized inverse
${\rmn Sym}_{(\alpha\beta)} E_{\alpha\beta}^{-1}$
of the asymmetric Fisher matrix
is to be distinguished from the inverse
$E_{(\alpha\beta)}^{-1}$
of the symmetrized Fisher matrix.
But it is not hard to show that
\be
\label{SymEinv}
  \Sym{(\alpha\beta)} E^{-1}_{\alpha\beta}
  = E_{\alpha\gamma}^{-1} \,
    E^{(\gamma\delta)} \,
    E_{\beta\delta}^{-1}
  \ .
\ee
Thus equations~(\ref{YaYbEEE}) and (\ref{YaYbEi}) are identical.
However, both equations are wrong.

The problem is that,
while the Fisher matrix $E$ remains well-behaved in the presence
of loud noise, with near zero eigenvalues,
its inverse $E^{-1}$ becomes almost singular.
Consider the example of some noisy mode,
for which the eigenvalue of the Fisher matrix is almost zero.
It may well happen that the asymmetric Fisher matrix $E^{\alpha\beta}$
is numerically non-singular,
but that, because of approximations or numerics,
the computed eigenvalue of the symmetrized Fisher matrix $E^{(\alpha\beta)}$
is exactly zero.
Equation~(\ref{YaYbEEE}) would then say
that the variance of the noisy mode is zero
(for if the determinant of the symmetrized Fisher matrix is zero,
$|E^{(\alpha\beta)}| = 0$,
while the determinant of the asymmetric Fisher matrix is finite,
$|E^{\alpha\beta}| \neq 0$,
then the determinant of the variance in eq.~[\ref{YaYbEEE}] is zero).
This is plainly absurd.

It is safer to take
the covariance of $\hat Y_\alpha$ to be approximately equal
to the inverse of the symmetrized Fisher matrix $E^{(\alpha\beta)}$,
\be
\label{YaYbE}
  \langle \Delta \hat Y_\alpha \Delta \hat Y_\beta \rangle
  = E^{-1}_{(\alpha\beta)}
  \ .
\ee
Here a noisy mode will always reveal itself
by its small eigenvalue.

\subsection{Convert to $\hat X_\alpha$}
\label{convert}

For the purpose of constructing uncorrelated quantities
to be plotted on a graph,
it is desirable to compute the prewhitened power spectrum $\hat X_\alpha$.

To compute $\hat X_\alpha$,
start from the estimate $\hat Z_\alpha$ given by equation~(\ref{Zmvnicek}),
transform this into
$\hat Y_\alpha = E^{-1}_{\alpha\beta} \, \hat Z^\beta$,
equation~(\ref{Z}),
thence into the power spectrum
$\hat\xi_\alpha = (H^{-1/2})_\alpha^\beta \, \hat Y_\beta$,
equation~(\ref{Y}),
and thence into the prewhitened power spectrum $\hat X_\alpha$,
equation~(\ref{Xrhat}).

The covariance of the prewhitened power $\hat X_\alpha$ is,
by construction, the same as that of $\hat Y_\alpha$,
equation~(\ref{YaYbE}),
\be
\label{XaXbE}
  \langle \Delta \hat X_\alpha \Delta \hat X_\beta \rangle
  =
  E^{-1}_{(\alpha\beta)}
\ee
the inverse of the symmetrized Fisher matrix of the prewhitened power.

The estimator $\hat X_\alpha$ of prewhitened power, equation~(\ref{Xrhat}),
is a nonlinear transformation of the estimator $\hat\xi_\alpha$ of power,
and is therefore biased if $\hat\xi_\alpha$ is unbiased.
However, the estimator $\hat X_\alpha$ is unbiased in the asymptotic
limit of a large quantity of data.

\subsection{Decorrelate}
\label{decorrelate}

One final step remains,
which is to process the measured prewhitened power spectrum $\hat X_\alpha$
into a set of decorrelated band-powers.
How to accomplish such decorrelation is described in Paper~4.

One possibility would be to decorrelate the power spectrum $\hat\xi(k)$ itself.
This is a bad idea, because the power spectrum is highly correlated
in the nonlinear regime,
so the decorrelation matrices would be broad, with large negative
off-diagonal entries,
making it impossible to interpret the decorrelated band-powers as
representing the power spectrum over some narrow band.

Another possibility would be to decorrelate the prewhitened power
$\hat X_\alpha$ not in Fourier space but rather
in the representation of eigenfunctions $\phi_\alpha$
of the prewhitened 4-point matrix $M$.
Again this seems not so good an idea,
in the first place because the physical meaning of this representation
is obscure,
and in the second place because the eigenfunctions can mix where
their eigenvalues $\mu_\alpha$ are degenerate.
Since $\mu_\alpha \approx \xi(k_\alpha)$,
such mixing in practice occurs between wavenumbers $k_\alpha$
where the power $\xi(k_\alpha)$ is the same,
which happens to either side of the peak in the power spectrum,
Figure~\ref{xik}.
Perhaps in the future a better understanding of the eigenfunctions
$\phi_\alpha$ will emerge,
amongst other things allowing mixing to avoided,
but in the meantime these problems remain.

The natural solution is to decorrelate the prewhitened power $\hat X(k)$
in Fourier space.
As seen in \S\ref{4+3point},
the covariance of the prewhitened power is encouragingly narrow
in Fourier space,
narrow enough that the decorrelation matrices will be narrow,
so that the decorrelated band-powers
can be interpreted as estimates of the prewhitened power
over narrow intervals of wavenumber $k$.
In contrast to the prewhitened power $\hat X_\alpha$ in the
$\phi_\alpha$-representation,
the prewhitened power $\hat X(k)$ in Fourier space has a clear interpretation,
and there is no problem arising from mixing of eigenfunctions.

\section{The Full FKP}
\label{full}

Sections~\ref{fisher} and \ref{power}
invoked not only the FKP approximation,
but also the simplifying approximation~(\ref{Bnice}) to
the prewhitened reduced covariance $\fB$.
How much more work would it take to compute the
minimum variance estimator and Fisher matrix of nonlinear power
making only the FKP approximation and no other approximation?
The question is of both didactic and practical interest.

\subsection{Fisher matrix}
\label{fisherfull}

The FKP approximation to the Fisher matrix of the power spectrum,
equation~(\ref{FmvFKPasym}),
looks simplest expressed in real space:
\be
\label{Frr}
  F(r_\alpha,r_\beta)
  =
    \!\int\! \fC^{-1}\!(r_\alpha,r_\beta;\nbar_i,\nbar_j)
    \deltaD(r_{ij}-r_\beta) \, \ddd r_i \ddd r_j
  \ .
\ee
The corresponding expression for the FKP approximation to the Fisher
matrix of the prewhitened power spectrum, equation~(\ref{Emvasym}), is
\be
\label{Err}
  E(r_\alpha,r_\beta)
  =
    \!\int\! \fB^{-1}\!(r_\alpha,r_\beta;\nbar_i,\nbar_j)
    \deltaD(r_{ij}-r_\beta) \, \ddd r_i \ddd r_j
  \ .
\ee
These are 5-dimensional (thanks to the Dirac delta-function)
integrals over pairs of volume elements $ij$
separated by $r_{ij} \equiv |\r_i-\r_j| = r_\beta$ in the survey.
The integrals are actually quite doable.
If, as is typical, the selection function $\nbar(\r)$ separates into
the product of an angular mask and a radial selection function,
then the 3-dimensional angular integrals can be done analytically
(Hamilton 1993),
leaving a double integral of
$\fC^{-1}\!(r_\alpha,r_\beta;\nbar_i,\nbar_j)$
or $\fB^{-1}\!(r_\alpha,r_\beta;\nbar_i,\nbar_j)$
over the radial directions.
The matrices $\fC(r_\alpha,r_\beta;\nbar_i,\nbar_j)$
or $\fB(r_\alpha,r_\beta;\nbar_i,\nbar_j)$,
discretized (\S\ref{discrete}) over a grid of separations
$r_\alpha$ and $r_\beta$, can be inverted numerically
for each pair of values of the selection functions $\nbar_i$ and $\nbar_j$.

The problem with equations~(\ref{Frr}) or (\ref{Err}) is that experience
(\S\S\ref{comp4+3point}, \ref{compE}) suggests that
discretization of the matrix
$\fC(r_\alpha,r_\beta;\nbar_i,\nbar_j)$
or $\fB(r_\alpha,r_\beta;\nbar_i,\nbar_j)$
in real space is liable to introduce ringing and aliasing in Fourier space,
defeating the aim of constructing an accurate Fisher matrix
of the power spectrum.

A possibly more robust procedure would be to follow more closely
the program described in \S\ref{fisher} and \S\ref{power}.
In Fourier space,
the FKP approximation to the prewhitened Fisher matrix,
equation~(\ref{Emvasym}), is
\ba
\lefteqn{
  E(k_\alpha,k_\beta)
  =
} &&
  \\ \nonumber
\lefteqn{
  \ 
    \!\int\! \fB^{-1}\!(k_\alpha,k_\gamma;\nbar_i,\nbar_j)
    j_0(k_\gamma r_{ij}) j_0(k_\beta r_{ij})
    \, \ddd r_i \ddd r_j {4\PI k_\gamma^2 \dd k_\gamma \over (2\PI)^3}
  \ .
} &&
\ea
This integral might be evaluated as follows
(since I have not actually carried through this program,
I cannot say for sure that it would work without a hitch).
Firstly,
compute the matrix of pair integrals
\be
\label{Rkk}
  R(r;k_\alpha,k_\beta) =
    \!\int\! \fB^{-1}\!(k_\alpha,k_\beta;\nbar_i,\nbar_j)
    \deltaD(r_{ij} - r) \, \ddd r_i \ddd r_j
\ee
for many pair separations $r_{ij} = r$.
These pair integrals $R(r;k_\alpha,k_\beta)$, equation~(\ref{Rkk}),
are analogous to the FKP-weighted pair integrals $R(r;\mu_\alpha)$,
equation~(\ref{R}).
Next, cosine transform (e.g.\ with FFTLog) the pair integrals
\be
  \tilde R(k;k_\alpha,k_\beta) =
    2 \int_0^\infty \cos(kr) R(r;k_\alpha,k_\beta) \, \dd r
\ee
analogously to $\tilde R(k;\mu_\alpha)$, equation~(\ref{Rk}).
Finally,
compute the prewhitened Fisher matrix $E(k_\alpha,k_\beta)$
by integrating
\ba
\label{Ekk}
\lefteqn{
  E(k_\alpha,k_\beta)
  =
} &&
  \\ \nonumber
\lefteqn{
  \ 
    {1 \over 2\PI k_\beta}
    \int_0^\infty \left[
      \tilde R(k_\gamma\!-\!k_\beta;k_\alpha,k_\gamma)
      - \tilde R(k_\gamma\!+\!k_\beta;k_\alpha,k_\gamma)
    \right]
    \, k_\gamma \dd k_\gamma
  \ .
} &&
\ea

In practice, the matrix
$\fB(k_\alpha,k_\beta;\nbar_i,\nbar_j)$
in equation~(\ref{Rkk}) must be inverted on a discrete grid of
wavenumbers $k$.
Similarly, the integral over $k_\gamma$ in equation~(\ref{Ekk})
should be done as a discrete sum.
Specifically,
if the matrices are discretized
(\S\ref{discrete}) on a logarithmic grid of wavenumbers,
so that $R(k;k_\alpha,k_\beta)$ discretizes to
$\bR_{k_\alpha k_\beta}\!(k) =
R(k;k_\alpha,k_\beta)
\,$\discretionary{}{}{}$4\PI (k_\alpha k_\beta)^{3/2}$\discretionary{}{}{}$\Delta\ln k/$\discretionary{}{}{}$(2\PI)^3$,
and the Fisher matrix
$E(k_\alpha,k_\beta)$
discretizes to
$\bE_{k_\alpha k_\beta} =
E(k_\alpha,k_\beta)
\,$\discretionary{}{}{}$4\PI (k_\alpha k_\beta)^{3/2}$\discretionary{}{}{}$\Delta\ln k/$\discretionary{}{}{}$(2\PI)^3$,
then equation~(\ref{Ekk}) becomes
\be
\label{Ekks}
  \bE_{k_\alpha k_\beta} =
    \sum_{k_\gamma}
    {(k_\beta k_\gamma)^{1/2} \over 2\PI}
    \left[
      \tilde \bR_{k_\alpha k_\gamma}\!(k_\gamma\!-\!k_\beta)
      - \tilde \bR_{k_\alpha k_\gamma}\!(k_\gamma\!+\!k_\beta)
    \right]
  \ .
\ee

It may be anticipated that,
as in \S\ref{compE},
equation~(\ref{Ekks}) will tend to overestimate the diagonal elements
$\bE_{k_\alpha k_\alpha}$ if the gridding of the matrix
is too coarse to resolve the diagonal properly.
Integrating the continuous Fisher matrix $E(k_\alpha,k_\beta)$,
equation~(\ref{Ekk}), over $k_\beta$ yields
\be
\label{Eint}
  \int_0^\infty\!\! E(k_\alpha,k_\beta) \, k_\beta \dd k_\beta
  = {1 \over \PI} \!\int_0^\infty \!\!\!\! \int_0^{k_\gamma}\!\!
    \tilde R(k;k_\alpha,k_\gamma) \, \dd k \, k_\gamma \dd k_\gamma
  \ .
\ee
Discretized, equation~(\ref{Eint}) becomes
\be
\label{Eints}
  \sum_\beta \left( {k_\beta \over k_\alpha} \right)^{1/2} \!
    \bE_{k_\alpha k_\beta}
  = {1 \over \PI} \sum_{k_\gamma}
    \left( {k_\gamma \over k_\alpha} \right)^{1/2}
    \!\!\!\int_0^{k_\gamma}\!\!
    \tilde \bR_{k_\alpha k_\gamma}\!(k) \, \dd k
  \ .
\ee
The integral over $k$ on the right hand side of equation~(\ref{Eints})
can be done conveniently as a sine transform (e.g.\ with \mbox{FFTLog})
of the pair integral
\be
  \int_0^{k_\gamma}\!
    \tilde \bR_{k_\alpha k_\gamma}\!(k) \, \dd k
  = 2 \int_0^\infty\! \sin(kr) {\bR_{k_\alpha k_\gamma}\!(r) \over r} \, \dd r
  \ .
\ee
If the sum on the left hand side of equation~(\ref{Eints})
exceeds the right hand side,
then reduce the diagonal element $\bE_{k_\alpha k_\alpha}$ so
that the sum is satisfied.
It is fine to evaluate the sum on the right hand side of equation~(\ref{Eints})
as a discrete sum over $k_\gamma$, rather than as a continuous integral,
because $\bR_{k_\alpha k_\gamma}\!(k)$, equation~(\ref{Rkk}),
inherits its behaviour
from $\fB_{k_\alpha k_\gamma}$,
which, if constructed, equation~(\ref{BML1}),
from the 4-point and 3-point matrices $M$ and $L$
as discussed in \S\ref{comp4+3point},
should behave correctly near the diagonal even if the resolution
is too coarse to resolve the diagonal.

\subsection{Estimate of power}
\label{powerfull}

The FKP approximation to the minimum variance estimator of the power spectrum,
equation~(\ref{ximvFKP}),
again looks simplest when expressed in real space:
\be
  \hat\xi_\alpha = F^{-1}_{\alpha\beta} \, \hat T^\beta
\ee
with
\be
\label{Tr}
  \hat T(r_\alpha) =
    \!\int\! \fC^{-1}\!(r_\alpha,r_{ij};\nbar_i,\nbar_j)
    \delta(\r_i) \delta(\r_j) \, \ddd r_i \ddd r_j
  \ .
\ee
As usual,
the prewhitened estimator $\hat Y \equiv H^{-1/2} \hat\xi$,
equation~(\ref{Y}),
is related to the estimator $\hat Z$ by $\hat Y = E^{-1} \hat Z$,
equation~(\ref{Z}).
The FKP approximation to the estimator $\hat Z$ is,
equation~(\ref{Zmv}),
\be
\label{Zr}
  \hat Z(r_\alpha) =
    \!\int\! {\fB^{-1}\!(r_\alpha,r_{ij};\nbar_i,\nbar_j)
    \delta(\r_i) \delta(\r_j) \over [1 + \xi(r_{ij})]^{1/2}}
    \, \ddd r_i \ddd r_j
  \ .
\ee
Equations~(\ref{Tr}) and (\ref{Zr}) are
6-dimensional integrals over pairs of volume elements $ij$ in the survey.
But once again one may anticipate that
discretization of the matrices
$\fC(r_\alpha,r_\beta;\nbar_i,\nbar_j)$
or $\fB(r_\alpha,r_\beta;\nbar_i,\nbar_j)$
in real space would introduce ringing and aliasing in Fourier space,
defeating the aim of constructing an accurate estimator of the power spectrum.

Again, it seems likely that it would be more robust to
work with prewhitened quantities in Fourier space.
In Fourier space,
the FKP approximation to
the estimator $\hat Z_\alpha$ is,
equation~(\ref{Zmv}),
\ba
\label{Zkf}
\lefteqn{
  \hat Z(k_\alpha) =
} &&
  \\ \nonumber
\lefteqn{
  \ 
    \!\int\! {\fB^{-1}\!(k_\alpha,k_\beta;\nbar_i,\nbar_j)
    j_0(k_\beta r_{ij})
    \delta(\r_i) \delta(\r_j)
    \over [1 + \xi(r_{ij})]^{1/2}}
    \, \ddd r_i \ddd r_j {4\PI k_\beta^2 \dd k_\beta \over (2\PI)^3}
  \ .
} &&
\ea
One way to evaluate this integral might be as follows.
Firstly, compute the matrix
$\hat S(r;k_\alpha,k_\beta)$
of integrals over pairs of overdensities
$\delta(\r_i) \delta(\r_j)$
at many separations $r_{ij} = r$
\ba
\label{Srkk}
\lefteqn{
  \hat S(r;k_\alpha,k_\beta) =
} &&
  \\ \nonumber
\lefteqn{
  \ 
    \!\int\! \fB^{-1}\!(k_\alpha,k_\beta;\nbar_i,\nbar_j)
    \deltaD(r_{ij}\!-\!r)
    \delta(\r_i) \delta(\r_j)
    \, \ddd r_i \ddd r_j
} &&
\ea
which may be compared to equation~(\ref{S}).
Next,
prewhiten
(compare eq.~[\ref{Zrm}])
\be
  \hat Z(r;k_\alpha,k_\beta) =
    {\hat S(r;k_\alpha,k_\beta)
    \over [1 + \xi(r)]^{1/2}}
\ee
and Fourier transform, e.g.\ with FFTLog,
(compare eq.~[\ref{Zkm}])
\be
\label{Zkkk}
  \hat Z(k;k_\alpha,k_\beta) =
    \int_0^\infty
    j_0(kr)
    \hat Z(r;k_\alpha,k_\beta)
    \, 4\PI r^2 \dd r
  \ .
\ee
Actually it suffices to do this Fourier transform for $k = k_\beta$ only.
Finally, the estimator $\hat Z(k_\alpha)$, equation~(\ref{Zkf}), is
\be
\label{Zka}
  \hat Z(k_\alpha) =
    \int_0^\infty
    \hat Z(k_\beta; k_\alpha,k_\beta)
    \, {4\PI k_\beta^2 \dd k_\beta \over (2\PI)^3}
  \ .
\ee
The integral in equation~(\ref{Zka}) should be done as a discrete sum.
If discretized (\S\ref{discrete}) on a logarithmic grid of wavenumbers,
so that
$\hat Z(k_\alpha)$ discretizes to
$\hat \vZ_{k_\alpha} =
\hat Z(k_\alpha)
$\discretionary{}{}{}$[4\PI k_\alpha^3$\discretionary{}{}{}$\Delta\ln k/$\discretionary{}{}{}$(2\PI)^3]^{1/2}$
and
$\hat Z(k;k_\alpha,k_\beta)$ discretizes to
$\hat \bZ_{k_\alpha k_\beta}\!(k) =
\hat Z(k;k_\alpha,k_\beta)
\,$\discretionary{}{}{}$4\PI (k_\alpha k_\beta)^{3/2}$\discretionary{}{}{}$\Delta\ln k/$\discretionary{}{}{}$(2\PI)^3$,
then equation~(\ref{Zka}) is
\be
  \hat \vZ_{k_\alpha} =
    \sum_{k_\beta} \hat \bZ_{k_\alpha k_\beta}\!(k_\beta)
    \left[ {4\pi k_\beta^3 \Delta\ln k \over (2\PI)^3} \right]^{1/2}
  \ .
\ee

\section{Recipes}
\label{recipe}

This section summarizes the results of previous sections
into logical sequences of practical steps needed to estimate
the prewhitened nonlinear power spectrum from an actual galaxy survey.
The end result is
a set of uncorrelated prewhitened nonlinear band-powers with error bars,
over some prescribed grid of wavenumbers $k$.

There are three versions of the recipe,
gourmet (\S\ref{gourmet}),
fine (\S\ref{fine}),
and fastfood (\S\ref{fast}).
All the methods use the FKP approximation, equation~(\ref{Cab}).
Thus one should imagine that there is also a haute-cuisine method,
which might be brute-force,
or it might be some clever procedure that apodizes edges.

First a disclaimer.
{\em
The methods described herein do not take into account redshift distortions,
whose effects on the power spectrum are at least as great as those of
nonlinearity.
There is no point in using these methods as they stand,
without also taking into account redshift distortions}.
However,
given that a full-blown procedure including redshift distortions
may well be based in part on the methods described,
it seems worthwhile to lay out the steps required to implement them.

\subsection{Gourmet}
\label{gourmet}

This version of the recipe is conceptually the simplest,
but it takes the most computing power
(a supercomputer would be handy).
The procedure is a direct implementation of
the FKP approximation to the minimum variance estimator of prewhitened power
and the associated Fisher matrix,
as described in \S\ref{full}.

Naturally, if one were going to the trouble of using the gourmet recipe,
then one would want to use the best possible model of the
3-point and 4-point correlation functions,
not just the hierarchical model with constant amplitudes.

Steps~1 and 2 below require knowledge of the selection function of a survey,
but no actual data.
Steps~3--5 require actual data from a galaxy survey.

\skipp
{\bf Step 1}.
Compute
the FKP approximation to the asymmetric Fisher matrix
$E^{\alpha\beta}$
of the prewhitened nonlinear power spectrum,
as described in \S\ref{fisherfull}, equations~(\ref{Rkk})--(\ref{Ekk}).
Equation~(\ref{Rkk})
involves a 5-dimensional integral over pairs $ij$ of volume elements
separated by $r_{ij} = r$ in the survey.
If the selection function $\nbar(\r)$ separates into
the product of an angular mask and a radial selection function,
then the 3-dimensional angular integrals can be done analytically
(Hamilton 1993),
leaving a double integral of
$\fB^{-1}\!(k_\alpha,k_\beta;\nbar_i,\nbar_j)$
over the radial directions.

\skipp
{\bf Step 2}.
The covariance matrix
$\langle \Delta\hat X_\alpha \Delta\hat X_\beta \rangle$
of the prewhitened power is equal, equation~(\ref{XaXbE}),
to the inverse
$E^{-1}_{(\alpha\beta)}$
of the symmetrized Fisher matrix.
Use this covariance matrix to construct decorrelation matrices $W$,
as described in Paper~4,
with the property that
$W \langle \Delta\hat X \Delta\hat X^\transpose \rangle W^\transpose$
is diagonal in Fourier space
(cf.\ Paper~4 eq.\ [20]).
The diagonal elements of this diagonal matrix
are the expected variances of the decorrelated band-powers
$\hat B = W \hat X$ to be computed in Step~5.

\skipp
{\bf Step 3}.
Compute the estimator $\hat Z_\alpha$
as described in \S\ref{powerfull}, equations~(\ref{Srkk})--(\ref{Zka}).

\skipp
{\bf Step 4}.
Transform $\hat Z^\alpha$ into the prewhitened power $\hat X_\alpha$
using equations~(\ref{Z}), (\ref{Y}), and (\ref{Xrhat}),
as stated in \S\ref{convert}.

\skipp
{\bf Step 5}.
Decorrelate the estimated prewhitened power spectrum $\hat X$
into a set of uncorrelated band-powers $\hat B = W \hat X$,
using the decorrelation matrices $W$ computed in Step~2.
Bear in mind that, as usual in ML fitting,
the error bars should of course be interpreted
as being attached to the model, the prior band-powers $B$,
rather than to the data, the estimated band-powers $\hat B$.

\subsection{Fine}
\label{fine}

This method adopts the approximation made
in \S\ref{fisher} and \S\ref{power}
that the prewhitened reduced covariance matrix $\fB$
takes the simplified form~(\ref{Bnice}).
According to the results of \S\ref{4+3point},
this approximation to $\fB$ should be quite good.
If it is, then the fine method should yield results close to the
gourmet method of \S\ref{gourmet},
at a considerable saving in computer time.

Steps~1--5 below do not require any actual data;
the steps can be used to determine in advance how well
the prewhitened power spectrum might be measured from a survey.
Steps~6--9 require actual data from a galaxy survey.

\skipp
{\bf Step 1}.
Compute a table of FKP-weighted pair integrals $R(r;\mu)$
at many separations $r$ and several FKP constants $\mu$.
Calculating the pair integrals $R(r;\mu)$ requires knowing the selection
function $\nbar(\r)$ of a galaxy survey, but does not require actual data.
This pair integral, commonly denoted $\langle R R \rangle$,
is commonly computed by Monte Carlo integration,
but I find it faster, more accurate,
and more convenient (since the program was already written)
to compute the integral directly,
using the procedures described by
Hamilton (1993).

\skipp
{\bf Step 2}.
Compute the prewhitened 4-point contribution $M \equiv H^{-1/2} K \, H^{-1/2}$
to the reduced covariance of the nonlinear power spectrum.
This involves adopting a prior power spectrum $\xi(k)$,
and a model of the 4-point correlation function $\eta_{ijkl}$.
For the hierarchical model,
the covariance matrices $K$ and $H$ are given by
equations~(\ref{Khier}) and (\ref{Ihier}).
Some numerical issues concerning the computation of the matrix $M$
are discussed in \S\ref{comp4+3point}.

\skipp
{\bf Step 3}.
Compute the eigenfunctions $\phi_\alpha$ and eigenvalues $\mu_\alpha^2$,
equation~(\ref{mu}),
by diagonalizing the prewhitened 4-point matrix $M$.

\skipp
{\bf Step 4}.
Compute the asymmetric Fisher matrix $E^{\alpha\beta}$,
equation~(\ref{Emvnice}),
of the prewhitened nonlinear power spectrum of the survey,
in the representation of eigenfunctions $\phi_\alpha$
of the prewhitened 4-point matrix $M$.
This is where the pair integral $R(r,\mu)$ computed in Step~1 is needed.
Numerical issues are discussed in \S\ref{compE}.

\skipp
{\bf Step 5}.
Same as Step~2 of the gourmet method:
from the inverse
$E^{-1}_{(\alpha\beta)}$
of the symmetrized Fisher matrix,
construct decorrelation matrices $W$
such that the covariance of the band-powers
$\hat B = W \hat X$ is diagonal.
Decorrelation is the subject of Paper~4.

\skipp
{\bf Step 6}.
Compute a table of FKP-weighted pair densities $\hat S(r;\mu)$,
equation~(\ref{S}),
at many separations $r$ and several FKP constants $\mu$.
Calculating $\hat S(r;\mu)$,
commonly denoted
$\langle D D \rangle - 2 \langle D R \rangle + \langle R R \rangle$,
requires actual data from a survey.

\skipp
{\bf Step 7}.
From $\hat S(r;\mu_\alpha)$,
compute the estimate $\hat Z^\alpha$, equation~(\ref{Zmvnicek}),
in the representation of eigenfunctions $\phi_\alpha$,
as described in \S\ref{compZ}.

\skipp
{\bf Step 8}.
Same as Step~4 of the gourmet method:
transform $\hat Z^\alpha$ into the prewhitened power $\hat X_\alpha$
using equations~(\ref{Z}), (\ref{Y}), and (\ref{Xrhat}),
as stated in \S\ref{convert}.
The transformations may be done in whatever representation
proves most convenient or numerically reliable.
Ultimately,
one wants the prewhitened power spectrum $\hat X(k)$ in Fourier space.

\skipp
{\bf Step 9}.
Same as Step~5 of the gourmet method:
decorrelate the prewhitened power spectrum $\hat X$
into a set of uncorrelated band-powers $\hat B = W \hat X$,
using the decorrelation matrices $W$ computed in Step~5 above.

\subsection{Fastfood}
\label{fast}

For some purposes a simplified, approximate version of the procedure
in \S\ref{fine} may be considered adequate.

The basic simplifying approximation here is that the covariance
$\langle \Delta \hat X(k_\alpha) \Delta \hat X(k_\beta) \rangle$
of the prewhitened power spectrum
may be considered to be diagonal in Fourier space
without further refinement.
The procedure then becomes the same as the FKP procedure
for Gaussian fluctuations,
with the differences that
(a) it is the prewhitened power spectrum $\hat X(k)$, equation~(\ref{Xr}),
rather than the power spectrum $\hat \xi(k)$ that is being estimated;
and
(b) the FKP constants $\mu(k)$ in the FKP pair-weightings
are modified from the Gaussian case where $\mu(k) = \xi(k)$.

\muefffig

Figure~\ref{mueff} shows the ratio $\mu(k)/\xi(k)$
of the effective FKP constant $\mu(k)$ to the nonlinear power spectrum $\xi(k)$
for several different power spectra.
The ratios plotted in Figure~\ref{mueff}
should be regarded as indicative rather than definitive,
because they depend on the validity of the hierarchical model
with constant amplitudes $R_b = - R_a$
(as required to satisfy the Schwarz inequality, \S\ref{hierarchical}),
which as discussed in \S\ref{4point} is certainly wrong at some level.

The effective FKP constant $\mu(k)$ shown in Figure~\ref{mueff}
is {\em not\/} the same thing as the eigenvalue $\mu_\alpha$
of $M$ at the nominal wavenumber $k_\alpha$ shown in Figure~\ref{diag}.
As discussed in \S\ref{eigenvalues},
the correspondence between eigenvalue $\mu_\alpha$ and nominal
wavenumber $k_\alpha$ is precise only for Gaussian fluctuations.

The effective FKP constant $\mu(k)$ in Figure~\ref{mueff} was calculated
by going through Steps~2--5 of the recipe in \S\ref{fine}
for the case of a perfect, noiseless ($\nbar \rightarrow \infty$) survey
of (large) volume $V$.
In this case
the Fisher matrix $E_{\alpha\beta}$
of the prewhitened power, equation~(\ref{Emvnice}), reduces to
\be
  E_{\alpha\beta} = {V \over 2} \, M^{-1}_{\alpha\beta}
\ee
whose inverse gives the covariance of the prewhitened power $\hat X$
\be
\label{XaXbm}
  \langle \Delta\hat X_\alpha \Delta\hat X_\beta \rangle
  = {2 \over V} \, M_{\alpha\beta}
  \ .
\ee
Decorrelating this covariance, equation~(\ref{XaXbm}), in Fourier space,
as described by in Paper~4,
yields band-powers $\hat B(k)$ whose covariance is by
construction diagonal.
The diagonal values of the diagonal covariance matrix of the band-powers
can be taken to define the effective FKP constants $\mu(k)$
\be
  \langle \Delta\hat B(k_\alpha) \Delta\hat B(k_\beta) \rangle
  = \1_{\alpha\beta} \, {2 \, \mu(k_\alpha)^2 \over V}
  \ .
\ee

With the effective FKP constants $\mu(k)$ taken as given by Figure~\ref{mueff},
the shortcut recipe is then as follows.

\skipp
{\bf Step 1}.
Compute the effective spatial volume
$V(k)$
of the survey for modes at wavenumber $k$
\be
\label{V}
  V(k)
    = 2 \, \mu(k)^2 R[0;\mu(k)]
    = \int {\ddd r \over [1 + \nbar(\r)^{-1} \mu(k)^{-1}]^2}
  \ .
\ee

\skipp
{\bf Step 2}.
The Fisher matrix~(\ref{Emvnice}) of the prewhitened power reduces to
\be
\label{Ek}
  E(k_\alpha,k_\beta)
  = \frac{1}{2} \int_0^\infty \!\!
    j_0(k_\alpha r) j_0(k_\beta r) R[r;\mu(k_\alpha)]
    \, 4 \PI r^2 \dd r
  \ .
\ee
If the prewhitened power is averaged over sufficiently broad shells
in $k$-space, then, by arguments similar to those in \S\ref{coarse},
the Fisher matrix is approximately diagonal
(compare eq.~[\ref{Eapprox}])
\be
\label{Ekapprox}
  E(k_\alpha,k_\beta) \approx
    (2\PI)^3 \deltaD(k_\alpha-k_\beta) \,
    {V(k_\alpha) \over 2 \, \mu(k_\alpha)^2}
\ee
where $V(k_\alpha)$ is the effective volume given by equation~(\ref{V}).
For the approximation~(\ref{Ekapprox}) to be valid,
the shells in $k$-space must be broad not
only compared to the inverse scale of the survey
(as in the Gaussian case),
but also compared to the width of the 4-point matrix $M$
plotted in Figures~\ref{Mw} and \ref{Ms}.
In the large $k$, hierarchical limit,
the width of the matrix $M$ in $k$-space is comparable to
an inverse correlation length, $\Delta k \sim \PI/r_0$.

\skipp
{\bf Step 3}.
The covariance
$\langle \Delta\hat X(k_\alpha) \Delta\hat X(k_\beta) \rangle$
of the prewhitened power equals the inverse of the Fisher matrix
$E(k_\alpha,k_\beta)$ given by equation~(\ref{Ekapprox})
\be
  \langle \Delta\hat X(k_\alpha) \Delta\hat X(k_\beta) \rangle
  \approx  
    (2\PI)^3 \deltaD(k_\alpha-k_\beta) \,
    {2 \, \mu(k_\alpha)^2 \over V(k_\alpha)}
  \ .
\ee
Define prewhitened band-powers $\hat B(k)$ to be the prewhitened
power spectrum $\hat X(k)$
averaged over broad (as in Step~2) shells of volume $V_k$ about $k$
\be
\label{tildeXk}
  \hat B(k) \equiv V_k^{-1} \int \hat\xi(k) \, \dd V_k
\ee
where $\dd V_k \equiv 4\PI k^2 \dd k/(2\PI)^3$.
The variance of the shell-averaged prewhitened band-powers is
\be
  \langle \Delta\hat B(k)^2 \rangle
  \approx {2 \, \mu(k_\alpha)^2 \over V(k) \, V_k}
\ee
which is $2 \, \mu(k_\alpha)^2$ divided by the effective phase volume,
the product of the effective spatial volume $V(k)$, equation~(\ref{V}),
with the Fourier volume $V_k$ of the shell in $k$-space.

\skipp
{\bf Step 4}.
Same as Step~6 of \S\ref{fine}:
compute FKP-weighted pair densities $\hat S(r;\mu)$,
equation~(\ref{S}).

\skipp
{\bf Step 5}.
Compute the estimator $\hat Z(k)$
\be
\label{Zk}
  \hat Z(k) =
    \int_0^\infty
    {j_0(k r) \hat S[r;\mu(k)] \over [1+\xi(r)]^{1/2}}
    \, 4\PI r^2 \dd r
\ee
which may be compared to equation~(\ref{Zkm}).

\skipp
{\bf Step 6}.
Same as Step~8 of \S\ref{fine}:
transform $\hat Z(k)$ to $\hat X(k)$
using equations~(\ref{Z}), (\ref{Y}), and (\ref{Xrhat}).
The estimator $\hat Y = E^{-1} Z$ is
\be
\label{Yk}
  \hat Y(k) = {2 \, \mu(k)^2 \over V(k)} \, \hat Z(k)
  \ .
\ee

\skipp
{\bf Step 7}.
Form prewhitented band powers $\hat B(k)$
by averaging $\hat X(k)$ over sufficiently broad shells in $k$-space,
equation~(\ref{tildeXk}).

\section{Conclusions}
\label{conclusions}

The main finding of this paper is that
the prewhitened nonlinear power spectrum $X_\alpha$
defined by equation~(\ref{Xr}) has surprisingly sweet properties.

Firstly,
the covariance of the prewhitened nonlinear power is substantially narrower
in Fourier space than the covariance of the nonlinear power spectrum itself,
Figures~\ref{M}--\ref{Ms}.

Secondly,
in the FKP approximation,
the 4-point and 3-point contributions $M$ and $L$
to the covariance of prewhitened power are almost simultaneously diagonal
(the 2-point contribution is by construction the unit matrix,
so is automatically diagonal),
Figures~\ref{Mw3} and \ref{Mw3s}.
Thus the eigenmodes of the covariance of prewhitened nonlinear power
form a set of almost uncorrelated modes
somewhat analogous to the Fourier modes of power in the Gaussian case.

Thirdly,
the eigenvalues $\mu_\alpha$ and $\lambda_\alpha$,
as defined by equations~(\ref{mu}) and (\ref{lambda}),
of the 4-point and 3-point prewhitened matrices $M$ and $L$
are almost equal, $\mu_\alpha \approx \lambda_\alpha$,
Figure~\ref{diag34},
which is similar to the Gaussian case where $\mu(k) = \lambda(k) = \xi(k)$.

The second and third points above together make it possible
to construct a near-minimum variance estimator, \S\ref{power},
and Fisher matrix, \S\ref{fisher}, of the prewhitened nonlinear power spectrum
similar to the FKP estimator and Fisher matrix of the linear power spectrum
in the Gaussian case.

Fourthly,
all the above properties hold for all power spectra tested,
including power law nonlinear power spectra $\xi(k) \propto k^n$
with indices $-2 < n < 0$ over the full range allowed by the hierarchical model,
and including realistic power spectra,
such as the observationally derived power spectrum of Peacock (1997),
and an observationally concordant $\Lambda$CDM model of Eisenstein \& Hu (1998),
nonlinearly evolved according to the Peacock \& Dodds (1996) formula.

Fifthly,
in the realistic cases of the Peacock (1997)
and Eisenstein \& Hu (1998) power spectra,
the prewhitened nonlinear power spectrum $X(k)$
appears to be curiously close to the linear power spectrum $\xi_L(k)$,
Figure~\ref{xik}.

This having been said,
it should be emphasized that the above properties
are all premised on the hierarchical model with constant
hierarchical amplitudes, \S\ref{hierarchical},
which as discussed in \S\ref{4point} and by Scoccimarro et al.\ (1999 \S3.3)
is certainly wrong at some level.
Clearly it will be important to test how well these results stand up
in $N$-body simulations.

In the meantime, the results of this paper raise questions.
Is there some physical reason underlying the seemingly unreasonably
pretty properties of the prewhitened nonlinear power spectrum?
In general, modes may be statistically uncorrelated
without being dynamically independent.
But the fact that the covariance of the prewhitened power is narrow
for all power spectra is suggestive:
do the eigenmodes of the covariance of prewhitened power
somehow encode the information in the linear power spectrum
that is ravelled by nonlinear evolution in the power spectrum itself?
And is there somehow a connection to the mapping between
linear and nonlinear power spectra found by Hamilton et al.\ (1991)?

I conclude with a repeat of the warning
that this paper has ignored redshift distortions, light-to-mass bias,
and evolution,
and it has assumed that the only sources of variance are cosmic variance
and shot-noise variance arising from Poisson sampling of galaxies.
In real galaxy surveys, all these problems must be grappled with.

\section*{Acknowledgements}

This work was supported by
NASA Astrophysical Theory Grant NAG~5-7128.
I thank Max Tegmark for much detailed comment on the original manuscript.
Discussions with participants at the 1997 Aspen summer workshop on
Precision Measurement of Large Scale Structure
informed the content of this paper.

\section*{APPENDIX A: Justification of Equation~(\ref{Cab})}

Equation~(\ref{Cab}) is the FKP approximation,
expressed in concise mathematical form.
This appendix offers further details justifying this equation.

The pair-pair covariance matrix $\fC_{ijkl}$ can be regarded as an operator
that acts on pair-functions $\Psi_{kl}$.
It is helpful to think of $\Psi_{kl}$ as a 2-particle wavefunction
(symmetric under pair exchange $k \leftrightarrow l$),
and $\fC_{ijkl}$ as a Hermitian operator that acts on the space of
such wavefunctions.
The pair-wavefunctions $\Psi_{kl}$ of interest in the present case
have translation and rotation symmetry,
which means that they have zero total momentum and zero total angular momentum.
In the Fourier representation such wavefunctions $\Psi_{kl}$
can be expressed in the form
\be
\label{psik}
  \Psi(\k_k,\k_l) = (2\PI)^3 \deltaD(\k_k + \k_l) \, \psi(k_k)
\ee
where $\psi(k_k)$ is a function of the scalar $k_k \equiv |\k_k|$.
In a general representation, equation~(\ref{psik}) is
\be
\label{psi}
  \Psi_{kl} = D_{kl}^\alpha \psi_\alpha
\ee
where $D_{kl}^\alpha$ is the operator introduced in \S\ref{gaussian},
equations~(\ref{Dr}) and (\ref{Dk}).

In the FKP approximation,
the selection functions $\nbar_i$ and $\nbar_j$ upon which the
pair-covariance $\fC_{ijkl}$ depends, equations~(\ref{Cquad}) and (\ref{N}),
are taken to be locally constant,
so that $\fC_{ijkl}$ also has translation and rotation symmetry,
i.e.\ it commutes with the operators of total momentum and total angular
momentum.
Thus in the FKP approximation,
$\fC_{ijkl}$ acting on a wavefunction $D_{kl}^\alpha \psi_\alpha$
with zero momentum and angular momentum yields another wavefunction
$D_{ij}^\beta \chi_\beta$
with zero momentum and angular momentum
\be
\label{Cpsi}
  \fC_{ij}^{\ \ kl} D_{kl}^\alpha \psi_\alpha
  = D_{ij}^\beta \chi_\beta
  \ .
\ee
Take $\psi_\alpha$ in equation~(\ref{Cpsi}) to be the elements of a
complete orthonormal basis of functions.
Then equation~(\ref{Cpsi}) implies that
\be
\label{Cabp}
  \fC_{ijkl} D^{kl}_\alpha
  = D_{ij}^\beta \fC_{\beta\alpha}
\ee
for some matrix $\fC_{\beta\alpha}$.
Equation~(\ref{Cabp}) is the desired equation~(\ref{Cab})
that was to be justified
(at least if the indices on $\fC_{\alpha\beta}$ are swapped
in eq.~[\ref{Cabp}], which is fine because $\fC_{\alpha\beta}$ is symmetric,
as proven below).
The wavefunctions $\psi_\alpha$ and $\chi_\beta$
in equation~(\ref{Cpsi}) are related by
\be
  \fC_\beta^\alpha \psi_\alpha = \chi_\beta
  \ .
\ee

A wavefunction of the form $D_{ij}^\alpha \psi_\alpha$
is unnormalized -- that is,
$\psi^\alpha D^{ij}_\alpha D_{ij}^\beta \psi_\beta$ diverges --
as is usual in quantum mechanics for a wavefunction that has
definite momentum, and must therefore be defined over infinite space.
The divergence can be tamed by regarding the wavefunction as being
defined instead over an extremely large but finite volume $V$.
Then
\be
\label{DD}
  D^{ij}_\alpha D_{ij}^\beta = \1_{\alpha\beta} \, V
\ee
which is most easily proven from the real space representation of
$D_{ij}^\alpha$, equation~(\ref{Dr}).
Equation~(\ref{DD}) should be interpreted with due care.
For example, equation~(\ref{DD}) should not be substituted into
equation~(\ref{FmvFKPasym}) for the Fisher matrix in the FKP approximation,
because the matrix $\fC^{-1 \alpha\gamma}(\nbar_i,\nbar_j)$
on the right hand side of equation~(\ref{FmvFKPasym})
varies with positions $i$ and $j$.

Operating on equation~(\ref{Cabp}) with $D^{ij}_\beta$ implies,
from equation~(\ref{DD}), that
($\nbar_i$ and $\nbar_j$ here are being regarded formally
as fixed constants in the huge volume $V$)
\be
\label{DCD}
  D^{ij}_\beta \fC_{ijkl} D^{kl}_\alpha
  = V \, \fC_{\beta\alpha}
  \ .
\ee
It is evident from this equation
that the pair exchange symmetry $ij \leftrightarrow kl$ of $\fC_{ijkl}$
implies that $\fC_{\alpha\beta}$ is similarly symmetric
\be
  \fC_{\alpha\beta} = \fC_{\beta\alpha}
  \ .
\ee
Equation~(\ref{DCD}) shows that,
modulo the normalization factor $V$,
the reduced matrix $\fC_{\alpha\beta}$ can be regarded as
the matrix elements of the operator $\fC_{ijkl}$
restricted to the class of wavefunctions that have zero total momentum
and zero total angular momentum.

\section*{APPENDIX B: FFTLog}

\subsection*{B.1 \ Introduction}

FFTLog computes the fast Fourier or Hankel (= Fourier-Bessel) transform
of a periodic sequence of logarithmically spaced points. 
FFTLog can be regarded as a natural analogue to the standard FFT,
in the sense that,
just as the normal FFT gives the exact (to machine precision)
Fourier transform of a linearly spaced periodic sequence,
so also FFTLog gives the exact Fourier or Hankel transform,
of arbitrary order $\mu$,
of a logarithmically spaced periodic sequence.
FFTLog shares with the normal FFT the problems of
ringing (response to sudden steps)
and aliasing (periodic folding of frequencies),
but under appropriate circumstances FFTLog may approximate
the results of a continuous Fourier or Hankel transform.

The FFTLog algorithm was originally proposed by Talman (1978).
However, it seems worthwhile here to present the algorithm in some detail.

The FFTLog code may be downloaded from
http:\discretionary{}{}{}$/\!/\discretionary{}{}{}$casa\discretionary{}{}{}.colorado\discretionary{}{}{}.edu\discretionary{}{}{}/$\sim$ajsh\discretionary{}{}{}/FFTLog/\ .

Consider the continuous Hankel (= Fourier-Bessel) transform pair
\be
\label{Hkq}
  \tilde a(k) = \int_0^\infty a(r) \, (k r)^q J_\mu(k r) \, k \, \dd r
\ee
\be
\label{Hrq}
  a(r) = \int_0^\infty \tilde a(k) \, (k r)^{-q} J_\mu(k r) \, r \, \dd k
  \ .
\ee
If the substitution
\be
\label{aA}
  a(r) = A(r) \, r^{-q}
  \quad \mbox{and} \quad
  \tilde a(k) = \tilde A(k) \, k^q
\ee
is made, then the Hankel transform pair~(\ref{Hkq}), (\ref{Hrq}),
becomes equivalent to the transform pair
\be
\label{Hk}
  \tilde A(k) = \int_0^\infty A(r) \, J_\mu(k r) \, k \, \dd r
\ee
\be
\label{Hr}
  A(r) = \int_0^\infty \tilde A(k) \, J_\mu(k r) \, r \, \dd k
  \ .
\ee
Although the Hankel transform~(\ref{Hkq}) with a power law bias
$(k r)^{\pm q}$
is thus equivalent in the continuous case
to the unbiased Hankel transform~(\ref{Hk}),
the transforms are different when they are discretized and made periodic;
for if $a(r)$ is periodic, then $A(r) = a(r) \, r^q$ is not periodic.
FFTLog evaluates discrete Hankel transforms~(\ref{Hkq}) and (\ref{Hrq})
with arbitrary power law bias.

Fourier sine and cosine transforms can be regarded as special
cases of Hankel transforms with $\mu = \pm 1/2$, since
\be
  J_{1/2}(x) = (2/\PI x)^{1/2} \sin(x)
\ee
\be
  J_{-1/2}(x) = (2/\PI x)^{1/2} \cos(x)
  \ .
\ee

As first noted by Siegman (1977),
if the product $kr$ in the Hankel transform is written as
$e^{\ln k + \ln r}$,
then the transform becomes a convolution integral in the
integration variable $\ln r$ or $\ln k$.
Convolution is equivalent to multiplication
in the corresponding Fourier transform space.
Thus the Hankel transform can be computed numerically
by the algorithm:
FFT $\rightarrow$ multiply by a function $\rightarrow$ FFT back.
This is the idea behind
a number of Fast Hankel Transform algorithms
(Candel 1981; Anderson 1982; Hansen 1985; Fanning 1996)
including FFTLog (Talman 1978).

An advantage of FFTLog, emphasized by Talman (1978), is that
the order $\mu$ of the Bessel function may be any arbitrary real number.
In particular,
FFTLog works for $1/2$-integral $\mu$,
so includes the cases of Fourier sine and cosine transforms,
and spherical Hankel transforms
involving the spherical Bessel functions
$j_\lambda(x) \equiv (\PI/2 x)^{1/2} J_{\lambda+{1/2}}(x)$.
%

\subsection*{B.2 \ Normal discrete Fourier transform}

First,
recall the essential properties of the standard discrete Fourier transform
of a periodic sequence of linearly spaced points.
Suppose that $a(r)$ is a continuous, in general complex-valued, function
that is periodic with period $R$,
\be
  a(r + R) = a(r)
  \ .
\ee
Without loss of generality,
take the fundamental interval to be $[ -R/2, R/2 ]$,
centred at zero.
Since $a(r)$ is periodic,
its continuous Fourier transform contains only discrete Fourier modes
$e^{2\PI \im m r/R}$
with integral wavenumbers $m$.
Suppose further that the function $a(r)$
is `smooth' in the specific sense that it is
some linear combination only of the $N$ lowest frequency Fourier modes,
$m = 0$, $\pm 1$, $...$, $\pm [N/2]$,
where $[N/2]$ denotes the largest integer greater than or equal to $N/2$,
\be
\label{ar}
  a(r) = {\sum_m}' c_m \, \e^{2\PI \im m r/R}
\ee
the outermost Fourier coefficients being equal,
$c_{-N/2} = c_{N/2}$,
in the case of even $N$.
The primed sum in equation~(\ref{ar})
signifies a sum over integral $m$ from $-[N/2]$ to $[N/2]$,
with the proviso that for even $N$ the outermost elements of the sum
receive only half weight:
\be
  {\sum_n}' x_n \equiv \sum_{n=-[N/2]}^{[N/2]} w_n x_n
\ee
with $w_n$ = 1 except that $w_{-N/2} = w_{N/2} = 1/2$ if $N$ is even.

The sampling theorem (e.g.\ Press et al.\ 1986 \S12.1)
asserts that,
given a function $a(r)$ satisfying equation~(\ref{ar}),
the Fourier coefficients $c_m$
can be expressed in terms of the values $a_n \equiv a(r_n)$
of the function $a(r)$
at the $N$ discrete points $r_n = n R / N$
for $n = 0$, $\pm 1$, $...$, $\pm [N/2]$.
For even $N$, the periodicity of $a(r)$ ensures that
$a_{-N/2} = a_{N/2}$.
Specifically, the sampling theorem asserts that
the Fourier coefficients in the expansion~(\ref{ar}) satisfy
\be
\label{cm}
  c_m = \frac{1}{N} {\sum_n}' a_n \, \e^{- 2\PI \im m n / N}
\ee
the discrete points $a_n$ themselves satisfying
\be
\label{an}
  a_n = {\sum_m}' c_m \, \e^{2\PI \im m n / N}
\ee
in accordance with equation~(\ref{ar}).

Equations~(\ref{cm}) and (\ref{an})
constitute a discrete Fourier transform pair
relating two periodic, linearly spaced sequences $a_n$ and $c_m$
of length $N$.
The standard FFT evaluates the discrete Fourier transform exactly
(that is, to machine precision).

\subsection*{B.3 \ Discrete Hankel transform}

Now suppose that the function $a(r)$,
instead of being periodic in ordinary space $r$,
is periodic in logarithmic space $\ln r$, with logarithmic period $L$,
\be
  a(r \e^L) = a(r)
  \ .
\ee
Take the fundamental interval to be
$[\ln r_0 - L/2,$\discretionary{}{}{}$\, \ln r_0 + L/2]$,
centred at $\ln r_0$.
As in \S B.2,
the periodicity of $a(r)$ implies that its Fourier transform
with respect to $\ln r$ contains only discrete Fourier modes
$e^{2\PI \im m \ln(r/r_0)/L}$
with integral wavenumbers $m$.
Suppose further, as in \S B.2 eq.~(\ref{ar}),
that $a(r)$ contains only the $N$ lowest frequency Fourier modes
\be
\label{arl}
  a(r) =
    {\sum_m}'
    c_m \,
    \e^{2\PI \im m \ln(r/r_0)/L}
\ee
with $c_{-N/2} = c_{N/2}$ for even $N$.
The sampling theorem asserts that the Fourier coefficients $c_m$ are given by
\be
\label{cml}
  c_m =
    \frac{1}{N} {\sum_n}'
    a_n \,
    \e^{- 2\PI \im m n/N}
\ee
where $a_n \equiv a(r_n)$ are the values
of the function $a(r)$
at the $N$ discrete points $r_n = r_0 \e^{n L / N}$
for $n = 0$, $\pm 1$, $...$, $\pm [N/2]$,
\be
\label{anl}
  a_n =
    {\sum_m}'
    c_m \,
    \e^{2\PI \im m n /N}
  \ .
\ee

The continuous Hankel transform $\tilde a(k)$, equation~(\ref{Hkq}),
of a function $a(r)$ of the form~(\ref{arl}) is
\be
\label{akl}
  \tilde a(k) = {\sum_m}'
    c_m
    \int_0^\infty \!
    \e^{2\PI \im m \ln(r/r_0)/L} \,
    (k r)^q J_\mu(k r) \, k \, \dd r
  \ .
\ee
The integrals on the right hand side of equation~(\ref{akl})
can be done analytically, in terms of
\be
\label{U}
  U_\mu(x) \equiv
  \int_0^\infty \! t^x J_\mu(t) \, \dd t
  =
  2^x \, {\Gamma[(\mu+1+x)/2] \over \Gamma[(\mu+1-x)/2]}
\ee
where $\Gamma(z)$ is the usual Gamma-function.
Thus equation~(\ref{akl}) reduces to
\be
\label{ak}
  \tilde a(k) = {\sum_m}'
    c_m u_m \,
    \e^{- 2\PI \im m \ln(k/k_0)/L}
\ee
where $u_m$ is
\be
\label{um}
  u_m(\mu,q) \equiv
    ( k_0 r_0 )^{- 2\PI \im m / L} \,
    U_\mu \left( q + {2\PI \im m \over L} \right)
  \ .
\ee
Notice that $u_m^\ast = u_{-m}$,
which ensures that $\tilde a(k)$ is real if $a(r)$ is real.
Equation~(\ref{ak}) gives the (exact) continuous Hankel transform $\tilde a(k)$
of a function $a(r)$ of the form~(\ref{ar}).
Like $a(r)$, the Hankel transform $\tilde a(k)$ is periodic in logarithmic
space $\ln k$, with period $L$.
The fundamental interval is
$[\ln k_0 - L/2,$\discretionary{}{}{}$\ln k_0 + L/2]$,
centred at $\ln k_0$, which may be chosen arbitrarily
(but see \S B.5 below).

The sampling theorem requires that $u_{-N/2} = u_{N/2}$ for even $N$,
which is not necessarily satisfied by equation~(\ref{um}).
However,
at the discrete points $k_n = k_0 \e^{n L / N}$
considered by the sampling theorem,
the contributions at $m = \pm N/2$ to the sum on the right hand side
of equation~(\ref{ak})
are $(-)^n c_{N/2} ( u_{N/2} + u_{N/2}^\ast )/2$,
whose imaginary part cancels out.
Thus the equality~(\ref{ak}) remains true
at the discrete points $k_n$
if $u_{\pm N/2}$ are replaced by their real parts,
\be
\label{ure}
  u_{\pm N/2} \rightarrow {\rmn Re} \, u_{N/2}
  \ .
\ee
With the replacement~(\ref{ure}),
the sampling theorem asserts that the coefficients
$c_m u_m$ in the sum~(\ref{ak}) are determined by the values
$\tilde a_n \equiv \tilde a(k_n)$ of the Hankel transform
at the $N$ discrete points $k_n = k_0 \e^{n L / N}$
for $n = 0$, $\pm 1$, $...$, $\pm [N/2]$
\be
\label{cmum}
  c_m u_m = \frac{1}{N} {\sum_n}'
    \tilde a_n \,
    \e^{2\PI \im m n /N}
\ee
\be
\label{hanl}
  \tilde a_n =
    {\sum_m}'
    c_m u_m \,
    \e^{- 2\PI \im m n /N}
  \ .
\ee

Putting together equations~(\ref{cml}), (\ref{anl}),
(\ref{cmum}) and (\ref{hanl})
yields the discrete Hankel transform pair
\be
\label{hn}
  \tilde a_n = {\sum_m}' a_m \, v^+_{m+n}(\mu,q)
\ee
\be
\label{hm}
  a_m = {\sum_n}' \tilde a_n \, v^-_{m+n}(\mu,q)
\ee
in which the forward discrete Hankel mode $v^+_n(\mu,q)$
is the discrete Fourier transform of $u_m(\mu,q)$ given by
equations~(\ref{um}) and (\ref{ure}),
\be
\label{vpn}
  v^+_n(\mu,q)
  = \frac{1}{N} {\sum_m}' u_m(\mu,q) \, \e^{- 2\PI \im m n / N}
\ee
while the inverse discrete Hankel mode $v^-_n(\mu,q)$ is the discrete
Fourier transform of the reciprocal $1/u_{-m}(\mu,q)$,
\be
\label{vmn}
  v^-_n(\mu,q)
  = \frac{1}{N} {\sum_m}' {1 \over u_{-m}(\mu,q)} \, \e^{- 2\PI \im m n / N}
  \ .
\ee
The Hankel transform matrices $v^+_{m+n}(\mu,q)$ and $v^-_{m+n}(\mu,q)$
are mutually inverse
\be
\label{vv}
  {\sum_l}' v^+_{m+l}(\mu,q) \, v^-_{l+n}(\mu,q) = \delta_{mn}
\ee
where $\delta_{mn}$ denotes the Kronecker delta.
The forward and inverse Hankel modes have the interesting property of being
self-similar;
that is, Hankel modes $v^+_{m+n}(\mu,q)$ [or $v^-_{m+n}(\mu,q)$]
with different indices $m$ consist of the same periodic sequence
$v^+_n(\mu,q)$ [or $v^-_n(\mu,q)$]
cyclically shifted by $m$ notches.

FFTLog evaluates the forward and inverse discrete Hankel transforms
given by equations~(\ref{hn}), (\ref{hm}),
exactly (to machine precision).

The reciprocal $1/u_{-m}(\mu,q)$ in equation~(\ref{vmn})
is equal to $u_m(\mu,-q)$, according to equations~(\ref{U}) and (\ref{um}),
\be
\label{ui}
  {1 \over u_{-m}(\mu,q)} = u_m(\mu,-q)
  \qquad (m \neq N/2)
\ee
{\em except\/} in the case $m = \pm N/2$ for even $N$,
when the replacement~(\ref{ure}) generally invalidates equation~(\ref{ui}).
However, in the special case where $u_{\pm N/2}$ are already real,
then equation~(\ref{ure}) leaves $u_{\pm N/2}$ unchanged,
and equation~(\ref{ui}) remains valid also at $m = \pm N/2$.
This special case is of particular interest,
and is discussed further in \S B.5 below.

In the continuous case,
the inverse Hankel transform is equal to the forward transform
with $q \rightarrow -q$, equations~(\ref{Hkq}) and (\ref{Hrq}).
In the discrete case this remains true for odd $N$,
but it is not generally true for even $N$ (the usual choice)
except in the important special case discussed in \S B.5.

In the general discrete case
(i.e.\ if the condition~[\ref{uend}] in \S B.5 is not satisfied),
the inverse discrete Hankel mode $v^-_n(\mu,q)$, equation~(\ref{vmn}),
differs from the forward Hankel mode $v^+_n(\mu,-q)$, equation~(\ref{vpn}),
only for even $N$
and only in the coefficient of the highest frequency Fourier component,
$1 / u_{-m}(\mu,q)$ versus $u_{m}(\mu,-q)$ for $m = \pm N/2$.
To the extent that the highest frequency Fourier coefficient $c_{\pm N/2}$
of a sequence $a_n$ is small,
the difference between its inverse discrete Hankel transform
and its forward transform with $q \rightarrow -q$ should be small.

It is possible for the inverse discrete Hankel transform to be singular,
if $u_{\pm N/2}$ is purely imaginary, so that its real part vanishes,
making $v^-_n(\mu,q)$ singular.
As discussed in \S B.5,
this singularity can be avoided by choosing a low-ringing value of $k_0 r_0$,
equation~(\ref{k0r0}).

The forward (inverse) discrete Hankel transforms are also singular
at special values of $\mu$ and $q$,
namely where $\mu+1+q$ (or $\mu+1-q$ in the inverse case) vanishes,
because $u_0(\mu,q) = U_\mu(q)$ is singular at these points.
This singularity reflects a real singularity in the corresponding
continuous Hankel transform
(unlike the singularity of the previous paragraph,
which is an avoidable artefact of discreteness).
The singularity in $u_0$ leads to an additive infinite constant in the
discrete Hankel transform.
In physical problems this additive infinite constant may somehow cancel out
(for example, in the difference between two Hankel transforms).
FFTLog's strategy in these singular cases is to evaluate the discrete
Hankel transform with the infinite constant set to zero,
and to issue a warning.

\subsection*{B.4 \ FFTLog algorithm}

The FFTLog algorithm
for taking the discrete Hankel transform, equation~(\ref{hn}), of a
sequence $a_n$ of $N$ logarithmically spaced points is:
\begin{itemize}
\item
FFT $a_n$ to obtain the Fourier coefficients $c_m$, equation~(\ref{cml});
\item
multiply by $u_m$ given by equations~(\ref{um}) and (\ref{ure})
to obtain $c_m u_m$;
\item
FFT $c_m u_m$ back
to obtain the discrete Hankel transform $\tilde a_n$, equation~(\ref{hanl}).
\end{itemize}

A variant of the algorithm is to sandwich the above operations
with power law biasing and unbiasing operations.
For example, one way to take the unbiased continuous Hankel transform
$\tilde A(k)$ of a function $A(r)$, equation~(\ref{Hk}),
is to bias $A(r)$ and $\tilde A(k)$ with power laws, equation~(\ref{aA}),
and take a biased Hankel transform, equation~(\ref{Hkq}).
The discrete equivalent of this is:
\begin{itemize}
\item
Bias $A_n$ with a power law to obtain
$a_n = A_n r_n^{-q}$,
equation~(\ref{aA});
\item
FFT $a_n$ to obtain the Fourier coefficients $c_m$, equation~(\ref{cml});
\item
multiply by $u_m$ given by equations~(\ref{um}) and (\ref{ure})
to obtain $c_m u_m$;
\item
FFT $c_m u_m$ back
to obtain the discrete Hankel transform $\tilde a_n$, equation~(\ref{hanl});
\item
Unbias $\tilde a_n$ with a power law to obtain
$\tilde A_n = \tilde a_n k_n^{-q}$,
equation~(\ref{aA}).
\end{itemize}
Although in the continuous limit the result would be identical
to an unbiased Hankel transform,
in the discrete case the result differs.
With a simple unbiased discrete Hankel transform,
it is the sequence $A_n$ that is taken to be periodic,
whereas in the algorithm above
it is not $A_n$ but rather $a_n$ that is periodic.

The inverse discrete Hankel transform is accomplished by the same
series of steps, except that $c_m$ is divided instead of multiplied
by $u_m$.

The FFTLog code is built on top of
the NCAR suite of FFT routines
(Swarztrauber 1979),
and a modified version of an implementation of the complex Gamma-function
from the gamerf package by Ooura (1996).

FFTLog includes driver routines for the specific cases
of the Fourier sine and cosine transforms.

\subsection*{B.5 \ Low-ringing condition on $k_0 r_0$}

The central values $\ln r_0$ and $\ln k_0$
of the periodic intervals in $\ln r$ and $\ln k$
may be chosen arbitrarily.
However, ringing of the discrete Hankel transform may be reduced,
for either even or odd $N$,
if the product $k_0 r_0$ is chosen in such a way that
the boundary points of the sequence $u_m$, equation~(\ref{um}), are equal
\be
\label{uend}
  u_{-N/2} = u_{N/2}
  \ .
\ee
Recall that the general procedure, for even $N$,
was to replace $u_{\pm N/2}$ by their real part, equation~(\ref{ure}).
The condition~(\ref{uend}) requires that $u_{\pm N/2}$ are already real.
The condition~(\ref{uend}) reduces ringing because it makes
the periodic sequence $u_m$
fold smoothly across the period boundary at $m = \pm N/2$.

In addition to reducing ringing, the condition~(\ref{uend})
means that equation~(\ref{ui}) remains true also at $m = \pm N/2$,
so is true for all $m$.
In this case the inverse Hankel mode $v^-_n(\mu,q)$, equation~(\ref{vmn}), is
equal to the forward Hankel mode $v^+_n(\mu,-q)$ with $q$ of the opposite sign
\be
  v^-_n(\mu,q) = v^+_n(\mu,-q)
  = \frac{1}{N} {\sum_m}' u_m(\mu,-q) \, \e^{- 2\PI \im m n / N}
  \ .
\ee

In other words, if condition~(\ref{uend}) is satisfied,
then the inverse discrete Hankel transform equals the forward discrete
Hankel transform with $q \rightarrow -q$.
This is like the continuous Hankel transform,
equations~(\ref{Hkq}), (\ref{Hrq}),
where the inverse transform equals the forward transform with
$q \rightarrow -q$.

The periodicity condition~(\ref{uend}) on $u_{\pm N/2}$ translates,
for real $\mu$ and $q$,
into a condition on $k_0 r_0$
\be
\label{k0r0}
  \ln (k_0 r_0 ) =
    {L \over N} \left\{ {1 \over \PI} \,
      \Arg \left[ U_\mu \left( q + {\PI \im N \over L} \right) \right]
      + {\rmn integer}
    \right\}
\ee
where $\Arg z \equiv {\rmn Im} \ln z$
denotes the argument of a complex number,
and integer is any integer.
In other words, to reduce ringing,
it may help to choose $k_0 r_0$
so as to satisfy the condition~(\ref{k0r0}).
This is not too much of a restriction,
since $L/N$ is the logarithmic spacing between points (= one notch),
so the low-ringing condition~(\ref{k0r0}) allows $k_0 r_0$ to be chosen
to lie within half a notch [= $L/(2 N)$]
of whatever number one chooses,
for example within half a notch of $k_0 r_0 = 1$.

FFTLog can be set to use automatically the low-ringing value of $k_0 r_0$
nearest to any input value of $k_0 r_0$.

How else does the choice of $k_0 r_0$ affect the Hankel transform?
Increasing the value of $\ln(k_0 r_0)$ by one notch $L/N$
cyclically shifts the discrete Hankel transform $\tilde a_n$,
equation~(\ref{hanl}),
by one notch to the left, $\tilde a_n \rightarrow \tilde a_{n-1}$.
In other words,
changing $\ln(k_0 r_0)$ by an integral number of notches
shifts the origin of the transform,
but leaves the transform otherwise unchanged,
as might have been expected.

In practice, since in most cases one is probably using the discrete
Hankel transform as an approximation to the continuous transform,
one would probably want to use
$k_0 r_0 \approx 1$ (or $2$, or $\PI$, according to taste).

\subsection*{B.6 \ Unitary Hankel transform}

The discrete Hankel transform
with both low-ringing $k_0 r_0$
and no power law bias, $q = 0$,
is of particular interest because it is unitary,
like the Fourier transform.
Indeed, being also real,
the low-ringing unbiased Hankel transform is orthogonal, i.e.\ self-inverse,
like the Fourier sine and cosine transforms.
This is like the continuous unbiased ($q = 0$) Hankel transform,
equations~(\ref{Hkq}), (\ref{Hrq}),
which is self-inverse.

The discrete Hankel modes
$v_m(\mu,0) = v^+_m(\mu,0) = v^-_m(\mu,0)$
in the low-ringing unbiased ($q = 0$) case
are periodic, orthonormal, and self-similar,
equation~(\ref{vv}),
\be
\label{vv0}
  {\sum_l}' v_{m+l}(\mu,0) \, v_{l+n}(\mu,0) = \delta_{mn}
  \ .
\ee


Like any orthogonal transformation,
the low-ringing unbiased ($q = 0$) Hankel transform
commutes with the operations of matrix multiplication, inversion,
and diagonalization
(for non-low-ringing or biased Hankel transforms, $q \neq 0$,
the operations do not commute).
That is, the Hankel transform of the product of two matrices
is equal to the product of their Hankel transforms, and so on.

All else being equal (which it may not be),
given a choice between applying an unbiased ($q = 0$) or biased ($q \neq 0$)
Hankel transform,
and between a low-ringing $k_0 r_0$, equation~(\ref{k0r0}), or otherwise,
one would be inclined to choose the low-ringing unbiased transform,
because of its orthogonality property.

\xirfig

\subsection*{B.7 \ Example}

Figure~\ref{xir}
shows the correlation function $\xi(r)$
computed by FFTLog for the nonlinear $\Lambda$CDM power spectrum
of Eisenstein \& Hu (1998) shown in Figure~\ref{xik}.
Two different resolutions are plotted on top of each other,
a low resolution case
with 96 points over the range $r = 10^{-3}$ to $10^3 \, h^{-1} \Mpc$,
and a high resolution case
with 768 points over the range $r = 10^{-6}$ to $10^6 \, h^{-1} \Mpc$.
Both cases used an unbiased ($q = 0$) transform
and a low-ringing value of $k_0 r_0$
(actually the choice of $k_0 r_0$ made little difference here).

The low and high resolution correlation functions
shown in Figure~\ref{xir}
agree well except near the edges
$r \approx 10^{-3}$ and $10^{3} \, h^{-1} \Mpc$;
in particular, the low resolution correlation function tends to
a positive constant $\approx 10^{-5}$ at $r \rightarrow 10^{3} \, h^{-1} \Mpc$,
whereas the high resolution correlation function is negative
and declining as a power law $\propto r^{-4}$ at large $r$.
The disagreement is caused by aliasing
(see \S B.8)
of small and large separations in the low resolution case.
Aliasing is almost eliminated in the high resolution case
because the range $r = 10^{-6}$ to $10^6 \, h^{-1} \Mpc$
over which the transform was computed is much broader than the range plotted.

The bottom panel of Figure~\ref{xir}
shows the ratio $\xi_{\rmn FFT}/\xi_{\rmn FFTLog}$
of the correlation function $\xi_{\rmn FFT}$
computed with a normal FFT (sine transform) with 1023 points over the range
$r = 0.125$ to $128  \, h^{-1} \Mpc$,
to the (high resolution) correlation function $\xi_{\rmn FFTLog}$
computed with FFTLog.
Even with 1023 points,
the FFT'd correlation function rings noticeably,
with an amplitude of about $\pm 5$~percent.

In this particular instance,
FFTLog outperforms the normal FFT on all counts:
it is more accurate, with fewer points, over a larger range,
and it shows no signs of ringing.
This does not mean that FFTLog is always better than FFT.
Rather, FFTLog is well matched to the problem at hand:
the cosmological power spectrum extends over many orders of magnitude
in wavenumber $k$, and varies smoothly in $\ln k$.

\subsection*{B.8 \ Ringing and aliasing}

FFTLog suffers from the same problems of
ringing (response to sudden steps)
and aliasing (periodic folding of frequencies)
as the normal FFT.

Usually one is interested in the discrete Fourier or Hankel transform
not for its own sake, but rather as an approximation to the continuous
transform. 
The usual procedure would be to apply the discrete transform
to a finite segment of the function $a(r)$ to be transformed.
For FFTLog, the procedure can be regarded as involving two steps:
truncating the function to a finite logarithmic interval,
which causes ringing of the transform;
followed by periodic replication of the function in logarithmic space,
which causes aliasing.

Figure~\ref{fftlogexample}
illustrates these steps for the
unbiased ($q = 0$) Hankel transform,
equation~(\ref{Hkq}),
of order $\mu = -1/2$
of a function that is Gaussian in the log 
\be
  a(r) = \exp[-(\ln r)^2/2]
  \ . 
\ee

Truncation of the function $a(r)$
leads to ringing of its transform $\tilde a(k)$
at high frequencies $k$,
as seen in the middle right panel of Figure~\ref{fftlogexample}.
The oscillations at large $k$ are actually uniformly spaced in $k$,
but appear bunched up because of the logarithmic plotting.

Periodic replication means taking a sum of copies shifted by integral periods.
From the definition~(\ref{Hkq}) of the continuous Hankel transform,
it can be seen that
periodically replicating a function $a(r)$ in logarithmic space $\ln r$
and then taking its continuous Hankel transform
is equivalent to Hankel transforming the function $a(r)$
and then periodically replicating the transform $\tilde a(k)$ in $\ln k$.
But truncating a function does not truncate its transform.
So whereas a truncated, periodically replicated function $a(r)$
contains contributions from only one period at each point $r$,
the periodically replicated transform
contains overlapping contributions from many periods at each point $k$.
This is aliasing.
In Figure~\ref{fftlogexample} aliasing is visible
as an enhancement of the periodically replicated transform $\tilde a(k)$
on the high $k$ side of the periodic interval.

Ringing and aliasing can be reduced by taking suitable precautions.

The ringing that results from taking the discrete transform of a finite segment
of a function can be reduced by arranging
that the function folds smoothly from large to small scales.
It may help to bias the function with a power law before transforming it,
as in the second algorithm in \S B.4.
It may also help to use a low-ringing value of $k_0 r_0$, \S B.5.

Aliasing can be reduced by enlarging the periodic interval.
Aliasing can be eliminated (to machine precision) if the interval
can be enlarged to the point where the transform $\tilde a(k)$
goes sensibly to zero at the boundaries of the period.
Note that it is not sufficient to enlarge the interval
to the point where $a(r)$ is sensibly zero at the period boundaries:
what is important is that the transform $\tilde a(k)$ goes to zero
at the boundaries.

\fftlogexamplefig

\end{document}